\documentclass[pdflatex,sn-mathphys-num]{sn-jnl}

\usepackage{graphicx}%
\usepackage{multirow}%
\usepackage{amsmath,amssymb,amsfonts}%
\usepackage{amsthm}%
\usepackage{mathrsfs}%
\usepackage[title]{appendix}%
\usepackage{xcolor}%
\usepackage{textcomp}%
\usepackage{manyfoot}%
\usepackage{booktabs}%
\usepackage{algorithm}%
\usepackage{algorithmicx}%
\usepackage{algpseudocode}%
\usepackage{listings}%
\usepackage{subcaption}
\usepackage{graphicx}

\theoremstyle{thmstyleone}%
\theoremstyle{thmstyletwo}%

\theoremstyle{thmstylethree}%
\graphicspath{{./}{figures/}}

\newcommand\ion[2]{#1$\;${%
\ifx\@currsize\normalsize\small \else
\ifx\@currsize\small\footnotesize \else
\ifx\@currsize\footnotesize\scriptsize \else
\ifx\@currsize\scriptsize\tiny \else
\ifx\@currsize\large\normalsize \else
\ifx\@currsize\Large\large
\fi\fi\fi\fi\fi\fi}\rmfamily{#2}\relax}%

\begin{document}

\title[Life After the Quasar]{Life After the Quasar:
Overmassive Black Holes and Remnant Ionised Bubbles in and Around Two z$\sim$6.6 Galaxies}


\author*[1]{\fnm{Romain A.} \sur{Meyer}}\email{romain.meyer@unige.ch}

\author[1,2]{\fnm{Pascal A.} \sur{Oesch}}

\author[1]{\fnm{Callum} \sur{Witten}}

\author[3]{\fnm{Richard S.} \sur{Ellis}}

\author[4,5]{\fnm{Sarah E. I.} \sur{Bosman}}

\author[5]{\fnm{Fred} \sur{Davies}}

\author[6]{\fnm{Alyssa B.} \sur{Drake}}

\author[7]{\fnm{Nicolas} \sur{Laporte}}

\author[8]{\fnm{Jorryt} \sur{Matthee}}

\author[5]{\fnm{Fabian} \sur{Walter}}

\affil*[1]{\orgdiv{Department of Astronomy}, \orgname{University of Geneva}, \orgaddress{\street{Chemin Pegasi 51}, \city{Versoix}, \postcode{1290}, \state{Geneva}, \country{Switzerland}}}
\affil[2]{\orgdiv{Cosmic DAWN Center, Niels Bohr Institute, University of Copenhagen}, \orgaddress{\street{Jagtvej 128}, \city{K\o benhavn N}, \postcode{DK-2200}, \country{Denmark}}}

\affil[3]{\orgdiv{Department of Physics and Astronomy}, \orgname{University College London}, \orgaddress{\street{Gower Street}, \city{London}, \postcode{WC1E 6BT}, \country{UK}}}

\affil[4]{\orgdiv{Institute for Theoretical Physics} \orgname{Heidelberg University}, \orgaddress{\street{Philosophenweg 12}, \city{Heidelberg}, \postcode{D-69117},\country{Germany}}}

\affil[5]{\orgname{Max Planck Institut f\"ur Astronomie}, \orgaddress{\street{K\"onigstuhl 17}, \city{Heidelberg}, \postcode{D-69117},\country{Germany}}}

\affil[6]{\orgdiv{Centre for Astrophysics Research (CAR)}, \orgname{ University of Hertfordshire}, \orgaddress{\city{Hatfield}, \postcode{ AL10 9AB}, \country{UK}}}

\affil[7]{\orgdiv{LAM}, \orgname{Aix Marseille Universit\'e, CNRS, CNES}, \orgaddress{ \city{Marseille}, \country{France}}}

\affil[8]{ \orgname{Institute of Science and Technology Austria (ISTA)}, \orgaddress{ \street{Am Campus 1}, \city{Klosterneuburg},  \postcode{3400} , \country{Austria}}}


\abstract{Supermassive black holes (SMBH, $M_{\rm{BH}} > 10^8 M_\odot$) powering luminous quasars already exist one billion years after the Big Bang, yet their connection to their star-forming host galaxies, their relation to the general galaxy population and their contribution to Reionisation remains deeply enigmatic \cite{Fan2023,Volonteri2021}. JWST is finding numerous Active Galactic Nuclei (AGN) in high-redshift galaxies with black hole masses that appear to be over-massive compared to their host's stellar mass \cite{Harikane2023, Maiolino2024_AGNsample, Juodzbalis2026}, but rarely as massive as those found in luminous quasars. Here we report JWST/NIRSpec observations revealing overmassive SMBH in two ultra-luminous Lyman-$\alpha$ emitters at $z\sim6.6$ that exhibit rare double-peaked Lyman-alpha profiles \cite{Songaila2018, Matthee2018}. The broad Balmer lines indicate black hole masses $M_{\rm{BH}}\simeq 2\times10^8 M_\odot$, matching that found in faint $z\sim 6-7$ quasars, and very high BH-to-stellar-mass ratio ($\sim 0.1-0.2$) that exceed the local relation by a factor $\sim$400-800. Stellar population modelling favours young ages ($<50$ Myr), inconsistent with the sustained average Eddington-rate accretion required to reach the observed BH masses by $z=6.6$. The double-peak Lyman-$\alpha$ profiles require a large ionised bubble and high photoionisation rate that is consistent with the ionising output of quasars powered by black holes of similar mass, thus constraining the cessation of the last quasar episode to $<1$ Myr. We interpret both systems as post-quasar galaxies in which AGN feedback has delayed stellar mass assembly, and propose that episodic quasar activity partially explains the unexpected prevalence of large ionised bubbles deep into the Epoch of Reionisation.}

\keywords{Supermassive Black Holes, Quasars, Reionisation, Lyman-alpha Galaxies}

\maketitle

The rapid growth of supermassive black holes (SMBH) in the first billion years of the Universe is a persistent challenge for early models of black hole (BH) formation and growth. The discovery of luminous quasars at $z>6$ powered by $10^8-10^{10} \ M_\odot$ BH implies that they grew from massive seeds ($>10^3-10^5\ M_\odot$) at, or close to, the Eddington rate for most of their lifetime \cite{Volonteri2021,Schneider2023,Fan2023}. Observations in the far-infrared regime find them to be ``overmassive" with respect to their host galaxies \cite[e.g. 1/10-1/100,][]{Izumi2019,Pensabene2020,Neeleman2021} compared to the local relation \cite{KormendyHo2013, Reines2015}, suggesting BH grow faster than their host at early times. JWST is now discovering numerous AGN in $z\gtrsim 5$ galaxies thanks to its transformational access to their rest-optical spectrum. Broad line AGN are detected with JWST in $\sim 10-30\%$ of star-forming galaxies and often have elevated BH-to-stellar mass ratios \cite{Maiolino2024_AGNsample,Juodzbalis2026}, similar to luminous quasars. It is thus highly likely that JWST-detected AGN (or a subpopulation thereof) are the missing link between $10^8-10^{10}$ $M_\odot$ BH in $z>6$ quasars and heavy BH seeds accreting at or above Eddington from z$\sim 15$ onwards \cite{Schneider2023, Trinca2024}. Linking the two populations is key to studying the quasar duty cycle (e.g. the fraction of time the BH is accreting close to Eddington, determining whether it grew continuously or in short super-Eddington bursts), if and how the BH-stellar mass scaling relation changes at high-redshift \cite{Lauer2007,Pacucci2023,Volonteri2023}, and the potential impact of AGN on reionisation.

A potential clue to the AGN nature of luminous high-redshift galaxies is the detection of Lyman-$\alpha$ emission deep into the Reionisation era. At high redshift, Lyman-$\alpha$ emission is expected to be detected only in intrinsically strong equivalent width emitters (with a considerable velocity offset $\sim 1000-2000\ \rm{km\ s}^{-1}$ due to the IGM damping wing) or in objects located in large ionised bubbles \cite{Mason2020,Mason2026}. The latter are usually explained by the presence of large galaxy overdensities that collectively enhance the local photo-ionisation rate \cite{Saxena2024, Witstok2024}. However, for some puzzling objects the observational data fail to reveal large enough overdensities \cite{Tang2024}. Instead, hard ionisation fields due to AGN activity could help ionise the IGM locally and therefore boost the transmission of Lyman-$\alpha$ photons \cite{Tang2024, Witstok2025}. 

This paper aims to demonstrate the connection between Lyman-$\alpha$ visibility and episodic quasar activity at $z>6$ using new JWST/NIRSpec Integral Field Unit (IFU) G235M/G395M observations of two ultra luminous ($L_{Ly\alpha}\gtrsim10^{43.5}\ [\rm{erg}\ \rm{s}^{-1}]$) $z\sim 6.6$ Lyman-$\alpha$ emitters with rare double-peaked Lyman-$\alpha$ profiles. NEPLA4 and COLA1 were selected using HSC NB921 imaging (isolating strong Lyman-$\alpha$ $z\sim 6.6$), and later spectroscopically confirmed with Keck and VLT \cite{Hu2016,Songaila2018,Matthee2018}. The presence of a blue Lyman-$\alpha$ peak, assuming that the trough is at systemic redshift, puts lower limits on the minimum size ($R_p\gtrsim 1 \ \rm{pMpc}$) of the ionised region surrounding these sources \citep[e.g.][]{Songaila2018,Matthee2018,Mason2020,Meyer2021}. However, simulations of reionisation predict that double-peak profiles should be much rarer than the data suggest \cite{Gronke2021,Garel2021}, raising questions about the nature of the necessary ionising sources.  The ionising output of these two objects is however not sufficient to keep the surrounding IGM sufficiently transparent \citep{Meyer2021}, requiring instead a large overdensity of galaxies or an additional ionising source such as an AGN \cite{Padmanabhan2021}. Our new JWST data reveal that both COLA1 and NEPLA4 harbour $\sim 2\times10^8\ M_\odot$ black holes, clearly supporting the latter hypothesis with important implications for the quasar duty cycle and Lyman-$\alpha$ visibility at $z\sim6.6$.

\section*{Results}\label{results}

\begin{figure*}[h]
    \centering
    \includegraphics[width=0.8\linewidth, trim=0 0 0 0.8cm,clip]{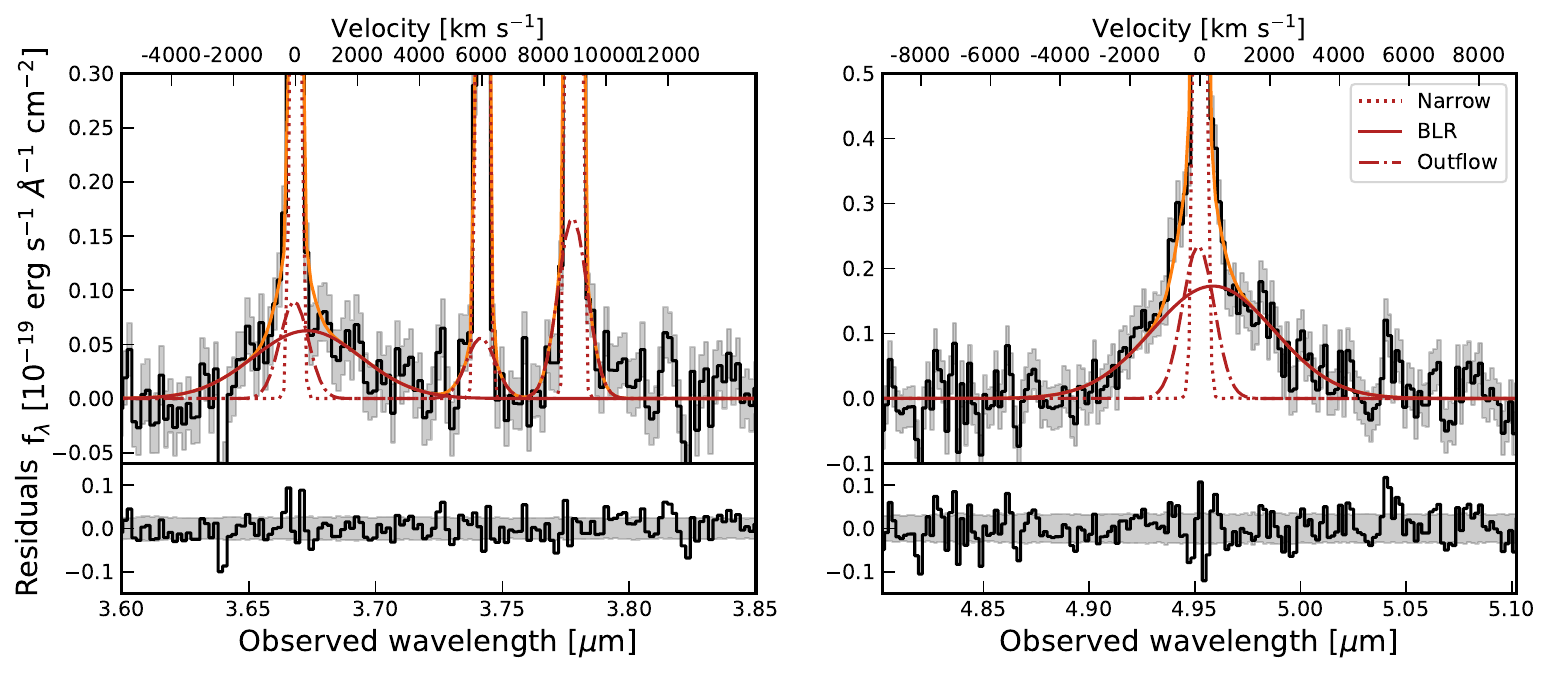}
    \includegraphics[width=0.8\linewidth, trim=0 0 0 0.8cm,clip]{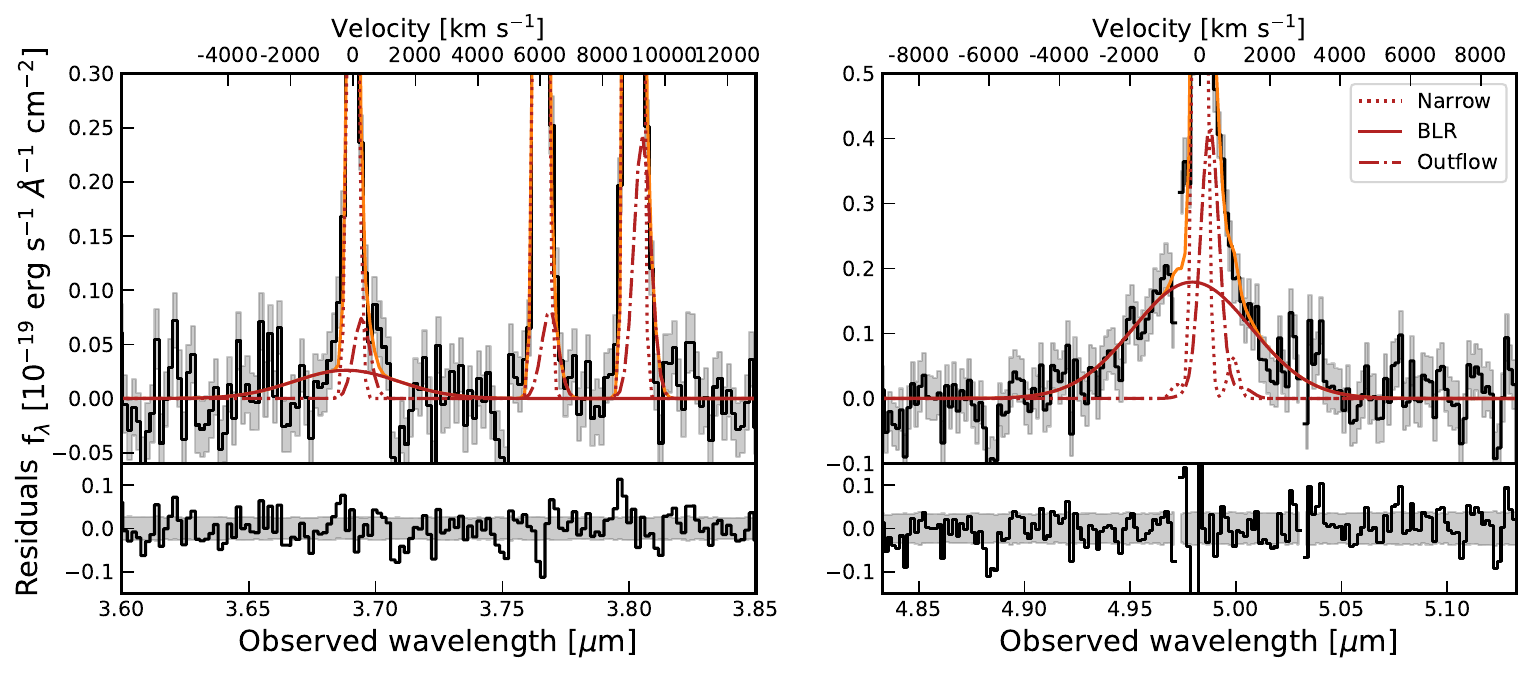}
    \caption{H$\beta$+[OIII] (left) and H$\alpha$+[NII] complexes (right) measured from the central 5 pixels ($r=0.''145)$ of NEPLA4 (top) and COLA1 (bottom). We show the best-fit model to the complexes including narrow lines, outflowing components and broad Balmer lines and the residuals in the bottom panels. Two pixels are missing in the COLA1 spectrum at $\lambda\simeq 4.98, 5.03\mu\rm{m}$ as they are masked by the outlier rejection postprocessing method from TEMPLATES \cite{Hutchison2024}. The broad ($\sim 4000\ \rm{km\ s}^{-1}$ Balmer components indicative of AGN are evident and distinctively narrower than the [OIII] lines.}
    \label{fig:broad_balmer_fit}
\end{figure*}

We obtained NIRSpec IFU G235 and G395M spectroscopy (Programme \#3767, PI: Meyer) of the UV-optical ($\lambda_{obs} = 1.7-5.3\mu\rm{m}$, $ 2300 <\lambda_{rest} < 6900 $\AA) of NEPLA4 and COLA1. The data reduction is detailed in Methods (Sec \ref{sec:data_red}) with the full spectra shown in Extended Figure \ref{fig:full_spectrum}. The data unambiguously reveal broad hydrogen Balmer emission lines (Fig. \ref{fig:broad_balmer_fit}) which implies the presence of a supermassive black hole at the centre of both COLA1 and NEPLA4 (see further Methods \ref{sec:BH_properties}). We note that COLA1 and NEPLA4 do not satisfy the colour selection criteria of Little Red Dots and show clear evidence for double-Gaussian profiles instead of exponential profiles (see Methods \ref{sec:not_lrds}). This ensures we can infer their BH mass from the standard single-epoch virial estimators using broad Balmer components.

\begin{figure*}[h]
    \centering
    \includegraphics[width=\linewidth]{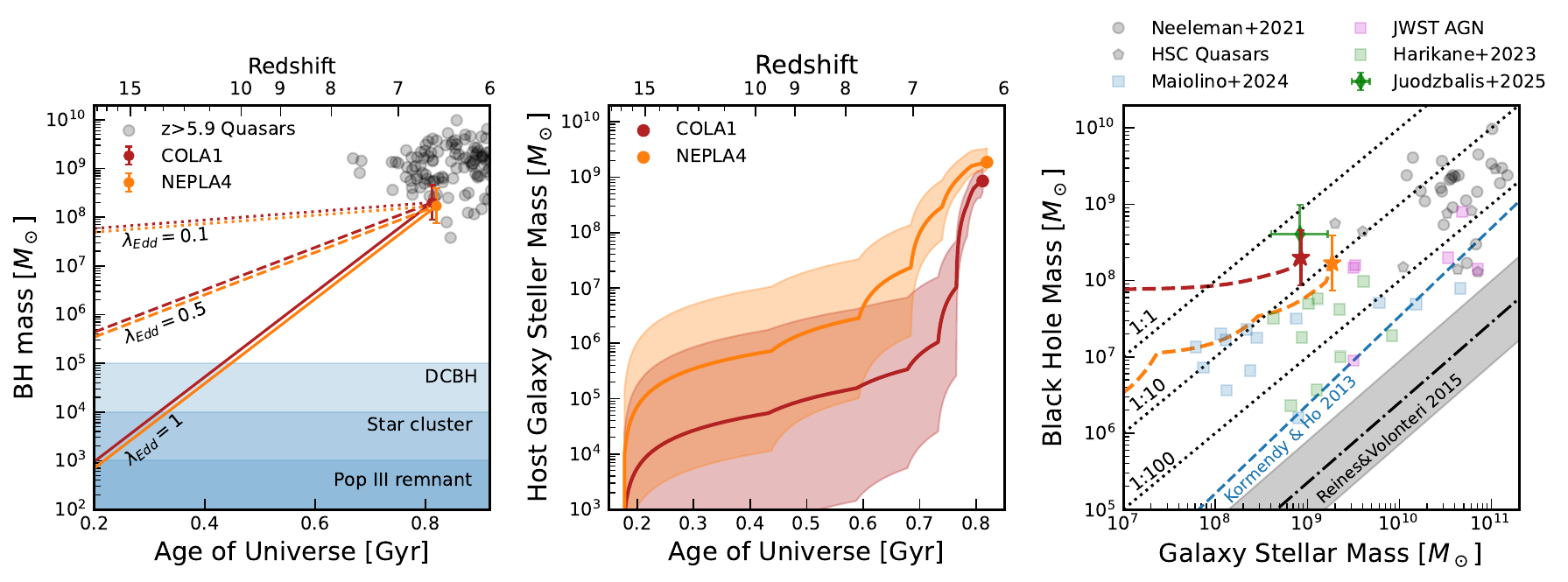}
    \caption{\textbf{Left:} Black hole mass growth history in COLA1/NEPLA assuming various Eddington ratios. The BH masses are derived using standard single-epoch virial estimators of the H$\alpha$ line, and include a fiducial systematic error of $0.358$ dex (see Methods). High-redshift quasars are indicated with black circles. \textbf{Middle:} Stellar mass growth history inferred from the best-fit \texttt{BAGPIPES} star-formation history. \textbf{Right:} Black hole to stellar mass ratio for NEPLA4 and COLA1, JWST AGN (coloured points \cite{Harikane2023,Maiolino2024_AGNsample,Juodzbalis2026}) and high-redshift quasars (black points, \cite{Izumi2019,Pensabene2020,Neeleman2021}). The inferred $M_{\rm{BH}}$-to-$M_*$ trajectory assuming an Eddington ratio of $1$ is shown with dashed lines. }
    \label{fig:bh_stellar_growth}
\end{figure*}

Assuming the single-epoch virial estimators are accurate for these two objects, the high black hole masses in COLA1 ($\log_{10} M_{BH}/[M_\odot] = 8.28\pm0.05 (\pm0.36)$) and NEPLA4 ($\log_{10} M_{BH}/[M_\odot] = 8.24\pm0.06(\pm0.36)$) rival that of known $z>6$ faint quasars \cite{Izumi2019, Banados2023}, yet their host galaxies have modest stellar masses (see Methods). Note that we indicate the systematic uncertainty from the single-epoch virial estimator \cite{DB2025} in parenthesis. From our best-fit BAGPIPES models to the observed photometry and spectroscopy (see Methods \ref{sec:sed_fitting}) we find stellar masses of $\log_{10} M_*[M_\odot] = 8.93^{+0.04}_{-0.04}$ and $9.27^{+0.02}_{-0.02}$ for COLA1 and NEPLA4, respectively. The resulting BH-to-stellar mass ratios are $0.22^{+0.04}_{-0.04}$ and $0.10^{+0.02}_{-0.01}$, which is $\sim400-800$ times higher than the local relation \cite{Reines2015}. We show the stellar mass to black hole mass ratio of NEPLA4 and COLA1 alongside JWST AGN and high-redshift quasars in Fig. \ref{fig:bh_stellar_growth}. They thus belong to the most overmassive BH found by JWST \cite{Ubler2023,Juodzbalis2024} and bridge the BH mass gap between the traditional categories of AGN, galaxies, and quasars at high redshift.

Another key feature of COLA1 and NEPLA4 is their contrasting BH and stellar growth history. Assuming that the black hole accreted at a constant fraction of the Eddington rate (on average), we can extrapolate its growth trajectory towards earlier times. As is the case for similar mass BHs at this redshift, we find that they must grow at nearly Eddington ($>0.8 \lambda_{Edd}$) rate from $z=15$ onwards if the seed BH is $<10^5\ M_\odot$, or additionally grow from lower mass seeds ($\sim 10^{3}\ M_\odot$) from $z\sim20$ onwards. Thus, for most of their lifetime, COLA1 and NEPLA4 have harboured actively growing BHs that would be the direct progenitors of luminous $z\sim6$ quasars. We show the extrapolated growth of the black holes for different Eddington ratios $\lambda_{Edd}=(0.1,0.5,1.0)$ in Figure \ref{fig:bh_stellar_growth} (left panel), alongside the galaxy stellar growth (Figure \ref{fig:bh_stellar_growth}, middle) resulting from the best-fit BAGPIPES non-parametric star-formation history (see Methods \ref{sec:sed_fitting}). The difference between the stellar and BH assembly histories is striking: whereas the SMBH must have accreted a significant amount of mass at early times, $\gtrsim 90\%$ of the stellar mass was only created in the previous $\sim 50$ Myr (Fig. \ref{fig:bh_stellar_growth}, middle panel). Currently, COLA1 and NEPLA4 are in a starburst phase, as evidenced by the observed H$\alpha$- and UV-based SFR, and the best-fit star-formation history (see further Methods), whereas the AGN is in a dormant state ($\lambda_{Edd}\sim 0.04-0.05$). This recent inversion between stellar and BH mass assembly is consistent with the high BH-to-stellar mass ratio, and further implies that, provided star-formation is not stopped, COLA1 and NEPLA4 will settle on the local BH-stellar mass on a quasi-horizontal track (Fig. \ref{fig:bh_stellar_growth}, right panel).

We now demonstrate that the detection of the double-peak Lyman-$\alpha$ profile in NEPLA4 and COLA1 requires a locally enhanced photo-ionisation rate that can only be explained if the recent shutdown of the quasar engine is as recent as $\Delta t_{QSO}\lesssim 1\ \rm{Myr}$. Indeed, the detection of the Lyman-$\alpha$ double-peak necessitates the presence of a highly ionised \ion{H}{II} bubble at least out to $r\gtrsim 0.3 \ \rm{pMpc}$ \ \cite{Songaila2018, Matthee2018, Meyer2021} and an extremely low residual neutral fraction inside the bubble ($x_{\rm{HI}}\sim10^{-6}$, \cite{Mason2020,Meyer2021}). The latter requirement is incompatible with the present ionising output of COLA1 and NEPLA4 (see Methods). However, if the BHs in NEPLA4/COLA1 were accreting at Eddington rate and unobscured by dust, they would appear as $M_{1450}\sim -23.2$ quasars (see Methods). They would rapidly create large ionised bubbles (proximity zones) with sizes $\sim 3-6\ \rm{pMpc}$ and boost the photo-ionisation rate in the ionised region high enough for Lyman-$\alpha$ blue peak photons to be transmitted \cite{Eilers2017,Eilers2020,Davies2020,Satyavolu2023}. Once the quasar shuts off, the photo-ionisation rate returns rapidly to equilibrium on timescales $\propto e^{-t \Gamma_{bkg}}$ where $\Gamma_{bkg}$ is the UV background photo-ionisation rate \cite{Davies2020}. The literature consensus on the $z\sim 6.5$ photoionisation rate is $\Gamma_{bkg} \sim 0.5-1.5 \times 10^{-13}\, \rm{s^{-1}}$ \cite{Keating2020, Garaldi2022, Lewis2022, Gaikwad2023}, corresponding to timescales of $t\sim0.2-0.6 \ \rm{Myr}$. Determining exactly at which point the photo-ionisation rate becomes too low to enable the transmission of the blue-peak photons is out of the scope of this study. We conservatively assume a timescale $t_{double-peak}=2.3 t \sim 0.5-1.5\ \rm{Myr}$ during which the blue peak Lyman-$\alpha$ photons are still visible, corresponding to a $90\%$ decline of the photo-ionisation from its ``quasar phase'' equilibrium value. Regardless of the exact values assumed above, the timescale to observe the double-peaks in COLA1 and NEPLA4 after a quasar phase is extremely short, of the order of a Myr or less.

\begin{figure*}
    \centering
    \includegraphics[width=1\linewidth, trim = 0 13cm 0 0.1cm, clip]{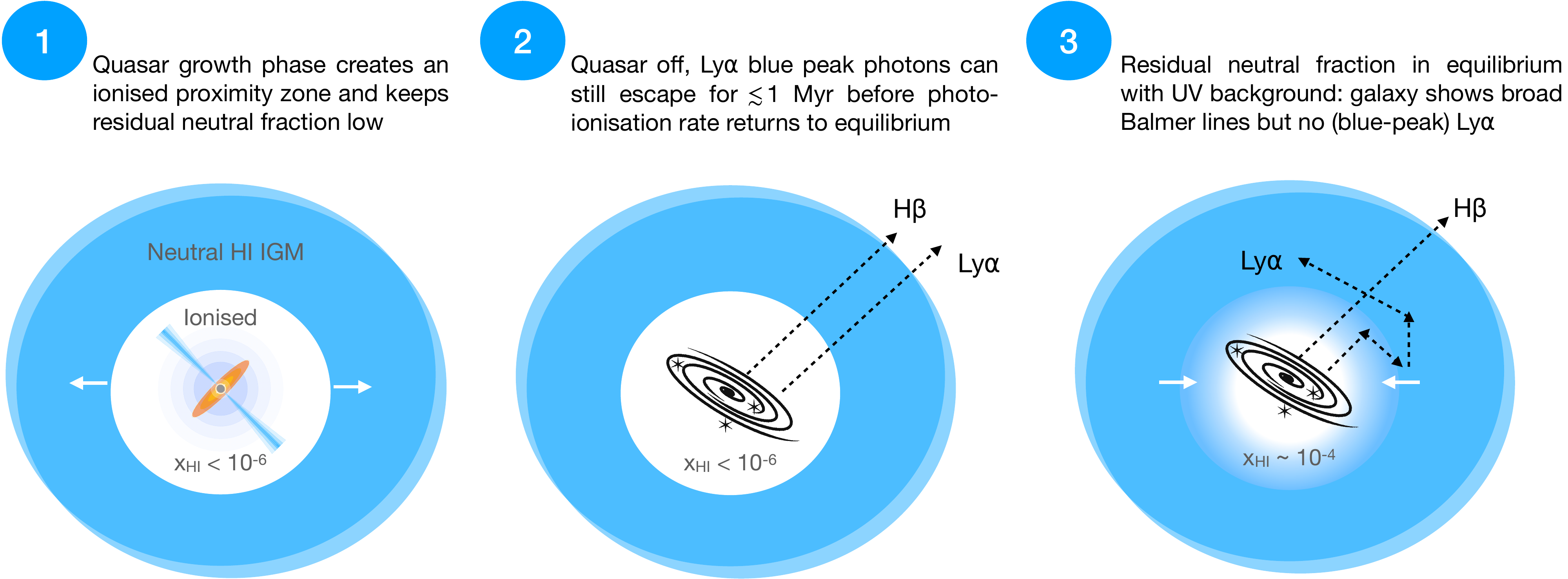}
    \caption{Schematic representation of the proposed evolutionary pathway leading to the detection of double-peaked Lyman$-\alpha$ emission in post-quasar galaxies.}
    \label{fig:schematic}
\end{figure*}

The combination of the BH and stellar mass growth trajectories with the detection of the double-peaked Lyman-$\alpha$ profile thus leads to the following evolutionary scenario (see Figure \ref{fig:schematic}). COLA1 and NEPLA4 have experienced efficient massive black hole growth at $z>6$ combined with a suppressed stellar mass growth, likely due to intense AGN feedback.  Until recently, these objects would appear as luminous quasars and have carved large ionising bubbles. A recent lull in the quasar accretion rate has led to the end of the quasar phase, returning to a lower luminosity AGN, and allowing the star-formation to increase rapidly, consistent with the observed star-formation rate, Eddington rate, and low mass loading factor and kinetic energy of the outflow (see Methods). For a short time ($\lesssim 1 \ \rm{Myr}$), a large ionised bubble surrounds the low-luminosity AGN and enables the escape of blue-peak photons. This is the phase at which we are observing NEPLA4 and COLA1. Finally, the photo-ionisation returns to the background equilibrium value, followed by the ionised bubble recombining on longer timescales. At this stage, the objects would appear as luminous starburst galaxies with diminishing Lyman-$\alpha$ visibility but still reveal broad Balmer lines in JWST spectroscopy \cite[e.g.][]{Maiolino2024_AGNsample,Juodzbalis2026}. COLA1 and NEPLA4 are thus direct observational evidence for a episodic quasar accretion \cite[e.g.][]{Davies2020} connecting quasars to luminous galaxies and JWST AGN in a single evolutionary pathway.

\begin{figure*}
    \centering
    \includegraphics[width=0.9\linewidth]{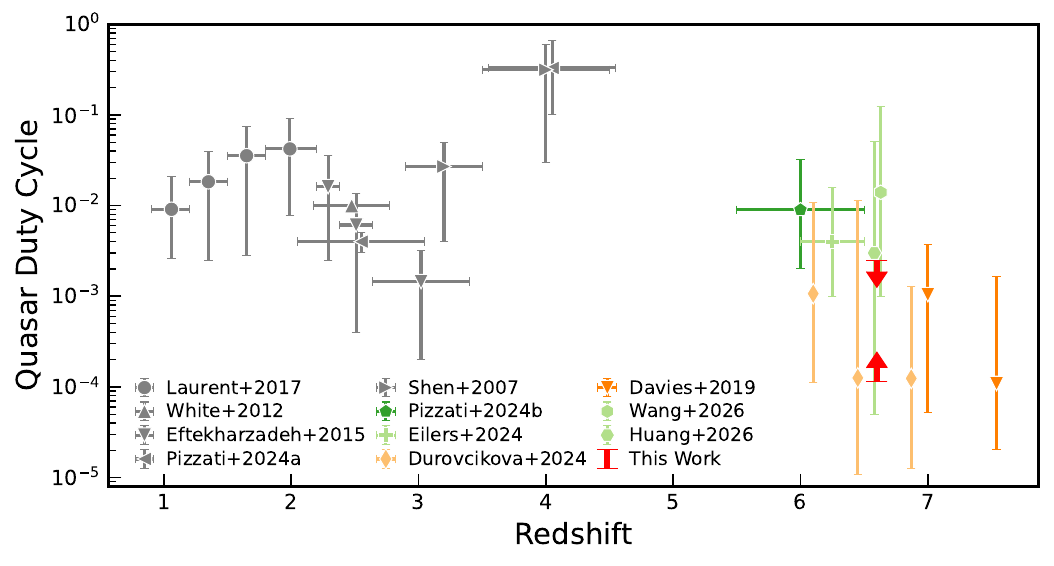}
    \caption{Quasar duty cycle constraints against redshift. Literature values from clustering at low-redshift (gray), quasar-quasar (green) and galaxy-quasar (pale green) clustering at intermediate-high redshifts, and proximity zone measurements at $z>6$ (orange shades) \cite{Shen2007,White2012, Eftekharzadeh2015,Laurent2017,Pizzati2024,Pizzati2024a,Eilers2024,Durovcikova2024,Davies2019,Wang2026,Huang2026}. Constraints from this work are shown in red for two extreme cases assuming: 1) all ultra-luminous LAEs host SMBH with $\gtrsim10^8\ M_\odot$ BH, or 2) lower limits on BH occupation fraction from SHELLQS luminous LAEs observed with JWST \cite{Matsuoka2025}. }
    \label{fig:duty_cycle}
\end{figure*}

An important consequence of our proposed evolutionary scenario is that we can use the relative number density of objects in the three phases to measure the duty cycle of $z\sim 6$ quasars (e.g. the fraction of time during which they accrete close to Eddington and appear as quasars). We use the relative number densities of quasars, ultra-luminous single- and double-peaked LAEs, we derive the following limits (see Methods) on the duty cycle of $z\simeq 6.6$ quasars: $0.11\times 10^{-3}  < f_{duty} < 2.4 \times 10^{-3}  \ \ \ \ (2\sigma \ \rm{limits})$. We find excellent agreement between our constraints and independent measurements from the literature, \cite{Shen2007,White2012, Eftekharzadeh2015,Laurent2017,Pizzati2024,Pizzati2024a,Eilers2024,Durovcikova2024,Davies2019,Wang2026,Huang2026}, further supporting the evolutionary picture proposed above (see Figure \ref{fig:duty_cycle}). Additionally, we estimate that faint quasars will undergo $N\sim 80$ episodic accretion events with typical timescales $1 \lesssim t_{\rm{ep}} [10^3\ \rm{Myr}] \lesssim 23$ (see Methods). This supports a picture where $z>6$ quasars grow through a succession of multiple episodes at high accretion rates, as suggested by other studies in the literature \citep[][]{Eilers2020, Davies2020, Satyavolu2023b}.

We now conclude by discussing the impact of episodic BH growth on the visibility of Lyman-$\alpha$ in the Reionisation Era. In the HEROES survey of ultra-luminous LAEs from which COLA1 and NEPLA4 were discovered \cite{Taylor2023}, the fraction of observed double-peaks is $3.5^{+3.4}_{-1.7} \%$. Low-redshift studies, where the IGM is not absorbing the blue-peak photons, show that the intrinsic fraction of double-peaked LAEs is around $\sim 30\%$ \cite{Kulas2012,Vitte2025}. Therefore the fraction of ultra-luminous Lyman-$\alpha$ emitters whose visibility is temporarily boosted by a previous AGN/quasar accretion phase is $\sim 11^{+6}_{-3} \%$. This is a conservative estimate where we have not considered longer timescales where the photo-ionisation would not enable blue-peak photons to escape but the large ionised bubble would still diminish the impact of IGM absorption and boost the Lyman-$\alpha$ red peak transmission.

\begin{table*}[]
    \centering
     \caption{Key Inferred properties of NEPLA4 and COLA1. $^{a}$ \cite{Songaila2018} $^{b}$ \cite{Matthee2018} $^{c}$ \cite{Nakajima2022} $^{d}$ \cite{Bian2018}}
\begin{tabular}{|l|c|c|}
\hline
Property & COLA1 & NEPLA4  \\ 
\hline
SMBH \& outflow & & \\
\hline
$L_{\rm{Bol}}\ [10^{45}\ \rm{erg\ s}^{-1}]$  & $1.17 \pm 0.01$ & $0.88 \pm 0.03$  \\
$ \log_{10} M_{BH}/[M_\odot]$ (H$\alpha)$ & $ 8.28 \pm 0.05 $ &$ 8.24 \pm 0.06 $\\ 
$ \log_{10} M_{BH}/[M_\odot]$ (H$\beta)$  & $ 7.87 \pm 0.09 $ &$ 8.07 \pm 0.06 $\\ 
$\lambda_{Edd}$ (H$\alpha$) & $ 0.050 \pm 0.004 $ & $ 0.039 \pm 0.004 $ \\ 
$\lambda_{Edd}$ (H$\beta$) & $ 0.155 \pm 0.044 $ & $ 0.080 \pm 0.019 $ \\ 
E(B-V)$^{BLR}$ & $ 0.94 \pm 0.76 $ &$ 0.17 \pm 0.52 $\\ 
BH/stellar mass ratio &$ 0.22 ^{+ 0.04}_{- 0.04} $ & $ 0.10 ^{+ 0.02}_{- 0.01} $ \\ 
$\dot M_{out,ion} [M_\odot \ \rm{yr^{-1}}] $  & $ 4.3 \pm 3.6 $ & $ 9.2 \pm 4.2 $ \\ 
$\eta$ & $ 0.08 \pm 0.06 $ & $ 0.18 \pm 0.08 $ \\ 
$P_{k,ion}$ [$10^{40}\rm{erg\ s^{-1}}$] & $ 31 \pm 26 $ & $ 223 \pm 118 $ \\ 
\hline
Galaxy properties & & \\
\hline
$M_{\rm{UV}}$ & $ -21.6^{a} $ & $ -21.8^{b} $ \\ 
$A_V^{obs} (H\alpha /H\beta)$ & $0.34^{+0.45}_{-0.34}$ & $<0.42 \ (2\sigma)$ \\ 
Z(R3)$^{c}$ & $ 7.72^{+ 0.07}_{-0.06} $ & $ 7.42 ^{+ 0.02}_{-0.02} $ \\ 
Z(O3H$\beta)^d$  & $ 7.48^{+ 0.08}_{-0.07} $ & $ 7.16^{+0.03}_{-0.03} $ \\ 
$R_e$ [kpc]  & $0.81\pm0.68$  & $0.60\pm 0.35$\\
$n_{\rm{Sersic}}$ &  $1.3 \pm 3.3$ &  $0.68 \pm  0.28 $ \\
$M_{dyn} [10^{10} M_\odot]$ & $ 2.6 \pm 2.2 $ & $ 1.06 \pm 0.43 $ \\ 
$T_e (r=0.''145)$ [K] & $ 8184 ^{+ 1838}_{- 803} $ & $ 9889 ^{+ 2432}_{- 1084} $ \\ 
$T_e (r=0.''3)$ [K] & $ < 7071\  (2\sigma) $ & $ 10016 ^{+ 2586}_{-1126} $ \\ 
$\tau_{dep}$ [Gyr$^{-1}$] & $ 0.46 \pm 0.40 $ & $ 0.21 \pm 0.09 $ \\ 
$\Delta v_{Ly\alpha} [\rm{km\ s}^{-1}]$ & $ 218\pm3 $ & $ 249\pm 7 $ \\ 
$f_{esc} (LyC)$ & $ 0.29 ^{+ 0.10}_{- 0.01} $ & $ 0.20 ^{+ 0.04}_{- 0.02} $ \\ 
$f_{esc} (Ly\alpha, obs)$  & $ 0.37 \pm 0.02 $ & $ 0.22 \pm 0.01 $ \\ 
$f_{esc} (Ly\alpha, corr)$  & $ 0.52 ^{+ 0.07}_{- 0.03} $ & $ 0.28 ^{+ 0.02}_{- 0.01} $ \\ 
$f_{Ly\alpha} [10^{-17} \rm{erg\ s\ cm^{-2}}]$ & $ 4.36 ^{+ 0.12}_{- 0.17} $ & $ 2.98 ^{+ 0.04}_{- 0.10} $ \\ 
$\log_{10} \xi_{ion}/[\rm{erg \ s^{-1}\ \rm{Hz^{-1}}}]$ & $ 25.45 ^{+ 0.04}_{- 0.04} $ & $ 25.37 ^{+ 0.02}_{- 0.02} $ \\ 
\hline
BAGPIPES & & \\
\hline
$A_v$ & $ 0.29 ^{+ 0.04}_{- 0.03} $ & $ 0.306 ^{+ 0.03}_{- 0.03} $ \\ 
$\log U$ & $ -0.07 ^{+ 0.05}_{- 0.07} $ & $ -0.09 ^{+ 0.06}_{- 0.07} $ \\ 
$\log_{10} M_* / [M_\odot ]$ & $ 8.93 ^{+ 0.04}_{- 0.04} $ & $ 9.27 ^{+ 0.02}_{- 0.02} $ \\ 
Z  & $ 0.185 ^{+ 0.003}_{- 0.002} $ & $ 0.132 ^{+ 0.002}_{- 0.002} $ \\
\hline
\end{tabular}
    \label{tab:key_properties}
\end{table*}

We thus propose that $\sim 11\%$ of Lyman-$\alpha$ emitters without apparent galaxy overdensities may be tracing relics of past quasar/AGN activity. This could explain the established trend of higher confirmation rate of Lyman-$\alpha$ in more luminous objects, as their SMBH are likely more massive \cite[e.g.][]{Stark2011,Pentericci2018}. A number of individual JWST also provide anecdotal evidence supporting our hypothesis. For example, COSz7, a luminous Lyman-break galaxy selected by \cite{Roberts-Borsani2016}, was then found to have an AGN \cite{Laporte2017,Ubler2024}. \cite{Tang2024} report the detection of a strong $z=8.38$ Lyman-$\alpha$ without a surrounding overdensity, but with hard ionisation fields that could be powered by an AGN. Similarly, GNz11 shows an exceptionally bright Lyman-$\alpha$ emission line and halo \cite{Bunker2023,Scholtz2023} and potential evidence for an AGN \cite{Maiolino2024}.  Finally, the most distant Lyman-$\alpha$ detection at $z\simeq 13$ has also been hypothesized to be linked to AGN activity \cite{Witstok2025,Cohon2025}. To what extent AGN activity is boosting the Lyman-$\alpha$ visibility at $z>6$ remains unclear as this depends on details of the AGN duty cycle and the fraction of obscured SMBH growth, but this would have a non-trivial impact for inferring the reionisation history and topology using Lyman-$\alpha$ emission lines. 

In conclusion, the relatively massive BHs, the star-formation history, and the presence of the large ionised bubbles suggest that NEPLA4 and COLA1 have recently ($\lesssim1\  \rm{Myr}$) transitioned from a quasar phase to a lower-luminosity AGN, with star-formation rapidly picking up after being inhibited by the rapid SMBH growth. The ionised bubbles enabling the detection of double-peaked Lyman-$\alpha$ are relics of the recently shut-off quasar activity ($\lesssim 1\ \rm{Myr}$), putting direct constraints on the duty cycle of high-redshift quasars with important implications for the visibility of Lyman-$\alpha$ in the reionisation era. The combination of existing and upcoming ground-based instruments (e.g., VLT/MOONS, ELT/MOSAIC, Subaru/PFS) with JWST spectroscopy is promising to systematically search for (double-peaked) Lyman-$\alpha$ emitters and unobscured AGN in deep fields. Establishing a larger census of rare objects such as COLA1 and NEPLA4 will open new opportunities to study the properties of quasar host galaxies, measure escape fraction via the double-peaked Lyman-$\alpha$ emission, and infer the AGN duty cycle in the first billion years of the Universe.

\section{Methods}

 \captionsetup[figure]{labelfont={bf},name={Extended Figure},labelsep=period}
 \setcounter{figure}{0}    

 \captionsetup[table]{labelfont={bf},name={Extended Table},labelsep=period}
 \setcounter{table}{0}    

\subsection{Cosmological model}
Throughout the paper, we use a concordance cosmology with $H_0=70\ \rm{km s}^{-1}\rm{Mpc}^{-1}$, $\Omega_m = 0.3, \Omega_\Lambda = 0.7$, and give magnitudes in the AB system \cite{OkeGunn1983}.

\subsection{JWST NIRSpec IFU Data}
\label{sec:data_red}
NEPLA4 and COLA1 were observed as part of Cycle 2 GO program \#3767 (PI: Meyer) with NIRSpec IFU using two medium gratings G235/F170LP and G395M/F290LP covering the rest-frame $2000-7000$ \AA \ spectrum of the galaxy. The targets were observed using 4-POINT-DITHER for a total 40 minutes on-source per grating using NRSIRS2RAPID readout with 40 groups. The data was reduced using the JWST pipeline (version 1.14.0, \texttt{pmap 1241}) using the standard and extra steps suggested by the TEMPLATES ERS team \cite{Rigby2023}. We include layered outlier rejection following \cite{Hutchison2024} as well as a custom outlier rejection in the spectral direction for every pixel following \cite{Loiacono2024}. We finally rescale the error array produced by the pipeline to the measured rms in each channel computed after masking the source and using three rounds of sigma-clipping. We reach a median rms of $0.99(2.0)\times10^{-21}\ \rm{erg\ s}^{-1} \rm{cm}^{-2}\ \AA^{-1}$ per $0.''1\times0.''1$ pixel per $18 \ $\AA \ channel in the G395M(G235M) IFU cubes.

\begin{figure*}[h]
    \centering
    \includegraphics[width=\textwidth]{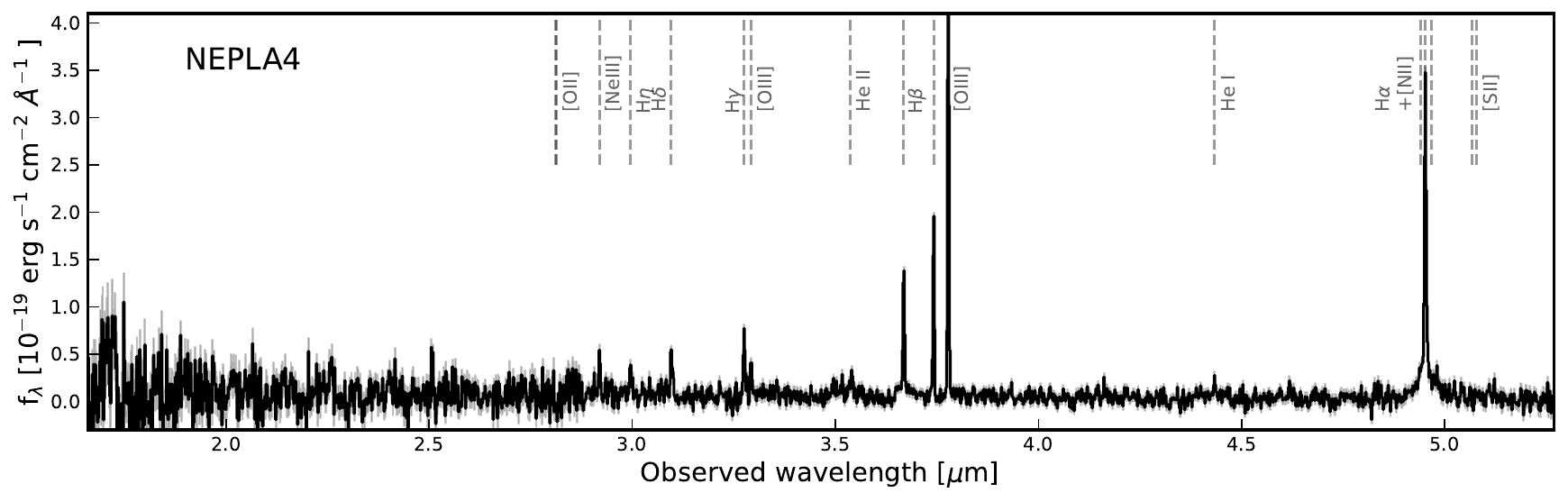}
    \includegraphics[width=\linewidth]{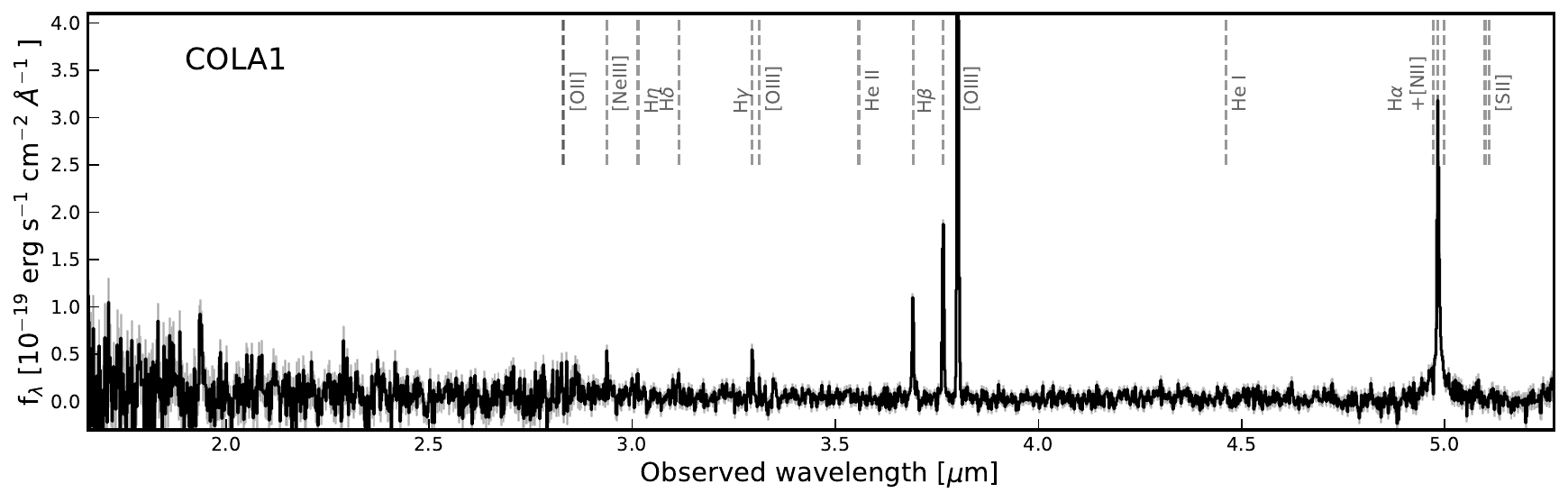}
    \caption{Aperture-integrated ($r=0.''3$) spectra of NEPLA4 and COLA1 in our combined G235M/F170LP and G395M/F290LP NIRSpec IFU observations.}
    \label{fig:full_spectrum}
\end{figure*}

To determine the optimal aperture for extracting the integrated spectra of COLA1 and NEPLA4, we make a mock narrow-band image of the [OIII] $5008\ \AA$ line (the brightest line in the spectrum). For this we stack channels within $\Delta v = 1.2\times \rm{FWHM}(OIII)$, using the best-fit redshift and FWHM from the fit to the central part of the galaxies (see Section \ref{sec:mbh}). The choice of $\Delta v = 1.2\times \rm{FWHM}(OIII)$ maximises the SNR of the line. We find that emission is very circular with an aperture of $r=0.''3$ encompassing all the SNR$>3$ pixels, which we thus adopt as our fiducial aperture size for the integrated galaxy spectrum. We have checked that the central brightest pixel of the continuum in the two grating is the same as that of  [OIII] $5008\ \AA$ emission. We extract both G235M and G395M spectra with the former rebinned to the same wavelength spacing at G395 to obtain a uniform spectrum. In the overlap between G235 and G395M the integrated spectrum are stitched using inverse-variance weighting. We show the integrated galaxy spectra in Extended Figure \ref{fig:full_spectrum}, revealing numerous emission lines in the rest-frame optical regime and confirming the redshifts at $z=6.59075$ and $z=6.54422$ respectively. Even in the integrated spectrum, the H$\alpha$ lines (and even H$\beta$ in NEPLA4) show broad components indicative of AGN activity.

\subsection{Balmer Line Complex Fitting}
\label{sec:mbh}
Following \cite{Ubler2024}, we extract the central pixel and the adjacent four pixels to capture the broad Balmer line emerging from the unresolved broad line region (BLR) of the BH whilst smoothing over the impact of the NIRSpec PSF undersampling in single IFU pixels \cite{Perna2023,Bianchin2024}. We check that the broad Balmer line emission is fully captured by these pixels (e.g. we do not find more broad line emission in the $r=0.''3$ spectrum), further supporting the BLR hypothesis as the origin of the broad lines. Unlike \cite{Ubler2024}, we do not find any offset between the peak of the [OIII] 5008 emission and that of the broad Balmer line emission. 

We fit simultaneously the H$\beta$+[OIII] 4960, 5008\ \AA and H$\alpha$+[NII] complexes using a varied combination of narrow and broad line components.  Before fitting the Balmer complexes, we first fit and subtract the rest-frame optical ($>3650\ \AA$, masking all the emission lines) continuum with a powerlaw. We perform three fits: a) narrow components for all lines only, b) narrow + 1 broad component (FWHM$>600\ \rm{km\ s^{-1}}$), and c) narrow + outflowing ($600< \rm{FWHM}/\ [\rm{km\ s^{-1}}] <1800$) + an additional broad component (Broad Line Region) for the Balmer lines (FWHM$>1800\ \rm{km\ s^{-1}}$). We assume that all narrow components share the same redshift (in practice this is mostly driven by the [OIII] emission which has the highest SNR). Broad components, when added, are tied in redshift space and share the same FWHM. We neglect the outflow component in [NII] as we do not have significantly high resolution and SNR to detect it. We fix the Balmer line ratio of the outflow to that of the narrow component, but let that of the broad BLR component free. Finally, the Balmer broad BLR lines have a tied FWHM but have free amplitudes (effectively enabling the attenuation to differ between the narrow and BLR components).  The ratio of the [OIII] and [NII] lines are fixed to theoretical value of $2.98$ and $2.96$, respectively, for all components. The number of free parameters is 6/11/15 for models a/b/c, respectively. We compute the Bayesian Information Criterion for each best-fit and find that a single broad component improves the BIC by $\Delta BIC = 1081(813)$ for NEPLA4 (COLA1), and the addition of BLR components further improves it by $\Delta BIC = 79(201)$. We show our best-fit model in Extended Figure \ref{fig:broad_balmer_fit}, the best-fit model parameters in Table \ref{tab:Balmer_line_properties}, as well as the best-fit rejected models in Extended Figures \ref{fig:COLA1_fits} and \ref{fig:NEPLA4_fits}.

\begin{table*}
    \centering
     \caption{Observed H$\beta$, [OIII], H$\alpha$ and [NII] line properties from the central 5 pixels and inferred AGN properties.  }
\begin{tabular}{|l|c|c|}
\hline
Property & COLA1 & NEPLA4  \\ 
\hline
Redshift & $ 6.59075 \pm 0.00003 $ & $ 6.54422 \pm 0.00002 $ \\ 
H$\alpha$ (broad) $[10^{-19}\ \rm{erg\ s}^{-1}\ \rm{cm}^{-2}] $ & $181.6 \pm 18.5$  & $138.3 \pm 17.8 $ \\ 
H$\alpha$ (narrow) $[10^{-19}\ \rm{erg\ s}^{-1}\ \rm{cm}^{-2}] $& $104.5 \pm 4.1$  & $116.4 \pm 2.1 $ \\ 
H$\alpha$ (outflow) $[10^{-19}\ \rm{erg\ s}^{-1}\ \rm{cm}^{-2}] $ & $45.4 \pm 4.7$  & $44.4 \pm 7.4 $ \\ 
H$\beta$ (broad) $[10^{-19}\ \rm{erg\ s}^{-1}\ \rm{cm}^{-2}] $ & $22.6 \pm 6.2$  & $38.5 \pm 5.9 $ \\ 
H$\beta$ (narrow) $[10^{-19}\ \rm{erg\ s}^{-1}\ \rm{cm}^{-2}] $  & $32.4 \pm 1.3$  & $40.0 \pm 1.0 $ \\ 
$[$NII$]$ 6584 (narrow) $[10^{-19}\ \rm{erg\ s}^{-1}\ \rm{cm}^{-2}] $ & $5.4 \pm 2.0$  & $0.1 \pm 0.7 $ \\ 
$[$OIII$]$ 5008 (narrow) $[10^{-19}\ \rm{erg\ s}^{-1}\ \rm{cm}^{-2}] $  & $210.1 \pm 3.1$  & $187.6 \pm 1.9 $ \\ 
$[$OIII$]$ 5008 (outflow) $[10^{-19}\ \rm{erg\ s}^{-1}\ \rm{cm}^{-2}] $& $346.5 \pm 121.1$  & $286.2 \pm 62.4 $ \\ 
FWHM H$\alpha$ (broad) $[\rm{km\ s}^{-1}] $ & $3947 \pm 263$  & $4114 \pm 305$ \\ 
FWHM H$\alpha$ (narrow) $[\rm{km\ s}^{-1}] $ & $345 \pm 3$  & $287 \pm 2$ \\ 
FWHM H$\alpha$ (outflow) $[\rm{km\ s}^{-1}] $ & $611 \pm 18$  & $1077 \pm 123$ \\ 
FWHM H$\beta$ (broad) $[\rm{km\ s}^{-1}] $ & $3947 \pm 263$  & $4114 \pm 305$ \\ 
L( H$\alpha$ , broad) $[10^{42}\ \rm{erg\ s}^{-1}] $ & $9.00 \pm 0.11 $ & $6.75 \pm 0.24$  \\ 
L( H$\alpha$ , outflow) $[10^{42}\ \rm{erg\ s}^{-1}] $ & $2.25 \pm 0.23 $ & $2.16 \pm 0.36$  \\ 
L( H$\beta$ , broad) $[10^{42}\ \rm{erg\ s}^{-1}] $ & $1.12 \pm 0.31 $ & $1.88 \pm 0.29$ \\ 
H$\alpha$ / H$\beta$ (narrow) & $ 3.23 \pm 0.18 $ &$ 2.91 \pm 0.09 $\\ 
H$\alpha$ / H$\beta$ (broad) & $ 9.25 \pm 2.69 $ &$ 3.73 \pm 0.75 $\\ 
$\Delta v$ (broad) [$\rm{km\ s}^{-1}$] & $ -220 \pm 107 $ &$ 371 \pm 120 $\\ 
$\Delta v$ (outflow)[$\rm{km\ s}^{-1}$]&$ 279 \pm 1326 $ &$ -51 \pm 720 $\\ 
\hline
\end{tabular}
\label{tab:Balmer_line_properties}
\end{table*}

\subsection{COLA1 and NEPLA4 are not Little Red Dots}
 \label{sec:not_lrds}
We have find strong evidence for two broad (FWHM(COLA1)$=3947 \pm 263\ \rm{km \ s}^{-1}$  and FWHM(NEPLA4)$=4114 \pm 305\ \rm{km \ s}^{-1}$ components in the Balmer lines in both objects. Such large FWHM exclude starburst-powered outflows \citep[e.g.][]{Davies2019_outflows,Llerena2023} and strongly imply an origin in the broad line region of an AGN. Before proceeding to discuss the implied BH masses and bolometric luminosities using standard estimators established in low-redshift quasars, we note that COLA1 and NEPLA4 are not LRDs \cite{Matthee2024} as their rest-optical slopes are not red enough. We fit their rest-frame UV ($\lambda_{rest}<3640 $\AA) and optical ($\lambda_{rest}>3640 $\AA) continuum with a powerlaw $f_\lambda \propto \lambda^\beta$, masking all emission lines in Table \ref{tab:integrated_fluxes} with a conservative $\Delta v \pm2000\ \text{km\ s}^{-1}$ range. Given their observed FWHM, the H$\alpha$ and H$\beta$ are masked over a $\Delta v \pm4000\ \text{km\ s}^{-1}$ range instead. We find $\beta_{UV}, \beta_{opt} = (-0.82\pm0.68,-1.64\pm0.28)$ for NEPLA4 and $\beta_{UV}, \beta_{opt} = (-1.75\pm0.80,-1.56\pm0.38)$ for COLA1. Clearly, the rest-frame optical colors of NEPLA4 and COLA1 are not red, and do not satisfy the usual UV-optical slope colour cuts  for LRDs \cite{Kokorev2024,Kocevski2025,deGraaff2025}. 

An alternative possibility to explain the broad lines could be Thomson scattering \cite{Rusakov2026, Chang2026}. To explore this hypothesis, we fit exponential profiles to the $H\alpha$/H$\beta$ profiles. We find that the best-fit profiles have much larger H$\alpha$ FWHM than H$\beta$ (specifically, we find FWHM(H$\beta$,COLA1) = $101^{+75}_{-2}\ \rm{km\ s}^{-1}$, FWHM(H$\alpha$,COLA1) = $786^{+69}_{-32}\ \rm{km\ s}^{-1}$ and FWHM(H$\beta$,NEPLA4) = $332^{+31}_{-48} \rm{km\ s}^{-1}$,  FWHM(H$\alpha$,NEPLA4) =$938^{+62}_{-50} \rm{km\ s}^{-1}$. This is at odds with the expectation that Thomson scattering wavelength-independent \cite{Rusakov2026, Juodzbalis2026, Chang2026}. We experiment tying the FWHM of the Balmer exponential profiles, and find significantly worse fits ($\Delta BIC = 22.4 / 31.5$ for COLA1/NEPLA4, respectively). Best-fit models with exponential profiles (with tied FWHM or not) can be found in Extended Figures \ref{fig:COLA1_fits} and \ref{fig:NEPLA4_fits}.

Finally, we note that the Balmer lines in NEPLA4 and COLA1  are dominated by the narrow line component, broad components have Balmer decrement consistent with $3.06$ (although only at $2\sigma$ for COLA1 given the broad H$\beta$ component), do not show a Balmer break \cite{Setton2025, Hviding2025} and do not show the typical absorption or P-Cygni profile found in many LRDs, and especially candidate objects for electron scattering \cite{Rusakov2026,Torralba-Torregrosa2026}. In summary, NEPLA4 and COLA1 are not LRDs by any classical definition and are very unlikely to have broad components originating from electron scattering. Importantly this implies that concerns about the bolometric luminosity and therefore the BH mass raised by recent multi-wavelength observations of LRDs \cite{Greene2026} do not apply for these objects, and throughout this text we assume they are actively star-forming galaxies with a modest contribution from a low-luminosity quasar/AGN.

\subsection{Black Holes Properties}
\label{sec:BH_properties}
We now proceed to infer black hole masses and bolometric luminosities using common estimators and scaling relations calibrated with low redshift quasars (see Table \ref{tab:key_properties} for a summary). We assume the median BLR value for a large sample of low-redshift broad line AGN of $3.06$ \cite{Dong2008} to derive dust attenuation in the BLR of NEPLA4 and COLA1 (the theoretical value in the NLR region is $\sim3.1$ \cite{GaskellFerland1984}).  Using the value of $3.06$ and a Calzetti attenuation law, we deduce an $E(B-V)$ values of $ 0.15 \pm 0.18 $ and $0.99\pm0.81$, e.g. both are compatible with little to no dust attenuation, in agreement with the narrow line values. We therefore take the conservative approach to assume no dust attenuation in what follows and note that this provides lower limits on black hole masses, bolometric luminosities, and associated quantities. From the H$\alpha$ broad line luminosity we derive the bolometric luminosity of COLA1(NEPLA4) using $L_{\rm{Bol}}=130\times L_{H\alpha} = 1.17\pm0.01(0.88\pm0.03)\times10^{45}\ \rm{erg\ s}^{-1}$ \cite{Stern2012}. Using the relations of \cite{DB2025} we derive the black hole mass from the H$\alpha$ and H$\beta$ lines. We find very high BH masses from the H$\alpha$ line  $\log_{10}M_{BH} =  8.28 \pm 0.05$, $8.24 \pm 0.06$. The masses derived from H$\beta$ \citep[][]{DB2020} are slightly lower by $0.2-0.4$ dex. Since the H$\beta$ line based masses are more uncertain due to having 1) lower SNR 2) potentially higher dust corrections, we adopt the H$\alpha$ masses in the rest of the paper. The masses rival that of faint quasars, but the bolometric luminosities are much lower. Consequently, the corresponding Eddington ratios are low $\lambda_{\rm{Edd}} = 0.05\pm0.004, 0.04\pm0.004$ for COLA1 and NEPLA4, respectively. Note that a factor $\times2.5$ uncertainty must be taken into account when converting the H$\alpha$ luminosity to the bolometric luminosity using the \cite{Stern2012} relation and consequently also needs to be propagated to the Eddington ratio.  

In summary, we find that NEPLA4 and COLA1 are likely the ``dormant" counterparts of high-luminosity z$\sim6$ quasars similar to the object reported by \cite{Juodzbalis2024} . If the SMBHs were accreting at the Eddington limit, their UV luminosities would be $M_{UV}^{AGN}(\lambda_{Edd}=1) \simeq -23.20 / -23.16$ for COLA1/NEPLA4, respectively. They would thus dominate the galaxy emission and appear as faint quasars similar to those selected in ground-based imaging surveys \cite[e.g.][]{Matsuoka2025}.

Using established bolometric corrections for the narrow [OIII] \cite{Pennell2017}, we would infer $L_{Bol}\simeq 2.94 (2.61)\times10^{46}$ for COLA1(NEPLA4) e.g. $\sim 34(22)\times$ higher than the one inferred from the broad H$\alpha$ line. Assuming the latter is correct, the NLR contribution to the [OIII] flux is limited to $\sim 3-5\%$.  We also find that, assuming standard bolometric to UV luminosity conversion \cite{Richards2006, Runnoe2012}, the low-luminosity quasars currently contribute at most $\sim 15\%$ of the UV continuum flux of their host, with both estimated UV magnitudes of the AGN at $M_{UV}^{AGN, now} \simeq 19.5$ against the observed $M_{UV}^{obs}\simeq -21.6(-21.8)$ for COLA1 (NEPLA4). Finally, the good correspondence between the H$\alpha$ and UV-based SFRs also suggests that the contamination of the AGN NLR to the galaxy spectra is minimal. In short, the AGN in COLA1 and NEPLA4 are in a dormant phase, offering a unique opportunity to study the host of high-redshift quasars without having to disentangle the quasar emission from its host.

\begin{table*}

    \caption{Line fluxes for the line covered (rest-frame $3700-7000\ \AA$) in our observations in the $r=0.''3$ aperture spectra. All fluxes are given in units of $10^{-19}\ \rm{erg\ s}^{-1}\ \rm{cm}^{-2} $. For non-detected lines we give the formal $2\sigma$ upper limit using the error array and the FWHM measured from the narrow component of H$\alpha$ line, except for [NII] which are fit in a complex with H$\alpha$.}
        \centering
\begin{tabular}{|c|c|c|c|c|}
\hline
Line & COLA1 & NEPLA4  \\ 
\hline
\text{[\ion{O}{II}]} $3727,3729$ & $< 11.2 $ & $ < 9.7 $  \\ 
\text{[\ion{Ne}{III}]} $3870$ & $ 18.7 \pm 2.3 $ & $ 16.0 \pm 2.5 $  \\ 
H8+\ion{He}{I} 3890 & $ < 5.3 $ & $ < 4.9 $  \\ 
\text{[\ion{Ne}{III}]} 3969 &  $ 13.3 \pm 6.3 $ & $ < 4.4 $  \\ 
H$\epsilon$ 3971 & $ < 5.2 $ & $ 11.8 \pm 4.2 $  \\ 
H$\delta$ 4103 & $ 10.5 \pm 2.5 $ & $ 19.6 \pm 3.2 $  \\ 
\ion{He}{I} 4145 & $ < 4.8 $ & $ < 2.7 $  \\ 
\text{[\ion{Fe}{II}]} 4234 & $ < 5.3 $ & $ < 4.2 $  \\ 
H$\gamma$ 4342 & $ 21.9 \pm 2.5 $ & $ 27.3 \pm 4.3 $  \\ 
\ion{He}{I} 4473 & $ 4.7 \pm 1.2 $ & $ < 4.0 $  \\ 
\ion{He}{II} 4687 & $ < 5.0 $ & $ < 4.0 $  \\ 
H$\beta$ 4863 & $ 44.7 \pm 2.4 $ & $ 51.7 \pm 1.8 $  \\ 
\text{[\ion{O}{III}]} 4364 & $ < 5.4 $ & $ 15.2 \pm 6.5 $  \\ 
\text{[\ion{O}{III}]} 4960 & $ 88.0 \pm 2.1 $ & $ 76.3 \pm 1.6 $  \\ 
\text{[\ion{O}{III}]} 5008 & $ 289.8 \pm 2.7 $ & $ 232.2 \pm 2.3 $  \\ 
\ion{He}{I} 5877 & $ 6.2 \pm 2.1 $ & $ 10.6 \pm 1.5 $  \\ 
\text{[\ion{O}{I}]} 6304 & $ < 10.5 $ & $ < 7.9 $  \\ 
\text{[\ion{O}{I}]} 6367 &  $ < 10.7 $ & $ < 6.3 $  \\ 
\text{[\ion{N}{II}]} 6550 & $ < 11.4 $ & $ 6.3 \pm 24.2 $  \\ 
H$\alpha$ 6565 & $ 149 \pm 6 $ & $ 151 \pm 3 $  \\ 
\text{[\ion{N}{II}]} 6585 & $ < 12.31 $ & $ 0.0 \pm 23.2 $  \\ 
\text{[\ion{S}{II}]} 6718  & $ < 12.19 $ & $ < 7.66 $  \\ 
\text{[\ion{S}{II}]} 6733  & $ < 12.40 $ & $ < 9.55 $  \\ 
\hline
\end{tabular}
\label{tab:integrated_fluxes}
\end{table*}

\subsection{Host galaxy properties inferred from nebular lines diagnostics}

An number of line ratio diagnostics confirm the presence of an AGN in the two objects studied. The H$\alpha$/[\ion{N}{II}] ratios and the $\log_{10}$ [\ion{O}{III}]/H$\beta$ ratios of COLA1 and NEPLA4 are consistent with star-formation dominated emission in the classical BPT diagram. However, numerous studies have pointed out that at high redshift, the classical BPT fails at identifying AGN \cite[e.g.][]{Maiolino2024_AGNsample, Mazzolari2024}. In particular, \cite{Mazzolari2024} propose new AGN diagnostics for high-z AGN relying mostly on the [\ion{O}{III}] 4364 line. We find a relatively low [\ion{O}{III}] 4364 / H$\gamma$ ratio and in COLA1 ($<0.25 $) and NEPLA4 ($0.55\pm0.25$) but also unusually high [\ion{O}{III}]/[\ion{O}{II}]$> 34(32) (2\sigma)$. NEPLA4 and COLA1 thus sit at the limit between star-forming galaxies and AGN in the \cite{Mazzolari2024} diagrams. We note that the very high O32 ratio are consistent with expectations for objects with high Lyman continuum leakage \cite{Izotov2018,Flury2022b}. From the ([\ion{O}{III}] 5008+4960) / [\ion{O}{III}] 4364 ratio we find an electron temperature of $8-9\times 10^3$ K in the centre of NEPLA4 and COLA1 \cite{Osterbrock2006}, indicating a contribution from an AGN.
We note that the AGN primarily contributes to boosting the [\ion{O}{III}] 4363 auroral line in the centre of the objects, whilst the other lines (e.g. [OIII]$5008, 4960 \ \AA$ line) are primarily powered by star-formation (see the discussion on the NLR bolometric estimator above). Indeed, in the best-fit BAGPIPES most lines are well reproduced by stellar emission only, except for [\ion{O}{III}] 4363, further confirming this interpretation.  

Turning to the galaxy properties, we use the $R3=$[\ion{O}{III}] / H$\beta$ to derive metallicities using standard calibrations  \cite{Bian2018,Nakajima2022} . We find metallicities 12+log(O/H) $\sim 7.5$ consistent with that of the average [\ion{O}{III}] emitter at the same redshift \cite{Matthee2023,Meyer2024}. Finally, from the narrow line H$\alpha$ emission (corrected for LyC escape, see below) we derive star-formation rates of $58\pm2, 50\pm1 \  \rm{M_\odot \ yr^{-1}}$ using a Chabrier IMF \cite{Chabrier2003} for COLA1(NEPLA4). The SFR of $\sim 24-28\ M_\odot \ \rm{yr}^{-1}$ derived from the UV luminosity \cite{Songaila2018,Matthee2018} using the \cite{Kennicutt2012} conversion are twice lower, indicating that a recent starburst is happening in those objects. This is further confirmed by the BAGPIPES SED modelling.

\subsection{Stellar masses and star-formation histories}
\label{sec:sed_fitting}
We fit the $1.7-5.3\ \mu\rm{m}$ NIRSpec/IFU spectroscopy of NEPLA4 simultaneously with ancillary photometry using BAGPIPES \cite{Carnall2018}. For NEPLA4 we use legacy SUBARU/HSC HEROES photometry \cite{Taylor2023} and for COLA1 we use both legaccy COSMOS photometry reported in \cite{Matthee2018} including ground-based data as well as Spitzer/IRAC as well as recent JWST/NIRCam imaging in the F115W, F150W, F200W and F356W filters \cite{Torralba-Torregrosa2024}. We use the $r=0.''3$ aperture spectrum and subtract the best-fit broad and outflowing components of the H$\alpha$, H$\beta$ and [OIII] lines from the spectrum to avoid biasing the fits. We find that the $r=0.''15$ best-fit broad components capture all the flux even in the $r=0.''3$ spectrum, further confirming that they are produced by the BLR of an AGN. 
We run BAGPIPES with a non-parametric star-formation history (SFH) with 10 bins logarithmically spaced between the age of the Universe at the redshift of observations and $z=20$. We force the two first bins to be (0,5) and (5,10) Myr to capture young burst of star-formation. We use a "bursty continuity"  prior with student t parameters (1,2) following \cite{Tacchella2022}. We use a flat prior for the ionisation parameter ($-4<\log U<0$), a log prior for the metallicity ($10^{-3}<Z / Z_\odot<2.5$). We set an attenuation prior $0<A_V<2$ using the \cite{Calzetti2000} dust law. 

We obtain tightly constrained fits with very small ages, high ionisation parameters, and low metallicity $\sim 0.1$ solar, indicating a very young starburst in contrast with the much older growth period of the SMBH. The final $A_V\sim 0.3$ are in agreement with the observed values within uncertainties, although we note that fixing the $A_V$ to the observed value (closer to 0) would result in lower stellar masses. Similarly, we note that since BAGPIPES does not include any AGN contribution to the SED, our resulting stellar masses are thus very conservative upper limits.
Even in our conservative fit, the stellar masses are relatively modest $\log_{10} M_*^{COLA1} / M_\odot = {8.93}^{+0.04}_{-0.04}$,$\log_{10} M_*^{NEPLA4} / M_\odot = {9.27}^{+0.02}_{-0.02}$, implying a BH-to-stellar mass ratios of $0.22^{+ 0.04}_{- 0.04} $ and $0.10 ^{+ 0.02}_{- 0.01} $ for COLA1 and NEPLA4. The resulting properties are tabulated in Table \ref{tab:key_properties} alongside the inferred black holes masses. The posterior distribution and best-fit models can be found in Extended Figures \ref{fig:COLA1_bagpipes} and \ref{fig:NEPLA4_bagpipes}.

\subsection{Dynamical Masses}
\label{sec:dyn_mass}
We estimate the dynamical masses of COLA1 and NEPLA4 using the standard estimator 
\begin{equation}
    M_{dyn} = K(n)K(q)\frac{\sigma^2R_e}{G}
\end{equation}
where $K(n), K(q)$ are morphological corrections defined as $K(n) = 8.87-0.831n + 0.0241 n^2$ where $n$ is the Sérsic index \cite{Cappellari2006} and $K(q) = [0.87+0.38e^{-3.71(1-q)}]^2$ with $q$ the axis ratio \citep{vanderWel2022}, $R_e$ the effective radius and $\sigma$ the stellar velocity dispersion. We measure $q,n,R_e$ by summing using inverse-variance weighting the line-free channels of the G395M IFU data of COLA1 and NEPLA4, and fitting a 2D Sérsic profile. The results are tabulated in Table \ref{tab:key_properties}. We find very small sizes, which for COLA1 is in agreement with the results from JWST imaging \cite{Torralba-Torregrosa2024}. The stellar velocity dispersion is estimated from the best-fit FWHM of the narrow [OIII] lines to which we subtract in quadrature the instrumental resolution at $\sim4 \ \rm{\mu m}$, and add $+0.1$ dex to correct from the ionised gas dispersion to the stellar velocity dispersion \cite{Bezanson2018}. We find dynamical masses of $M_{dyn}= 2.6\pm2.2, 1.1\pm0.4 \times 10^{10}\ M_\odot\ \rm{yr^{-1}}$ for COLA1, NEPLA4, respectively. The implied BH-to-dynamical mass ratio are high, e.g. $\sim 0.2(0.1)$ for COLA1(NEPLA4), highly above the expected fundamental relation \cite{KormendyHo2013} and most other JWST AGN \cite[e.g.][]{Maiolino2024_AGNsample, Juodzbalis2026}. Note that the overmassiveness of these two objects is also apparent when considering the BH-to-dynamical mass ratio.

\subsection{Outflow energetics}
Before turning to the host galaxy properties, we first study the (intermediate) broad components ($600 < \rm{FWHM} < 1800 \rm{\ km\ s }^{-1}$) which we interpret as AGN-powered outflows. Following \cite[][and references therein]{Ubler2023} we estimate the ionised gas mass outflow rate from H$\alpha$
\begin{equation}
     \dot M_{ion, out}  (H\alpha) [M_\odot\ \rm{yr}^{-1}] =  \frac{1.36 m_H}{\gamma_{H\alpha}} L_{H\alpha} \left( \frac{v_{max}}{1000} \right)\left( \frac{n_e}{1000} \right)^{-1}R_{out}^{-1}
\end{equation}
   
where $\gamma_{H\alpha}=3.25\times10^{-25} \rm{erg\ s^{-1} cm^{-3}}$ is the $H\alpha$ emissivity at $T_e=10^4\ \rm{K}$, $v_{max}$ is the maximum outflow velocity in units of $\rm{km\ s}^{-1}$, $n_e$ the electron density in $\rm{cm}^{-3}$ and $R_{out}$ the extent of the outflow in kpc. We assume $n_e =1000 \rm{cm}^{-3}$ and $R_{out} = R_e$ from our S\'ersic fit (see Section \ref{sec:dyn_mass}).

We find mass outflow rates of $\dot M_{ion, out} (H\alpha)= 4.7\pm3.6, 9.2\pm4.2 \ M_\odot\ \rm{yr}^{-1}$ for COLA1, NEPLA4 respectively. The corresponding mass loading factors using the H$\alpha$-based SFR are relatively low $\eta=0.08\pm0.06, 0.18\pm0.08$. The kinetic power of the outflow $P_{k,ion}$ corresponds to $\sim 0.03\%$ and $\sim 0.25\%$ of the bolometric luminosity, in good agreement with the established trend for AGN of similar bolometric luminosities at lower-redshift \cite[e.g.][]{Fiore2017}. The low kinetic power and loading factor suggest that negative feedback is \textit{currently} not important in NEPLA4 and COLA1, consistent with the recent starburst phase and decline of the quasar activity.

\subsection{Ground-based Lyman-$\alpha$ spectroscopy and inferred escape fraction}
\begin{figure}
    \centering
    \includegraphics[width=\linewidth]{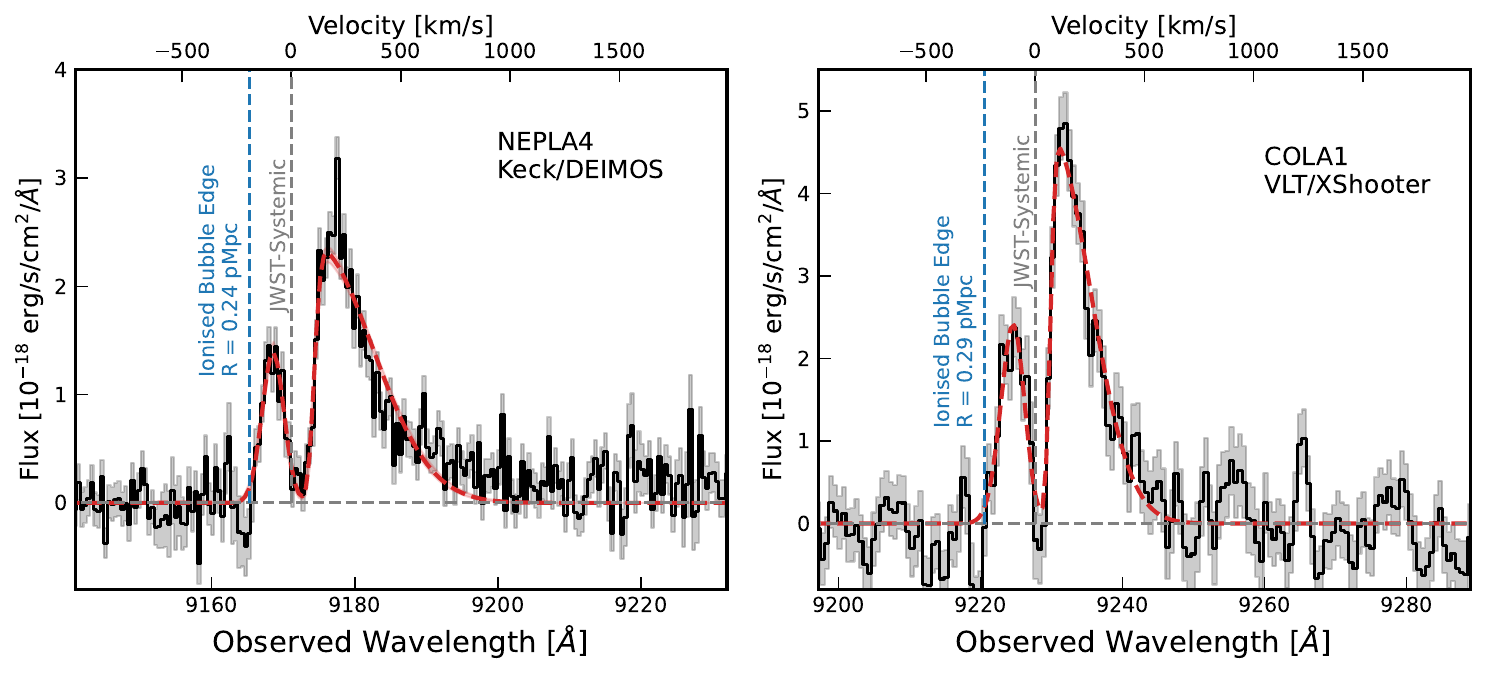}
    \caption{Lyman-$\alpha$ spectra in observed wavelength. Systemic redshifts from JWST are indicated in gray (vacuum-air corrected). The edge of the ionised bubble, defined using the wavelength where the blue peak goes to zero, is indicated in dashed blue. The best-fit double skewed Gaussian model is shown in dotted red. }
    \label{fig:lyman_alpha}
\end{figure}

We use existing ground-based spectroscopy of the Lyman-$\alpha$ profile of COLA1 and NEPLA4. For COLA1 we use the publicly available XShooter spectrum from \cite{Matthee2018}, whereas for NEPLA4 we re-reduce the original Keck/DEIMOS data published \cite{Songaila2018} using Pypeit version $1.18.0$ \cite{pypeit:joss_pub,pypeit:zenodo} with standard parameters. We show the two Lyman-$\alpha$ profiles in Extended Figure \ref{fig:lyman_alpha}. We find that the JWST systemic redshift from the nebular lines aligns very well with the absorption trough of the double-peaked profiles \cite[as already reported for COLA1 by ][]{Torralba-Torregrosa2024}. We note that our best-fit systemic redshift is closer to that of the Module B NIRCam/WFSS redshift of \cite[][]{Torralba-Torregrosa2024}, who adopt the module A redshift due to better consistency between the [\ion{O}{III}] 4960,5008 and H$\beta$ redshifts in that module. This has little incidence on the results, except for the minimum ionised bubble size which changes by $\sim 30\%$ between the two redshift solutions. 

We fit both profiles with two skewed gaussians and find best-fit peak separation and Lyman-$\alpha$ luminosities in agreement with previous works (see Table \ref{tab:key_properties}). We then use the Lyman-$\alpha$ double-peak separation from the best-fit model to derive the Lyman Continuum escape fraction following the empirical relation from \cite{Izotov2018} 
\begin{equation}
    f_{esc}(LyC) =  0.095 - \frac{1.05\times10^2}{\Delta v_{LyA}} + \frac{3.34\times10^{4}}{\Delta v_{LyA}^2}
\end{equation}
where $\Delta v_{LyA}$ is the peak separation in $\rm{km\ s}^{-1}$. We propagate the uncertainty from our best-fit models on the escape by resampling the covariance matrix and re-measuring the double-peak separation. We find escape fractions of $f_{esc}(LyC, NEPLA4) = 0.20^{+0.04}_{-0.02}$ and  $f_{esc}(LyC, COLA1) = 0.29^{+0.10}_{-0.01}$. We then use the H$\alpha$ luminosities to derive the ionising efficiencies and Lyman$\alpha$ escape fraction. We do not correct the H$\alpha$ fluxes for dust attenuation since the Balmer decrement in the narrow line, integrated fluxes, are consistent with no dust attenuation. Assuming an intrinsic ratio of H$\alpha$/Lyman-$\alpha =8.7$, we find LyC-corrected Lyman-$\alpha$ escape fractions of $f_{esc}(Ly\alpha,NEPLA4) = 0.28^{+0.02}_{-0.01}$ and $f_{esc}(Ly\alpha,COLA1) = 0.52^{+0.07}_{-0.03}$. In both cases the Lyman-$\alpha$ escape fractions are superior to the Lyman continuum escape fraction, consistent with the theoretical expectation and empirical results at low redshift \cite[e.g.][]{Verhamme2017}. Additionally, the consistent escape fractions, H$\alpha$ and Lyman-$\alpha$ fluxes confirm the scenario where the narrow lines are predominantly emitted by star-forming regions in the galaxy. Note that throughout the paper, quantities based on H$\alpha$ fluxes (SFR, $\xi_{ion}$) are corrected for the best-fit LyC escape fraction.

We now derive the effective ionising output of the objects (at the time of observation). The ionising efficiencies (corrected for the measured LyC escape fraction) are $\log_{10}\xi_{ion} / [erg\ s^{-1} Hz^{-1}] = 25.45^{+0.04}_{-0.04}, 25.37^{+0.02}_{-0.02}$ for NEPLA4, COLA1, respectively. This is in excellent agreement with the average for $z\sim 6-7$ galaxies \cite[e.g.][]{Simmonds2024,Pahl2025}, thus indicating that NEPLA4 and COLA1 are not extreme ionising sources. Note that the literature typically reports ionising efficiencies assuming an escape fraction of $0\%$. The above values should be corrected by $0.08$ and $0.14$ dex for a direct comparison if required. For COLA1 our results are in excellent agreement with that derived from the H$\beta$ line observed in NIRCam WFSS by \cite{Torralba-Torregrosa2024}.  These values are close to that assumed in the analysis of \cite{Meyer2021}, which concluded that the ionising output of the galaxy itself is not sufficient to maintain the IGM sufficiently ionised at $\sim 0.3$ pMpc to enable the escape of blue peak photons. 

\subsection{Producing enough ionising photons to match the Lyman-$\alpha$ blue-peak transmission}

With the systemic redshift in hand, we can now precisely measure the minimum size of the ionised bubble required to enable the transmission of blue peak Lyman-$\alpha$ photons. We measure the maximal extent of the blue peak as the location where the blue peak flux drops to $0$ or below, which translates to a \textit{minimum} ionised bubble  size of $R_{min,\alpha }= 0.24(0.29)\ \rm{pMpc}$, in agreement with previous works \cite{Matthee2018,Torralba-Torregrosa2024,Meyer2021}. Using the Strömgren sphere approximation, this size is consistent with the cumulative ionising output of COLA1 and NEPLA4 if they have been forming star at a constant rate and had a constant high escape fraction for $\gtrsim 100\ \rm{Myr}$ \cite{Meyer2021,Torralba-Torregrosa2024}. Given the very young stellar population inferred from the SED and spectra (see Methods), it is unlikely that COLA1 and NEPLA4 carved out such large ionised bubbles. We also note that JWST NIRCam WFSS observations recently reported the detection of several UV-faint galaxies around COLA1 (although an average number for such a UV bright galaxy) \cite{Torralba-Torregrosa2024}. However, this mild overdensity only makes a small contribution to the local photo-ionisation rate that is dwarfed by that of COLA1 itself \cite{Torralba-Torregrosa2024}, thus not changing our argument.

More problematic than the edge of the bubble is the transmission of the blue-peak photons, which is governed by the residual neutral fraction in the ionised bubble. \cite{Mason2020} find that an extremely low neutral fraction $x_{\rm{HI}}\simeq 10^{-6}$ is necessary to explain the transmission of the blue peak photons at the level observed. However, the ionising output of COLA1 and NEPLA4 is not sufficient to keep the residual hydrogen at this level. Following the analysis in \cite{Meyer2021}, we compute the photo-ionisation rate at a distance $r$ from the source as
\begin{equation}
    \Gamma(r) =  \frac{\alpha_g \sigma_{\rm{912}}}{\alpha_g+3} \frac{f_{\rm{esc}} Q_{\rm{ion}} }{4\pi r^2} \rm{e}^{-r/\lambda_{\rm {mfp}}} 
\end{equation}
where $\alpha_g=2$ is the galaxy far-UV slope, $\sigma_{912}$ is the ionising photon cross-section at $912$ \AA, $\lambda_{\rm{mfp}}= 6\left( \frac{1+z}{7}\right) \rm{Mpc}$ \citep{Worseck2014} the ionising photon mean free path. We use the escape fraction and the ionising output measured from the narrow H$\alpha$ line from this work. The residual IGM neutral fraction in the bubble can then be computed using the fluctuating Gunn-Peterson approximation \cite[e.g.][]{Becker2015_rev}
\begin{equation}
    x_{\rm {HI} } = (\Gamma(r) + \Gamma_{\rm{bkg}}) C_{\rm {HII} } \Delta n_H(z) \alpha_B(T) \text{\ \ \ \ .}
\end{equation}
We assume a clumping factor $C_{\rm {HII} }=2$, overdensity factor $\Delta=1$, a UV background $\Gamma_{\rm{bkg}} = 10^{-13}\ \rm{s}^{-1}$ and a Case-B recombination coefficient $\alpha_B(T=10^4\ K) = 2.6\times 10^{-13} (T/10^4)^{-0.8}\ \rm{cm}^3\ \rm{s}^{-1}$. 

We find that the equilibrium neutral fraction around COLA1 and NEPLA4 (their total ionising output is similar) is $x_{\rm{HI}}(r=0.15\ \rm{Mpc}) \simeq 1.4\times 10^{-5},x_{HI}(r=0.3\ \rm{Mpc}) \simeq 3.6\times 10^{-5}$. This a factor $14-36$ higher than what the transmission of the blue peak photons requires. Instead, the required photo-ionisation rate and ionised bubble could be easily sustained when the AGN in COLA1 and NEPLA4 is accreting close to Eddington. Their much UV higher luminosities (see Methods; $M_{UV}^{AGN}(\lambda_{Edd}=1) \simeq -23.2$, e.g a factor $\sim 4.5$ times higher), coupled with a Lyman continuum escape fraction approaching unity in AGN and a harder ionising spectrum \citep[e.g.][]{Cristiani2016,Grazian2018,Romano2019} would easily boost the photo-ionisation rate by the factor $\gtrsim 15-30$ required to explain the transmission of blue peak photons. Once the quasar phase stops, the neutral fraction returns to equilibrium  on a timescale $t \sim \Gamma_\mathrm{bg}^{-1} \approx 1\ \rm{Myr}$ \cite{Davies2020}. This is rather short but, as shown below, consistent with the other quasar duty cycle constraints at $z\sim 6$.

\subsection{Quasar duty cycle constraints}

We take the $z=6.6$ quasar number density from \cite{Wang2019} who measure $\phi(-25.5<QSO,M_{UV}<-27.6) = 0.39\pm0.11 \ \rm{Gpc}^{-3}$ at $z\simeq 6$. For (double-peaked) ultra-luminous Lyman-$\alpha$ emitters we use the HEROES survey \cite{Taylor2025}, a sample of Narrow-Band selected \textit{and} spectroscopically confirmed ultra-luminous LAEs ($\log L_{Ly\alpha} \gtrsim 43.5$) in various deep fields that include NEPLA4 and COLA1. The spectroscopic confirmations are important as they enable us to establish the double-peak statistics (2/56, e.g. only COLA1 and NEPLA4 are double-peaked in their $z=6.6$ sample).
We note that at lower-redshift, only $\sim 1/3$ of Lyman-$\alpha$ emitters show a double-peak due to absorption by CGM gas and inflows \cite{Kulas2012,Kerutt2022,Vitte2025}. Since we are interested in the number of objects with a recently quiescent AGN and a remnant large ionised bubble regardless of the intrinsic Lyman-$\alpha$ profile, we correct the number density of double-peaked Lyman-$\alpha$ emitters (DOPLA) in HEROES survey by a factor 3.3 to find $\phi(DOPLA + SMBH) = 1.1^{+0.7}_{-0.6}\times10^{-7}\ \rm{Mpc}^{-3}$. 

Finally, the density of luminous LAEs hosting a dormant SMBH but without a large ionised bubble can be bounded in two ways. On the one hand, we can first assume that all ultra-luminous LAEs in HEROES host a SMBH comparable to that of COLA1 and NEPLA4, finding $\phi(LAE+SMBH) = 9.4^{+5.5}_{-3.7}\times10^{-7}\ \rm{cMpc}^{-3}$. On the other hand, we can use the recent constraints from SHELLQS that place a lower limit on the number of SMBH in their ultra-luminous LAEs $\phi(LAE+SMBH)>2\times10^{-8}\ \rm{cMpc^{-3}}$ \citep[][]{Matsuoka2025}. We then compute the quasar duty cycle, i.e. the ratio of the number density of quasars to that of inactive SMBH, as $f_{duty} = \frac{\phi(QSO)}{\phi(QSO) + \phi(LAE+SMBH) + \phi(DOPLA + SMBH)}$ and we find 
\begin{equation}
   0.11\times 10^{-3}  < f_{duty}(z=6.6) < 2.4 \times 10^{-3}  \ \ \ \ (2\sigma \ \rm{limits})
\end{equation}

We show in Extended Figure \ref{fig:duty_cycle} the good agreement between our constraints and that inferred from clustering and damping wing measurements at $z\sim 6-7$. At $z=6.6$, this corresponds to an average quasar lifetime of $0.09 < t_{\rm{QSO}} [\rm{Myr}] < 1.84$, in agreement with previous studies \citep[e.g.][]{Eilers2017,Eilers2020, Davies2020, Satyavolu2023, Eilers2024}. As already pointed out in these studies, our results are in strong tension with a constant near or at Eddington accretion required to grow a $10^8\ M_\odot$ SMBH by $z=6.6$ that would imply $t_{\rm{QSO}}\sim 1\ \rm{Gyr}$ and $f_{duty}\sim 0.5$. Episodic quasar accretion at high accretion rates has already been proposed to reconcile proximity zone and clustering measurements with the presence of $>10^8 \ M_\odot$ BH at $z>6.6$ \citep[e.g.][]{Eilers2020, Davies2020,Satyavolu2023b}. 

Our unique method enables us to measure the number and duration of these episodic accretion events. Given the relative number densities of double-peaked Lyman-$\alpha$ emitters and quasars, this also implies the object spends $\sim 80\ \rm{Myr}$ in a double-peak phase similar to COLA1 and NEPLA4. Assuming an average time of $1 \ \rm{Myr}$ for this phase (see above), we derive a number of episodic accretion events $N\sim 80$ and an episodic accretion timescale $1 \lesssim t_{\rm{ep}} [10^3\ \rm{Myr}] \lesssim 23$. These are sligthly shorter timescales than those proposed in the literature \citep[e.g.][]{Davies2020,Satyavolu2023b}, which is consistent with our sample targeting lower-luminosity quasars powered by lower-mass BHs. 

\backmatter





\subsection*{Acknowledgements}

RAM, CW, PO acknowledge support from the Swiss State Secretariat for Education, Research and Innovation (SERI) under contract number MB22.00072, as well as the Swiss National Science Foundation (SNSF) through project grant 200020\_207349. SEIB is supported by the Deutsche Forschungsgemeinschaft (DFG) under Emmy Noether grant number BO 5771/1-1. The Cosmic Dawn Center (DAWN) is funded by the Danish National Research Foundation under grant DNRF140.
The data were obtained from the Mikulski Archive for Space Telescopes at the Space Telescope Science Institute, which is operated by the Association of Universities for Research in Astronomy, Inc., under NASA contract NAS 5-03127 for JWST. These observations are associated with program \#3767. Some of the data presented herein were obtained at Keck Observatory, which is a private 501(c)3 non-profit organization operated as a scientific partnership among the California Institute of Technology, the University of California, and the National Aeronautics and Space Administration. The Observatory was made possible by the generous financial support of the W. M. Keck Foundation. The authors wish to recognize and acknowledge the very significant cultural role and reverence that the summit of Maunakea has always had within the Native Hawaiian community. We are most fortunate to have the opportunity to conduct observations from this mountain.

\subsection*{Data Availability}
The JWST (\#GO~3767, PI: Meyer), XShooter (PID: , PI: Matthee) and Keck (PID:H291, Semester 2019B, PI:Hu) data used in this paper are all publicly available from the JWST, ESO and Keck online archives.

\subsection*{Code Availability}
The JWST data were processed with the JWST calibration pipeline (\url{https://jwst-pipeline.readthedocs.io}) and the TEMPLATES ERS outlier rejection algorithm \cite{Hutchison2024} \url{https://github.com/aibhleog/baryon-sweep}.

\subsection*{Author Contribution}
R.A.M led the observation programme, data reduction, data analysis, and manuscript preparation. 
R.S.E, S.E.I.B, A.D., N.L, F.W all contributed to the preparation of the observation programme. 
P.A.O, C.W provided consulting on the SED fitting, IFU analysis of the data and manuscript preparation.
F.D. provided the new reduction of the NEPLA4 Keck/DEIMOS spectrum.
J.M provided the JWST photometry of COLA1.
All authors contributed to the discussion of the results and the manuscript preparation.

\begin{appendices}

\section{Extended Data \& Figures}\label{extended_data_fig}

\begin{figure}
    \centering
    \includegraphics[width=0.92\linewidth]{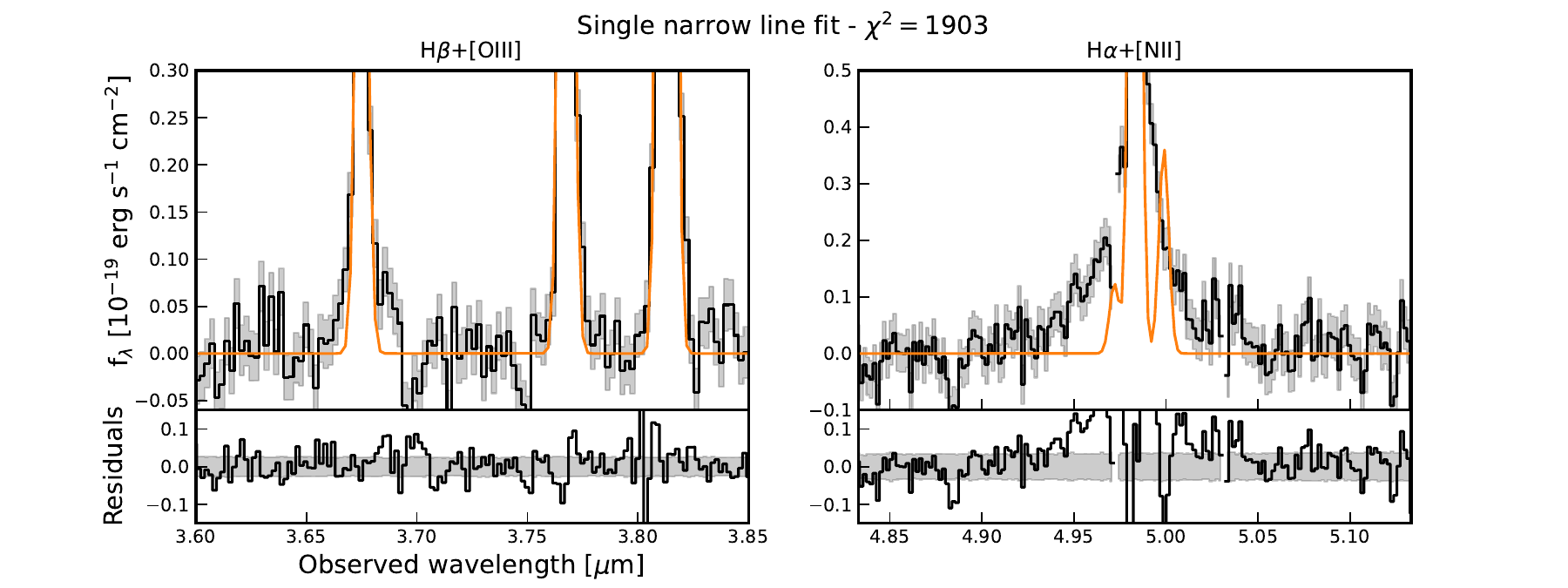}
    \includegraphics[width=0.92\linewidth]{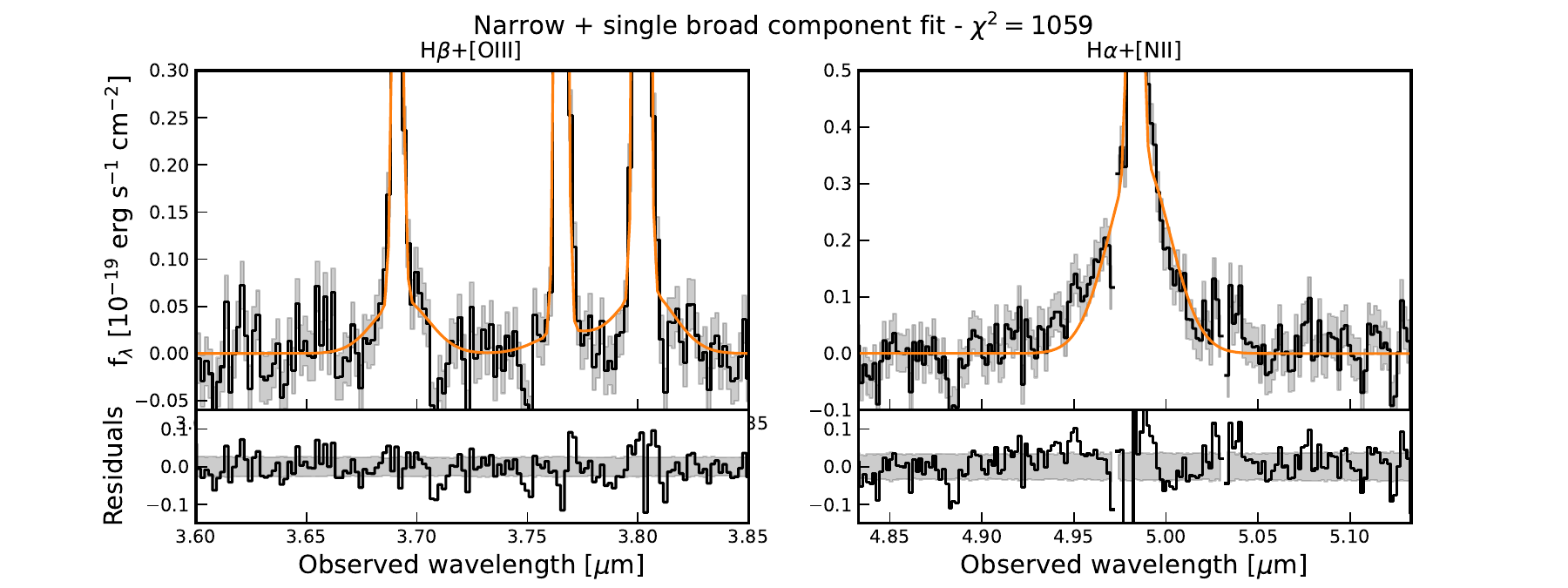}
    \includegraphics[width=0.92\linewidth]{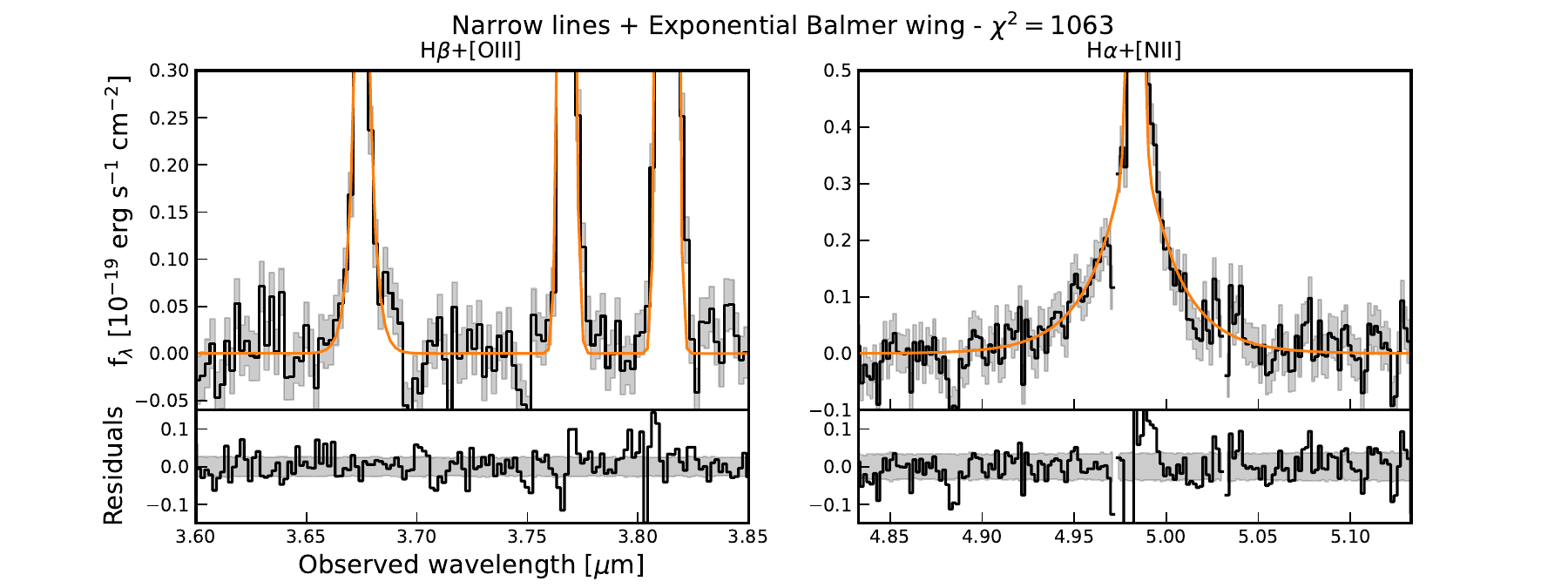}
    \includegraphics[width=0.92\linewidth]{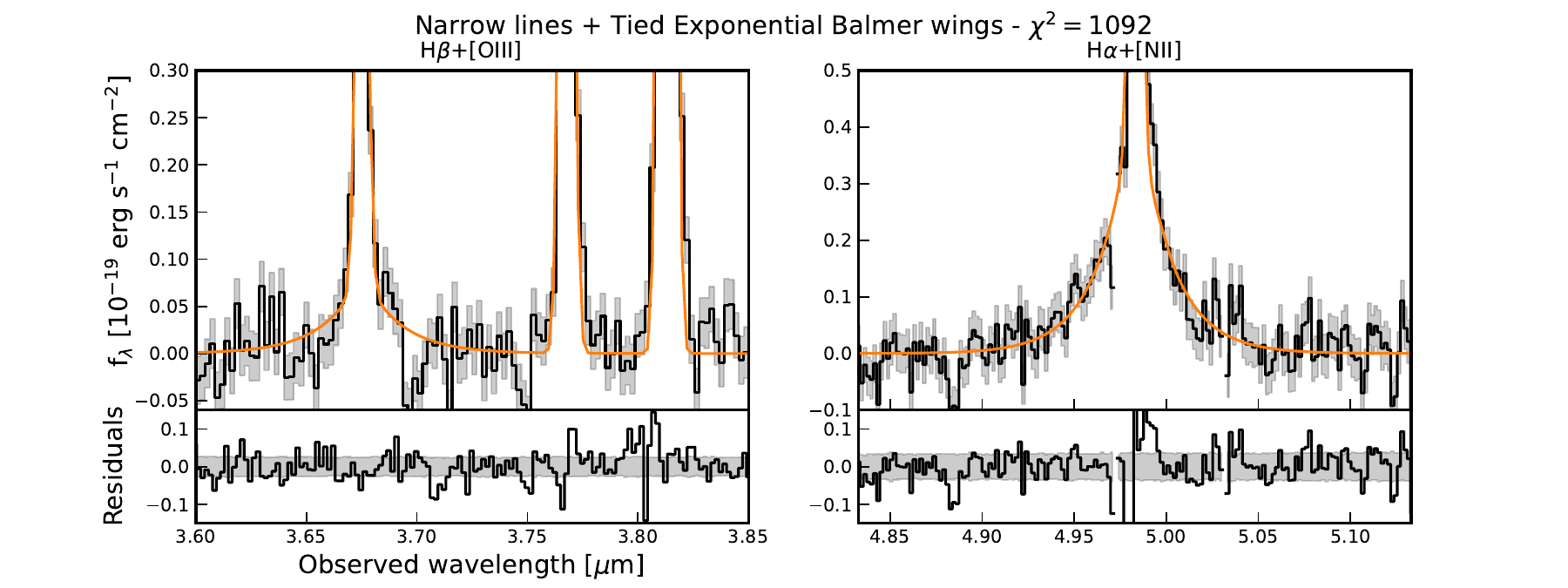}
    \caption{Rejected best-fit models for the Balmer complexes in COLA1  using a variety of narrow lines, single broad component and exponential wing for the Balmer components (with and tied and FWHM). All models perform worse than our fiducial model with a narrow component, outflowing component and an additional broad Balmer component.}
    \label{fig:COLA1_fits}
\end{figure}

\begin{figure}
    \centering
    \includegraphics[width=0.92\linewidth]{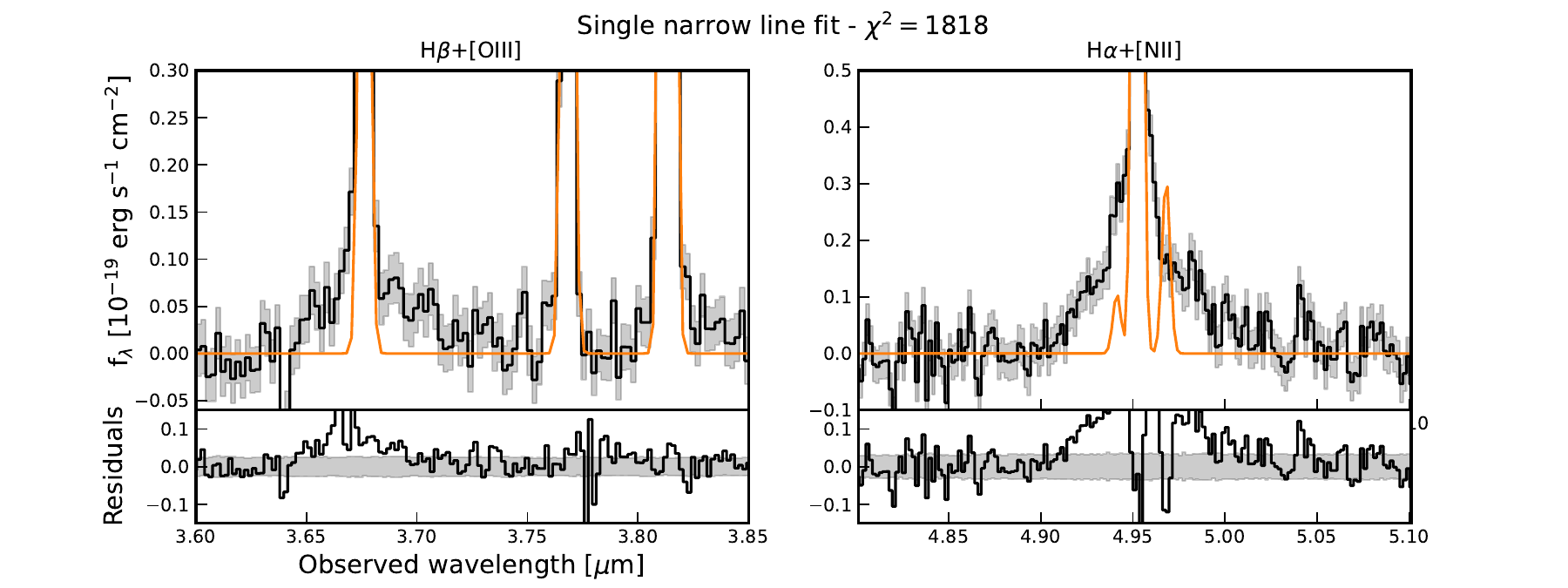}
    \includegraphics[width=0.92\linewidth]{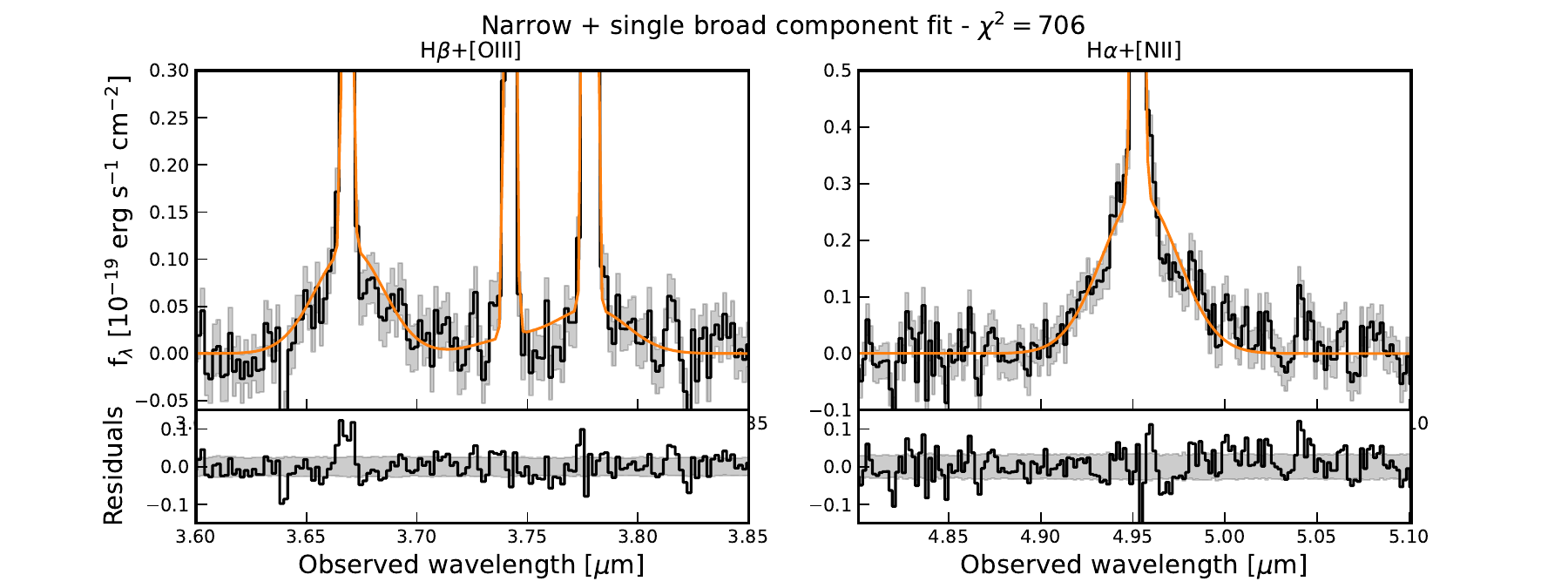}
    \includegraphics[width=0.92\linewidth]{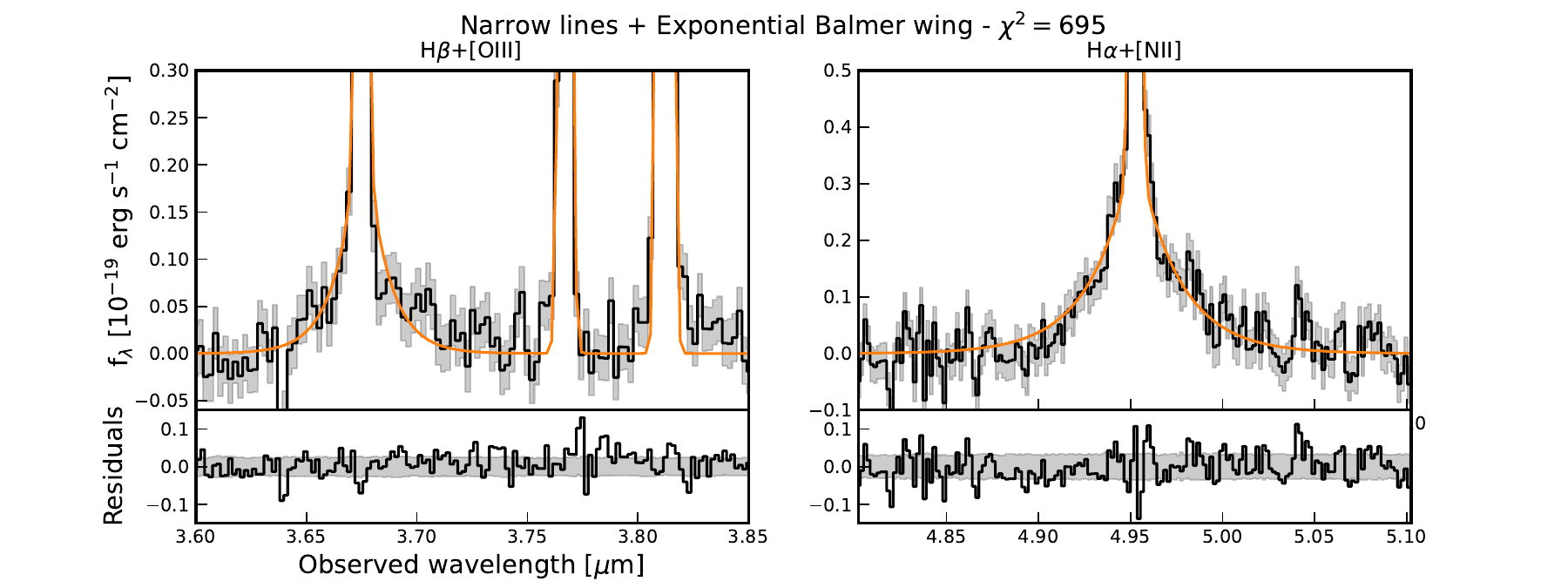}
    \includegraphics[width=0.92\linewidth]{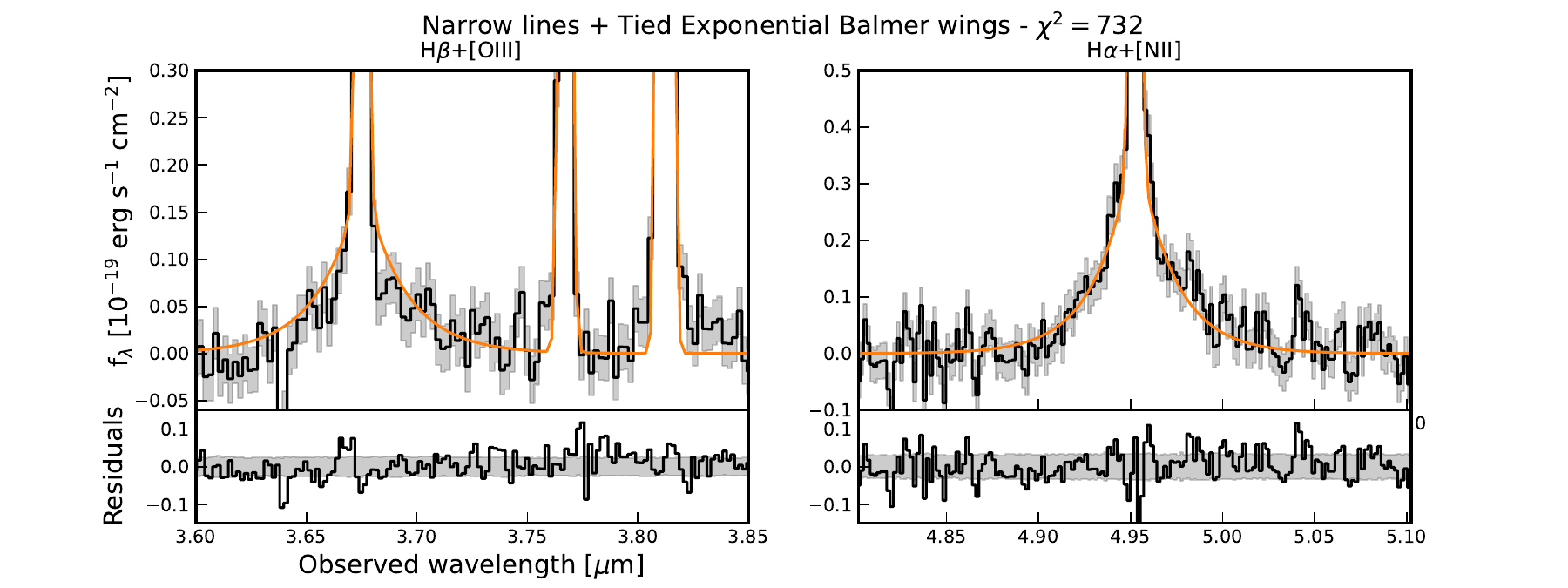}
    \caption{Rejected best-fit models for the Balmer complexes in COLA1  using a variety of narrow lines, single broad component and exponential wing for the Balmer components (with and tied and FWHM). All models perform worse than our fiducial model with a narrow component, outflowing component and an additional broad Balmer component.}
    \label{fig:NEPLA4_fits}
\end{figure}

\begin{figure}
    \centering
    \includegraphics[width=0.8\linewidth]{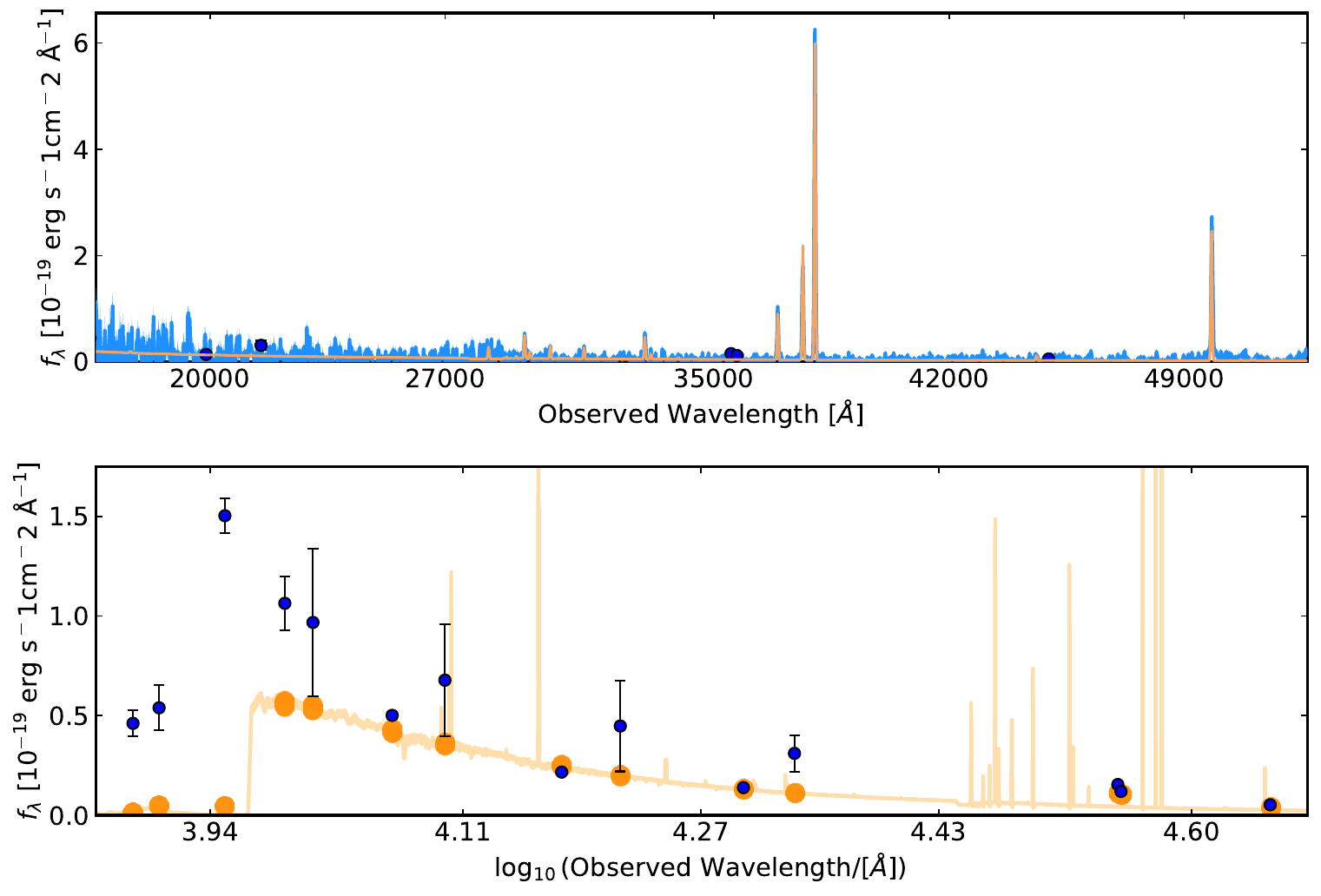}
    \includegraphics[width=0.85\linewidth]{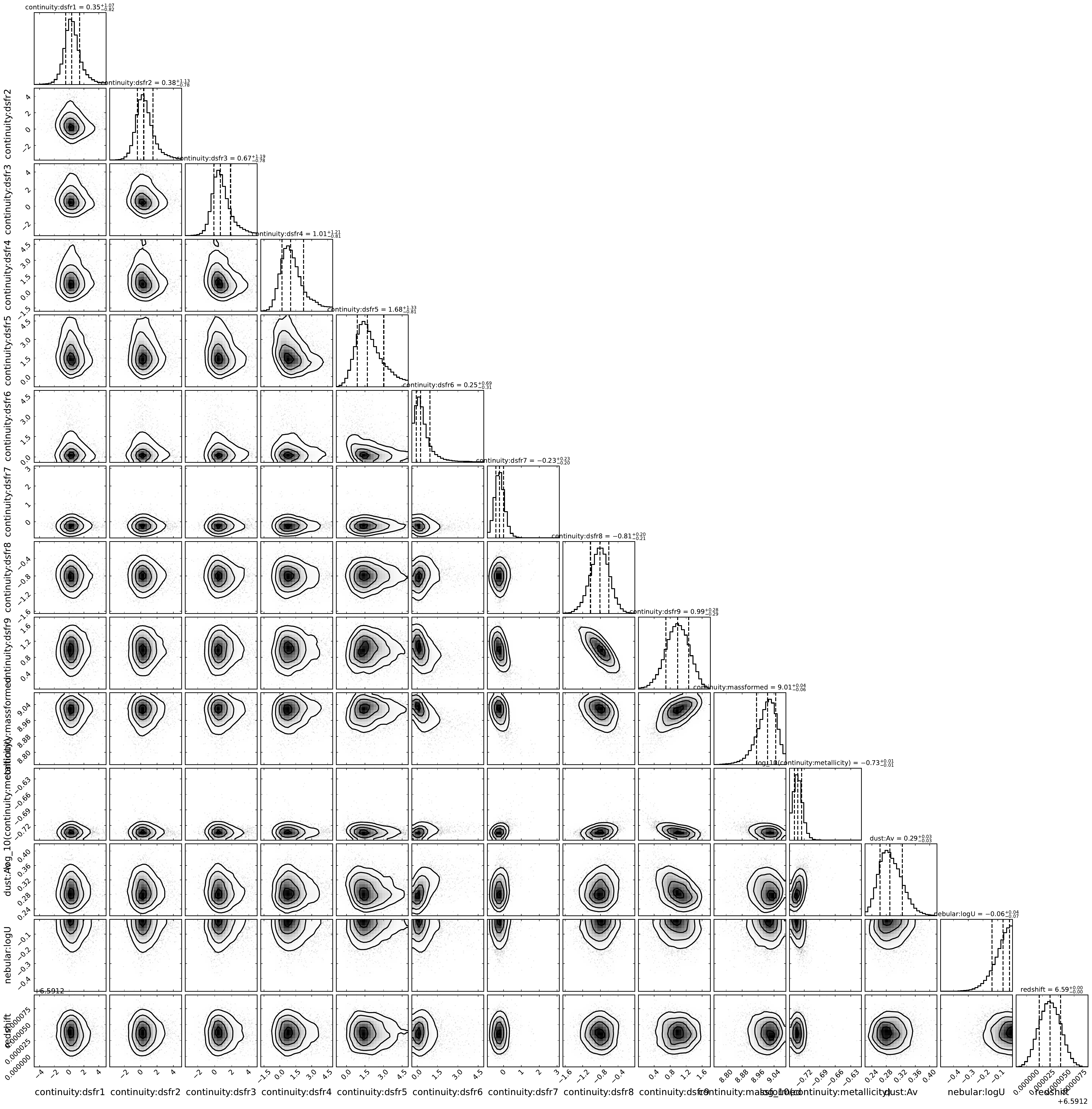}
    \caption{COLA1 observed photometry and spectroscopy and best-fit BAGPIPES model  (top two panels). The model parameter posterior distribution is shown in the lower panel.}
    \label{fig:COLA1_bagpipes}
\end{figure}

\begin{figure}
    \centering
    \includegraphics[width=0.8\linewidth]{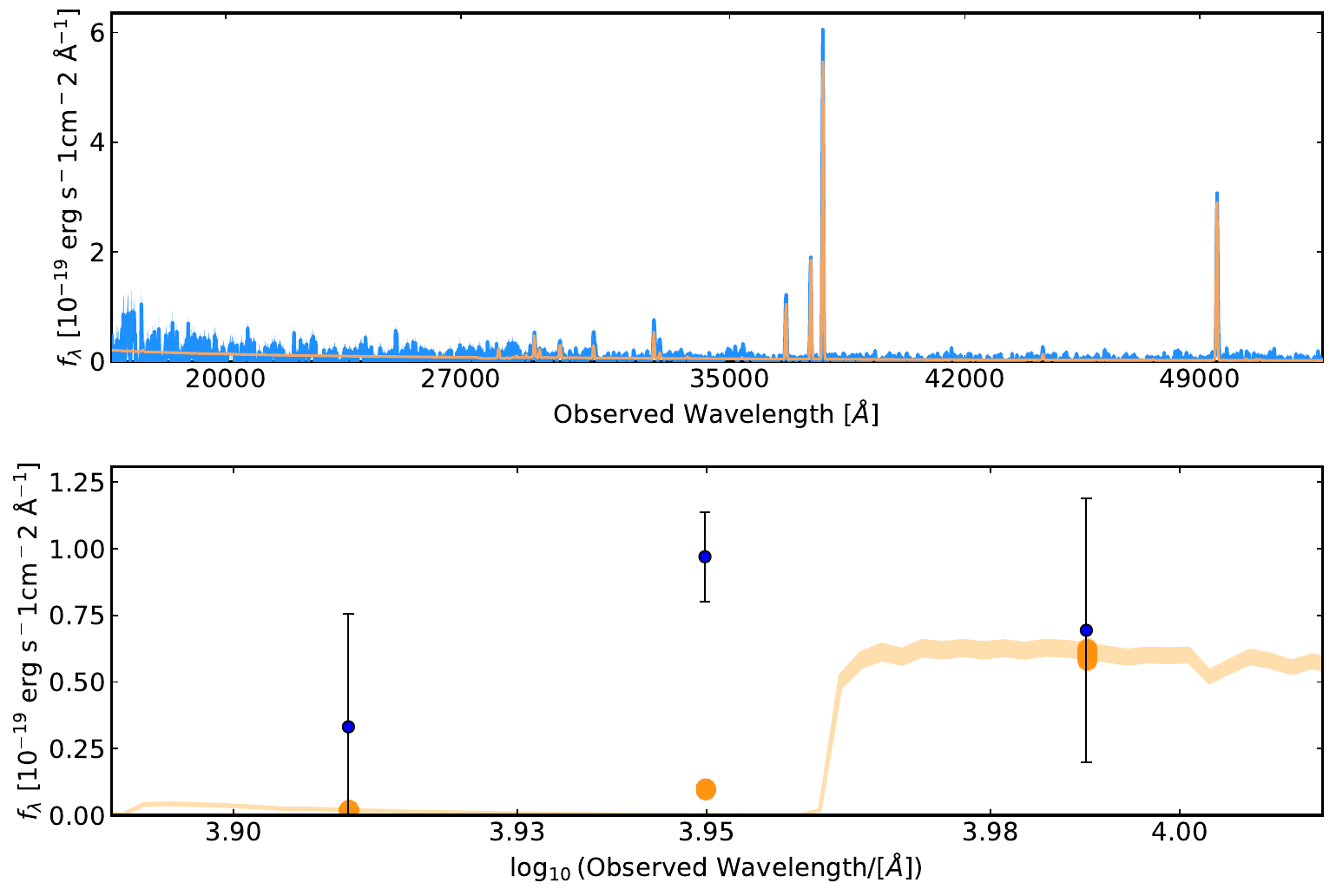}
    \includegraphics[width=0.85\linewidth]{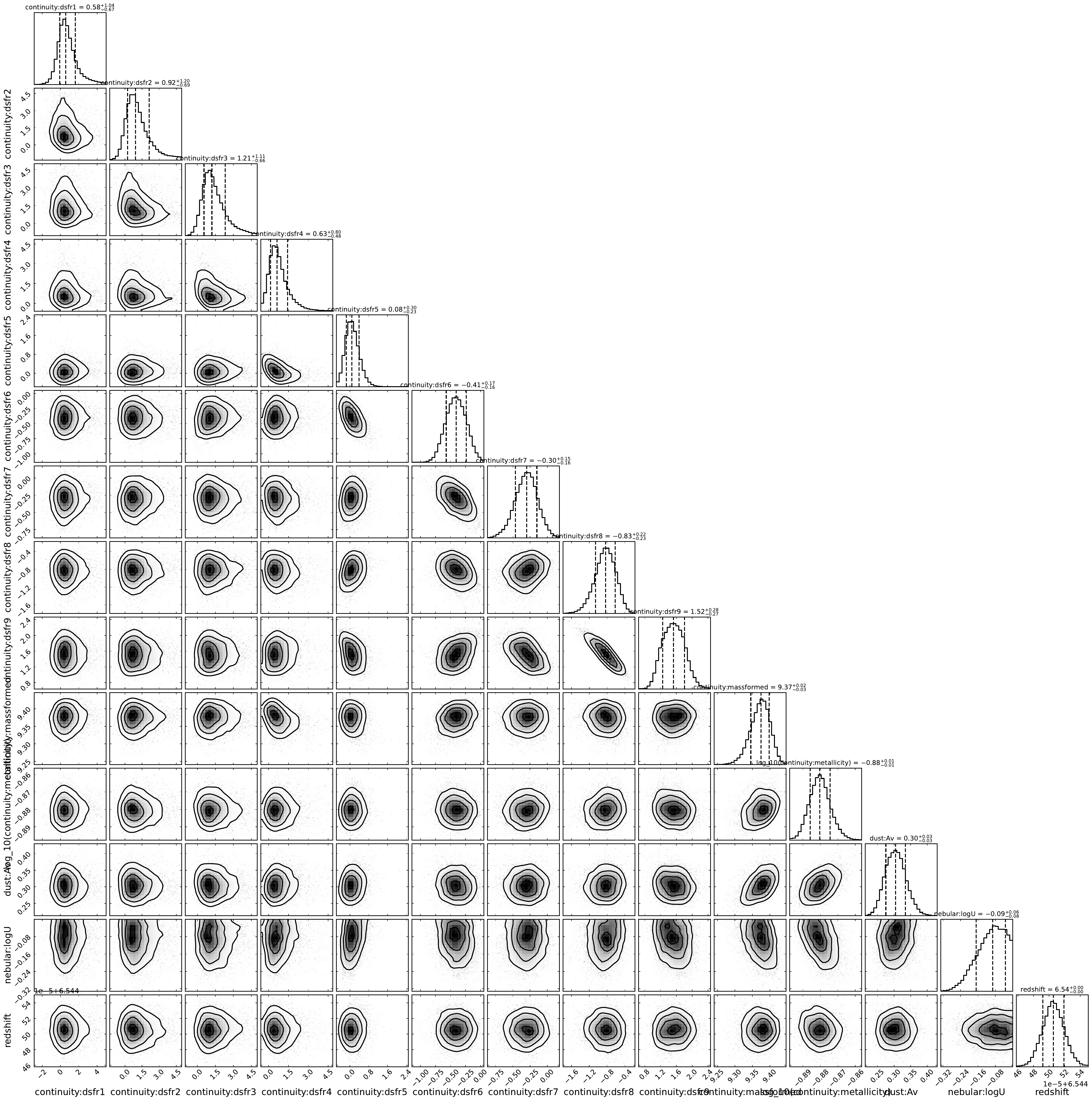}
    \caption{NEPLA4 observed spectroscopy and photometry and best-fit BAGPIPES model (top two panels). The model parameter posterior distribution is shown in the lower panel.}
    \label{fig:NEPLA4_bagpipes}
\end{figure}

We present here the rejected alternative models for the Balmer complexes of COLA1 and NEPLA4 in Extended Figures \ref{fig:COLA1_fits} and \ref{fig:NEPLA4_fits}, respectively. We then present the best-fit BAGPIPES models and parameter posterior distribution in Extended Figures \ref{fig:COLA1_bagpipes} and \ref{fig:NEPLA4_bagpipes}.

\end{appendices}


\bibliography{sn-article}


\begin{thebibliography}{121}
\ifx \bisbn   \undefined \def \bisbn  #1{ISBN #1}\fi
\ifx \binits  \undefined \def \binits#1{#1}\fi
\ifx \bauthor  \undefined \def \bauthor#1{#1}\fi
\ifx \batitle  \undefined \def \batitle#1{#1}\fi
\ifx \bjtitle  \undefined \def \bjtitle#1{#1}\fi
\ifx \bvolume  \undefined \def \bvolume#1{\textbf{#1}}\fi
\ifx \byear  \undefined \def \byear#1{#1}\fi
\ifx \bissue  \undefined \def \bissue#1{#1}\fi
\ifx \bfpage  \undefined \def \bfpage#1{#1}\fi
\ifx \blpage  \undefined \def \blpage #1{#1}\fi
\ifx \burl  \undefined \def \burl#1{\textsf{#1}}\fi
\ifx \doiurl  \undefined \def \doiurl#1{\url{https://doi.org/#1}}\fi
\ifx \betal  \undefined \def \betal{\textit{et al.}}\fi
\ifx \binstitute  \undefined \def \binstitute#1{#1}\fi
\ifx \binstitutionaled  \undefined \def \binstitutionaled#1{#1}\fi
\ifx \bctitle  \undefined \def \bctitle#1{#1}\fi
\ifx \beditor  \undefined \def \beditor#1{#1}\fi
\ifx \bpublisher  \undefined \def \bpublisher#1{#1}\fi
\ifx \bbtitle  \undefined \def \bbtitle#1{#1}\fi
\ifx \bedition  \undefined \def \bedition#1{#1}\fi
\ifx \bseriesno  \undefined \def \bseriesno#1{#1}\fi
\ifx \blocation  \undefined \def \blocation#1{#1}\fi
\ifx \bsertitle  \undefined \def \bsertitle#1{#1}\fi
\ifx \bsnm \undefined \def \bsnm#1{#1}\fi
\ifx \bsuffix \undefined \def \bsuffix#1{#1}\fi
\ifx \bparticle \undefined \def \bparticle#1{#1}\fi
\ifx \barticle \undefined \def \barticle#1{#1}\fi
\bibcommenthead
\ifx \bconfdate \undefined \def \bconfdate #1{#1}\fi
\ifx \botherref \undefined \def \botherref #1{#1}\fi
\ifx \url \undefined \def \url#1{\textsf{#1}}\fi
\ifx \bchapter \undefined \def \bchapter#1{#1}\fi
\ifx \bbook \undefined \def \bbook#1{#1}\fi
\ifx \bcomment \undefined \def \bcomment#1{#1}\fi
\ifx \oauthor \undefined \def \oauthor#1{#1}\fi
\ifx \citeauthoryear \undefined \def \citeauthoryear#1{#1}\fi
\ifx \endbibitem  \undefined \def \endbibitem {}\fi
\ifx \bconflocation  \undefined \def \bconflocation#1{#1}\fi
\ifx \arxivurl  \undefined \def \arxivurl#1{\textsf{#1}}\fi
\csname PreBibitemsHook\endcsname

\bibitem[\protect\citeauthoryear{{Fan} et~al.}{2023}]{Fan2023}
\begin{barticle}
\bauthor{\bsnm{{Fan}}, \binits{X.}},
\bauthor{\bsnm{{Ba{\~n}ados}}, \binits{E.}},
\bauthor{\bsnm{{Simcoe}}, \binits{R.A.}}:
\batitle{{Quasars and the Intergalactic Medium at Cosmic Dawn}}.
\bjtitle{\araa}
\bvolume{61},
\bfpage{373}--\blpage{426}
(\byear{2023})
\doiurl{10.1146/annurev-astro-052920-102455}
{\href{https://arxiv.org/abs/2212.06907}{{arXiv:2212.06907}}}
{[astro-ph.GA]}
\end{barticle}
\endbibitem

\bibitem[\protect\citeauthoryear{{Volonteri} et~al.}{2021}]{Volonteri2021}
\begin{barticle}
\bauthor{\bsnm{{Volonteri}}, \binits{M.}},
\bauthor{\bsnm{{Habouzit}}, \binits{M.}},
\bauthor{\bsnm{{Colpi}}, \binits{M.}}:
\batitle{{The origins of massive black holes}}.
\bjtitle{Nature Reviews Physics}
\bvolume{3}(\bissue{11}),
\bfpage{732}--\blpage{743}
(\byear{2021})
\doiurl{10.1038/s42254-021-00364-9}
{\href{https://arxiv.org/abs/2110.10175}{{arXiv:2110.10175}}}
{[astro-ph.GA]}
\end{barticle}
\endbibitem

\bibitem[\protect\citeauthoryear{{Harikane} et~al.}{2023}]{Harikane2023}
\begin{barticle}
\bauthor{\bsnm{{Harikane}}, \binits{Y.}},
\bauthor{\bsnm{{Zhang}}, \binits{Y.}},
\bauthor{\bsnm{{Nakajima}}, \binits{K.}},
\bauthor{\bsnm{{Ouchi}}, \binits{M.}},
\bauthor{\bsnm{{Isobe}}, \binits{Y.}},
\bauthor{\bsnm{{Ono}}, \binits{Y.}},
\bauthor{\bsnm{{Hatano}}, \binits{S.}},
\bauthor{\bsnm{{Xu}}, \binits{Y.}},
\bauthor{\bsnm{{Umeda}}, \binits{H.}}:
\batitle{{A JWST/NIRSpec First Census of Broad-line AGNs at z = 4-7: Detection of 10 Faint AGNs with M $_{BH}$ {}10$^{6}$-{}10$^{8}$ M $_\odot$ and Their Host Galaxy Properties}}.
\bjtitle{\apj}
\bvolume{959}(\bissue{1}),
\bfpage{39}
(\byear{2023})
\doiurl{10.3847/1538-4357/ad029e}
{\href{https://arxiv.org/abs/2303.11946}{{arXiv:2303.11946}}}
{[astro-ph.GA]}
\end{barticle}
\endbibitem

\bibitem[\protect\citeauthoryear{{Maiolino} et~al.}{2024}]{Maiolino2024_AGNsample}
\begin{barticle}
\bauthor{\bsnm{{Maiolino}}, \binits{R.}},
\bauthor{\bsnm{{Scholtz}}, \binits{J.}},
\bauthor{\bsnm{{Curtis-Lake}}, \binits{E.}},
\bauthor{\bsnm{{Carniani}}, \binits{S.}},
\bauthor{\bsnm{{Baker}}, \binits{W.}},
\bauthor{\bsnm{{de Graaff}}, \binits{A.}},
\bauthor{\bsnm{{Tacchella}}, \binits{S.}},
\bauthor{\bsnm{{{\"U}bler}}, \binits{H.}},
\bauthor{\bsnm{{D'Eugenio}}, \binits{F.}},
\bauthor{\bsnm{{Witstok}}, \binits{J.}},
\bauthor{\bsnm{{Curti}}, \binits{M.}},
\bauthor{\bsnm{{Arribas}}, \binits{S.}},
\bauthor{\bsnm{{Bunker}}, \binits{A.J.}},
\bauthor{\bsnm{{Charlot}}, \binits{S.}},
\bauthor{\bsnm{{Chevallard}}, \binits{J.}},
\bauthor{\bsnm{{Eisenstein}}, \binits{D.J.}},
\bauthor{\bsnm{{Egami}}, \binits{E.}},
\bauthor{\bsnm{{Ji}}, \binits{Z.}},
\bauthor{\bsnm{{Jones}}, \binits{G.C.}},
\bauthor{\bsnm{{Lyu}}, \binits{J.}},
\bauthor{\bsnm{{Rawle}}, \binits{T.}},
\bauthor{\bsnm{{Robertson}}, \binits{B.}},
\bauthor{\bsnm{{Rujopakarn}}, \binits{W.}},
\bauthor{\bsnm{{Perna}}, \binits{M.}},
\bauthor{\bsnm{{Sun}}, \binits{F.}},
\bauthor{\bsnm{{Venturi}}, \binits{G.}},
\bauthor{\bsnm{{Williams}}, \binits{C.C.}},
\bauthor{\bsnm{{Willott}}, \binits{C.}}:
\batitle{{JADES: The diverse population of infant black holes at 4 < z < 11: Merging, tiny, poor, but mighty}}.
\bjtitle{\aap}
\bvolume{691},
\bfpage{145}
(\byear{2024})
\doiurl{10.1051/0004-6361/202347640}
{\href{https://arxiv.org/abs/2308.01230}{{arXiv:2308.01230}}}
{[astro-ph.GA]}
\end{barticle}
\endbibitem

\bibitem[\protect\citeauthoryear{{Juod{\v{z}}balis} et~al.}{2026}]{Juodzbalis2026}
\begin{barticle}
\bauthor{\bsnm{{Juod{\v{z}}balis}}, \binits{I.}},
\bauthor{\bsnm{{Maiolino}}, \binits{R.}},
\bauthor{\bsnm{{Baker}}, \binits{W.M.}},
\bauthor{\bsnm{{Lake}}, \binits{E.C.}},
\bauthor{\bsnm{{Scholtz}}, \binits{J.}},
\bauthor{\bsnm{{D'Eugenio}}, \binits{F.}},
\bauthor{\bsnm{{Trefoloni}}, \binits{B.}},
\bauthor{\bsnm{{Isobe}}, \binits{Y.}},
\bauthor{\bsnm{{Tacchella}}, \binits{S.}},
\bauthor{\bsnm{{Bunker}}, \binits{A.J.}},
\bauthor{\bsnm{al.}}:
\batitle{{JADES: comprehensive census of broad-line AGN from Reionization to Cosmic Noon revealed by JWST}}.
\bjtitle{\mnras}
(\byear{2026})
\doiurl{10.1093/mnras/stag086}
{\href{https://arxiv.org/abs/2504.03551}{{arXiv:2504.03551}}}
{[astro-ph.GA]}
\end{barticle}
\endbibitem

\bibitem[\protect\citeauthoryear{{Songaila} et~al.}{2018}]{Songaila2018}
\begin{barticle}
\bauthor{\bsnm{{Songaila}}, \binits{A.}},
\bauthor{\bsnm{{Hu}}, \binits{E.M.}},
\bauthor{\bsnm{{Barger}}, \binits{A.J.}},
\bauthor{\bsnm{{Cowie}}, \binits{L.L.}},
\bauthor{\bsnm{{Hasinger}}, \binits{G.}},
\bauthor{\bsnm{{Rosenwasser}}, \binits{B.}},
\bauthor{\bsnm{{Waters}}, \binits{C.}}:
\batitle{{Complex Ly{\ensuremath{\alpha}} Profiles in Redshift 6.6 Ultraluminous Ly{\ensuremath{\alpha}} Emitters}}.
\bjtitle{\apj}
\bvolume{859}(\bissue{2}),
\bfpage{91}
(\byear{2018})
\doiurl{10.3847/1538-4357/aac021}
{\href{https://arxiv.org/abs/1805.00490}{{arXiv:1805.00490}}}
{[astro-ph.GA]}
\end{barticle}
\endbibitem

\bibitem[\protect\citeauthoryear{{Matthee} et~al.}{2018}]{Matthee2018}
\begin{barticle}
\bauthor{\bsnm{{Matthee}}, \binits{J.}},
\bauthor{\bsnm{{Sobral}}, \binits{D.}},
\bauthor{\bsnm{{Gronke}}, \binits{M.}},
\bauthor{\bsnm{{Paulino-Afonso}}, \binits{A.}},
\bauthor{\bsnm{{Stefanon}}, \binits{M.}},
\bauthor{\bsnm{{R{\"o}ttgering}}, \binits{H.}}:
\batitle{{Confirmation of double peaked Ly{\ensuremath{\alpha}} emission at z = 6.593. Witnessing a galaxy directly contributing to the reionisation of the Universe}}.
\bjtitle{\aap}
\bvolume{619},
\bfpage{136}
(\byear{2018})
\doiurl{10.1051/0004-6361/201833528}
{\href{https://arxiv.org/abs/1805.11621}{{arXiv:1805.11621}}}
{[astro-ph.GA]}
\end{barticle}
\endbibitem

\bibitem[\protect\citeauthoryear{{Schneider} et~al.}{2023}]{Schneider2023}
\begin{barticle}
\bauthor{\bsnm{{Schneider}}, \binits{R.}},
\bauthor{\bsnm{{Valiante}}, \binits{R.}},
\bauthor{\bsnm{{Trinca}}, \binits{A.}},
\bauthor{\bsnm{{Graziani}}, \binits{L.}},
\bauthor{\bsnm{{Volonteri}}, \binits{M.}},
\bauthor{\bsnm{{Maiolino}}, \binits{R.}}:
\batitle{{Are we surprised to find SMBHs with JWST at z {\ensuremath{\geq}} 9?}}
\bjtitle{\mnras}
\bvolume{526}(\bissue{3}),
\bfpage{3250}--\blpage{3261}
(\byear{2023})
\doiurl{10.1093/mnras/stad2503}
{\href{https://arxiv.org/abs/2305.12504}{{arXiv:2305.12504}}}
{[astro-ph.GA]}
\end{barticle}
\endbibitem

\bibitem[\protect\citeauthoryear{{Izumi} et~al.}{2019}]{Izumi2019}
\begin{barticle}
\bauthor{\bsnm{{Izumi}}, \binits{T.}},
\bauthor{\bsnm{{Onoue}}, \binits{M.}},
\bauthor{\bsnm{{Matsuoka}}, \binits{Y.}},
\bauthor{\bsnm{{Nagao}}, \binits{T.}},
\bauthor{\bsnm{{Strauss}}, \binits{M.A.}},
\bauthor{\bsnm{{Imanishi}}, \binits{M.}},
\bauthor{\bsnm{{Kashikawa}}, \binits{N.}},
\bauthor{\bsnm{{Fujimoto}}, \binits{S.}},
\bauthor{\bsnm{{Kohno}}, \binits{K.}},
\bauthor{\bsnm{{Toba}}, \binits{Y.}},
\bauthor{\bsnm{{Umehata}}, \binits{H.}},
\bauthor{\bsnm{{Goto}}, \binits{T.}},
\bauthor{\bsnm{{Ueda}}, \binits{Y.}},
\bauthor{\bsnm{{Shirakata}}, \binits{H.}},
\bauthor{\bsnm{{Silverman}}, \binits{J.D.}},
\bauthor{\bsnm{{Greene}}, \binits{J.E.}},
\bauthor{\bsnm{{Harikane}}, \binits{Y.}},
\bauthor{\bsnm{{Hashimoto}}, \binits{Y.}},
\bauthor{\bsnm{{Ikarashi}}, \binits{S.}},
\bauthor{\bsnm{{Iono}}, \binits{D.}},
\bauthor{\bsnm{{Iwasawa}}, \binits{K.}},
\bauthor{\bsnm{{Lee}}, \binits{C.-H.}},
\bauthor{\bsnm{{Minezaki}}, \binits{T.}},
\bauthor{\bsnm{{Nakanishi}}, \binits{K.}},
\bauthor{\bsnm{{Tamura}}, \binits{Y.}},
\bauthor{\bsnm{{Tang}}, \binits{J.-J.}},
\bauthor{\bsnm{{Taniguchi}}, \binits{A.}}:
\batitle{{Subaru High-z Exploration of Low-Luminosity Quasars (SHELLQs). VIII. A less biased view of the early co-evolution of black holes and host galaxies}}.
\bjtitle{\pasj}
\bvolume{71}(\bissue{6}),
\bfpage{111}
(\byear{2019})
\doiurl{10.1093/pasj/psz096}
{\href{https://arxiv.org/abs/1904.07345}{{arXiv:1904.07345}}}
{[astro-ph.GA]}
\end{barticle}
\endbibitem

\bibitem[\protect\citeauthoryear{{Pensabene} et~al.}{2020}]{Pensabene2020}
\begin{barticle}
\bauthor{\bsnm{{Pensabene}}, \binits{A.}},
\bauthor{\bsnm{{Carniani}}, \binits{S.}},
\bauthor{\bsnm{{Perna}}, \binits{M.}},
\bauthor{\bsnm{{Cresci}}, \binits{G.}},
\bauthor{\bsnm{{Decarli}}, \binits{R.}},
\bauthor{\bsnm{{Maiolino}}, \binits{R.}},
\bauthor{\bsnm{{Marconi}}, \binits{A.}}:
\batitle{{The ALMA view of the high-redshift relation between supermassive black holes and their host galaxies}}.
\bjtitle{\aap}
\bvolume{637},
\bfpage{84}
(\byear{2020})
\doiurl{10.1051/0004-6361/201936634}
{\href{https://arxiv.org/abs/2002.00958}{{arXiv:2002.00958}}}
{[astro-ph.GA]}
\end{barticle}
\endbibitem

\bibitem[\protect\citeauthoryear{{Neeleman} et~al.}{2021}]{Neeleman2021}
\begin{barticle}
\bauthor{\bsnm{{Neeleman}}, \binits{M.}},
\bauthor{\bsnm{{Novak}}, \binits{M.}},
\bauthor{\bsnm{{Venemans}}, \binits{B.P.}},
\bauthor{\bsnm{{Walter}}, \binits{F.}},
\bauthor{\bsnm{{Decarli}}, \binits{R.}},
\bauthor{\bsnm{{Kaasinen}}, \binits{M.}},
\bauthor{\bsnm{{Schindler}}, \binits{J.-T.}},
\bauthor{\bsnm{{Ba{\~n}ados}}, \binits{E.}},
\bauthor{\bsnm{{Carilli}}, \binits{C.L.}},
\bauthor{\bsnm{{Drake}}, \binits{A.B.}},
\bauthor{\bsnm{{Fan}}, \binits{X.}},
\bauthor{\bsnm{{Rix}}, \binits{H.-W.}}:
\batitle{{The Kinematics of z {\ensuremath{\gtrsim}} 6 Quasar Host Galaxies}}.
\bjtitle{\apj}
\bvolume{911}(\bissue{2}),
\bfpage{141}
(\byear{2021})
\doiurl{10.3847/1538-4357/abe70f}
{\href{https://arxiv.org/abs/2102.05679}{{arXiv:2102.05679}}}
{[astro-ph.GA]}
\end{barticle}
\endbibitem

\bibitem[\protect\citeauthoryear{{Kormendy} and {Ho}}{2013}]{KormendyHo2013}
\begin{barticle}
\bauthor{\bsnm{{Kormendy}}, \binits{J.}},
\bauthor{\bsnm{{Ho}}, \binits{L.C.}}:
\batitle{{Coevolution (Or Not) of Supermassive Black Holes and Host Galaxies}}.
\bjtitle{\araa}
\bvolume{51}(\bissue{1}),
\bfpage{511}--\blpage{653}
(\byear{2013})
\doiurl{10.1146/annurev-astro-082708-101811}
{\href{https://arxiv.org/abs/1304.7762}{{arXiv:1304.7762}}}
{[astro-ph.CO]}
\end{barticle}
\endbibitem

\bibitem[\protect\citeauthoryear{{Reines} and {Volonteri}}{2015}]{Reines2015}
\begin{barticle}
\bauthor{\bsnm{{Reines}}, \binits{A.E.}},
\bauthor{\bsnm{{Volonteri}}, \binits{M.}}:
\batitle{{Relations between Central Black Hole Mass and Total Galaxy Stellar Mass in the Local Universe}}.
\bjtitle{\apj}
\bvolume{813}(\bissue{2}),
\bfpage{82}
(\byear{2015})
\doiurl{10.1088/0004-637X/813/2/82}
{\href{https://arxiv.org/abs/1508.06274}{{arXiv:1508.06274}}}
{[astro-ph.GA]}
\end{barticle}
\endbibitem

\bibitem[\protect\citeauthoryear{{Trinca} et~al.}{2024}]{Trinca2024}
\begin{botherref}
\oauthor{\bsnm{{Trinca}}, \binits{A.}},
\oauthor{\bsnm{{Valiante}}, \binits{R.}},
\oauthor{\bsnm{{Schneider}}, \binits{R.}},
\oauthor{\bsnm{{Juod{\v{z}}balis}}, \binits{I.}},
\oauthor{\bsnm{{Maiolino}}, \binits{R.}},
\oauthor{\bsnm{{Graziani}}, \binits{L.}},
\oauthor{\bsnm{{Lupi}}, \binits{A.}},
\oauthor{\bsnm{{Natarajan}}, \binits{P.}},
\oauthor{\bsnm{{Volonteri}}, \binits{M.}},
\oauthor{\bsnm{{Zana}}, \binits{T.}}:
{Episodic super-Eddington accretion as a clue to Overmassive Black Holes in the early Universe}.
arXiv e-prints,
2412--14248
(2024)
\doiurl{10.48550/arXiv.2412.14248}
{\href{https://arxiv.org/abs/2412.14248}{{arXiv:2412.14248}}}
{[astro-ph.GA]}
\end{botherref}
\endbibitem

\bibitem[\protect\citeauthoryear{{Lauer} et~al.}{2007}]{Lauer2007}
\begin{barticle}
\bauthor{\bsnm{{Lauer}}, \binits{T.R.}},
\bauthor{\bsnm{{Faber}}, \binits{S.M.}},
\bauthor{\bsnm{{Richstone}}, \binits{D.}},
\bauthor{\bsnm{{Gebhardt}}, \binits{K.}},
\bauthor{\bsnm{{Tremaine}}, \binits{S.}},
\bauthor{\bsnm{{Postman}}, \binits{M.}},
\bauthor{\bsnm{{Dressler}}, \binits{A.}},
\bauthor{\bsnm{{Aller}}, \binits{M.C.}},
\bauthor{\bsnm{{Filippenko}}, \binits{A.V.}},
\bauthor{\bsnm{{Green}}, \binits{R.}},
\bauthor{\bsnm{al.}}:
\batitle{{The Masses of Nuclear Black Holes in Luminous Elliptical Galaxies and Implications for the Space Density of the Most Massive Black Holes}}.
\bjtitle{\apj}
\bvolume{662}(\bissue{2}),
\bfpage{808}--\blpage{834}
(\byear{2007})
\doiurl{10.1086/518223}
{\href{https://arxiv.org/abs/astro-ph/0606739}{{arXiv:astro-ph/0606739}}}
{[astro-ph]}
\end{barticle}
\endbibitem

\bibitem[\protect\citeauthoryear{{Pacucci} et~al.}{2023}]{Pacucci2023}
\begin{barticle}
\bauthor{\bsnm{{Pacucci}}, \binits{F.}},
\bauthor{\bsnm{{Nguyen}}, \binits{B.}},
\bauthor{\bsnm{{Carniani}}, \binits{S.}},
\bauthor{\bsnm{{Maiolino}}, \binits{R.}},
\bauthor{\bsnm{{Fan}}, \binits{X.}}:
\batitle{{JWST CEERS and JADES Active Galaxies at z = 4-7 Violate the Local M $_{{\textbullet}}$-M $_{{\ensuremath{\star}}}$ Relation at >3{\ensuremath{\sigma}}: Implications for Low-mass Black Holes and Seeding Models}}.
\bjtitle{\apjl}
\bvolume{957}(\bissue{1}),
\bfpage{3}
(\byear{2023})
\doiurl{10.3847/2041-8213/ad0158}
{\href{https://arxiv.org/abs/2308.12331}{{arXiv:2308.12331}}}
{[astro-ph.GA]}
\end{barticle}
\endbibitem

\bibitem[\protect\citeauthoryear{{Volonteri} et~al.}{2023}]{Volonteri2023}
\begin{barticle}
\bauthor{\bsnm{{Volonteri}}, \binits{M.}},
\bauthor{\bsnm{{Habouzit}}, \binits{M.}},
\bauthor{\bsnm{{Colpi}}, \binits{M.}}:
\batitle{{What if young z > 9 JWST galaxies hosted massive black holes?}}
\bjtitle{\mnras}
\bvolume{521}(\bissue{1}),
\bfpage{241}--\blpage{250}
(\byear{2023})
\doiurl{10.1093/mnras/stad499}
{\href{https://arxiv.org/abs/2212.04710}{{arXiv:2212.04710}}}
{[astro-ph.GA]}
\end{barticle}
\endbibitem

\bibitem[\protect\citeauthoryear{{Mason} and {Gronke}}{2020}]{Mason2020}
\begin{barticle}
\bauthor{\bsnm{{Mason}}, \binits{C.A.}},
\bauthor{\bsnm{{Gronke}}, \binits{M.}}:
\batitle{{Measuring the properties of reionized bubbles with resolved Ly{\ensuremath{\alpha}} spectra}}.
\bjtitle{\mnras}
\bvolume{499}(\bissue{1}),
\bfpage{1395}--\blpage{1405}
(\byear{2020})
\doiurl{10.1093/mnras/staa2910}
{\href{https://arxiv.org/abs/2004.13065}{{arXiv:2004.13065}}}
{[astro-ph.GA]}
\end{barticle}
\endbibitem

\bibitem[\protect\citeauthoryear{{Mason} et~al.}{2026}]{Mason2026}
\begin{barticle}
\bauthor{\bsnm{{Mason}}, \binits{C.A.}},
\bauthor{\bsnm{{Chen}}, \binits{Z.}},
\bauthor{\bsnm{{Stark}}, \binits{D.P.}},
\bauthor{\bsnm{{Yi Lu}}, \binits{T.}},
\bauthor{\bsnm{{Topping}}, \binits{M.}},
\bauthor{\bsnm{{Tang}}, \binits{M.}}:
\batitle{{Constraints on the z {\ensuremath{\sim}} 6{\ensuremath{-}}13 intergalactic medium from JWST spectroscopy of Lyman-alpha damping wings in galaxies}}.
\bjtitle{\aap}
\bvolume{705},
\bfpage{114}
(\byear{2026})
\doiurl{10.1051/0004-6361/202553820}
{\href{https://arxiv.org/abs/2501.11702}{{arXiv:2501.11702}}}
{[astro-ph.GA]}
\end{barticle}
\endbibitem

\bibitem[\protect\citeauthoryear{{Saxena} et~al.}{2024}]{Saxena2024}
\begin{barticle}
\bauthor{\bsnm{{Saxena}}, \binits{A.}},
\bauthor{\bsnm{{Bunker}}, \binits{A.J.}},
\bauthor{\bsnm{{Jones}}, \binits{G.C.}},
\bauthor{\bsnm{{Stark}}, \binits{D.P.}},
\bauthor{\bsnm{{Cameron}}, \binits{A.J.}},
\bauthor{\bsnm{{Witstok}}, \binits{J.}},
\bauthor{\bsnm{{Arribas}}, \binits{S.}},
\bauthor{\bsnm{{Baker}}, \binits{W.M.}},
\bauthor{\bsnm{{Baum}}, \binits{S.}},
\bauthor{\bsnm{{Bhatawdekar}}, \binits{R.}},
\bauthor{\bsnm{{Bowler}}, \binits{R.}},
\bauthor{\bsnm{{Boyett}}, \binits{K.}},
\bauthor{\bsnm{{Carniani}}, \binits{S.}},
\bauthor{\bsnm{{Charlot}}, \binits{S.}},
\bauthor{\bsnm{{Chevallard}}, \binits{J.}},
\bauthor{\bsnm{{Curti}}, \binits{M.}},
\bauthor{\bsnm{{Curtis-Lake}}, \binits{E.}},
\bauthor{\bsnm{{Eisenstein}}, \binits{D.J.}},
\bauthor{\bsnm{{Endsley}}, \binits{R.}},
\bauthor{\bsnm{{Hainline}}, \binits{K.}},
\bauthor{\bsnm{{Helton}}, \binits{J.M.}},
\bauthor{\bsnm{{Johnson}}, \binits{B.D.}},
\bauthor{\bsnm{{Kumari}}, \binits{N.}},
\bauthor{\bsnm{{Looser}}, \binits{T.J.}},
\bauthor{\bsnm{{Maiolino}}, \binits{R.}},
\bauthor{\bsnm{{Rieke}}, \binits{M.}},
\bauthor{\bsnm{{Rix}}, \binits{H.-W.}},
\bauthor{\bsnm{{Robertson}}, \binits{B.E.}},
\bauthor{\bsnm{{Sandles}}, \binits{L.}},
\bauthor{\bsnm{{Simmonds}}, \binits{C.}},
\bauthor{\bsnm{{Smit}}, \binits{R.}},
\bauthor{\bsnm{{Tacchella}}, \binits{S.}},
\bauthor{\bsnm{{Williams}}, \binits{C.C.}},
\bauthor{\bsnm{{Willmer}}, \binits{C.N.A.}},
\bauthor{\bsnm{{Willott}}, \binits{C.}}:
\batitle{{JADES: The production and escape of ionizing photons from faint Lyman-alpha emitters in the epoch of reionization}}.
\bjtitle{\aap}
\bvolume{684},
\bfpage{84}
(\byear{2024})
\doiurl{10.1051/0004-6361/202347132}
{\href{https://arxiv.org/abs/2306.04536}{{arXiv:2306.04536}}}
{[astro-ph.GA]}
\end{barticle}
\endbibitem

\bibitem[\protect\citeauthoryear{{Witstok} et~al.}{2024}]{Witstok2024}
\begin{botherref}
\oauthor{\bsnm{{Witstok}}, \binits{J.}},
\oauthor{\bsnm{{Maiolino}}, \binits{R.}},
\oauthor{\bsnm{{Smit}}, \binits{R.}},
\oauthor{\bsnm{{Jones}}, \binits{G.C.}},
\oauthor{\bsnm{{Bunker}}, \binits{A.J.}},
\oauthor{\bsnm{{Helton}}, \binits{J.M.}},
\oauthor{\bsnm{{Johnson}}, \binits{B.D.}},
\oauthor{\bsnm{{Tacchella}}, \binits{S.}},
\oauthor{\bsnm{{Saxena}}, \binits{A.}},
\oauthor{\bsnm{{Arribas}}, \binits{S.}},
\oauthor{\bsnm{{Bhatawdekar}}, \binits{R.}},
\oauthor{\bsnm{{Boyett}}, \binits{K.}},
\oauthor{\bsnm{{Cameron}}, \binits{A.J.}},
\oauthor{\bsnm{{Cargile}}, \binits{P.A.}},
\oauthor{\bsnm{{Carniani}}, \binits{S.}},
\oauthor{\bsnm{{Charlot}}, \binits{S.}},
\oauthor{\bsnm{{Chevallard}}, \binits{J.}},
\oauthor{\bsnm{{Curti}}, \binits{M.}},
\oauthor{\bsnm{{Curtis-Lake}}, \binits{E.}},
\oauthor{\bsnm{{D'Eugenio}}, \binits{F.}},
\oauthor{\bsnm{{Eisenstein}}, \binits{D.J.}},
\oauthor{\bsnm{{Hainline}}, \binits{K.}},
\oauthor{\bsnm{{Hausen}}, \binits{R.}},
\oauthor{\bsnm{{Kumari}}, \binits{N.}},
\oauthor{\bsnm{{Laseter}}, \binits{I.}},
\oauthor{\bsnm{{Maseda}}, \binits{M.V.}},
\oauthor{\bsnm{{Rieke}}, \binits{M.}},
\oauthor{\bsnm{{Robertson}}, \binits{B.}},
\oauthor{\bsnm{{Scholtz}}, \binits{J.}},
\oauthor{\bsnm{{Shivaei}}, \binits{I.}},
\oauthor{\bsnm{{Williams}}, \binits{C.C.}},
\oauthor{\bsnm{{Willmer}}, \binits{C.N.A.}},
\oauthor{\bsnm{{Willott}}, \binits{C.}}:
{JADES: Primeval Lyman-$\mathrm{\alpha}$ emitting galaxies reveal early sites of reionisation out to redshift $z \sim 9$}.
arXiv e-prints,
2404--05724
(2024)
\doiurl{10.48550/arXiv.2404.05724}
{\href{https://arxiv.org/abs/2404.05724}{{arXiv:2404.05724}}}
{[astro-ph.GA]}
\end{botherref}
\endbibitem

\bibitem[\protect\citeauthoryear{{Tang} et~al.}{2024}]{Tang2024}
\begin{barticle}
\bauthor{\bsnm{{Tang}}, \binits{M.}},
\bauthor{\bsnm{{Stark}}, \binits{D.P.}},
\bauthor{\bsnm{{Topping}}, \binits{M.W.}},
\bauthor{\bsnm{{Mason}}, \binits{C.}},
\bauthor{\bsnm{{Ellis}}, \binits{R.S.}}:
\batitle{{JWST/NIRSpec Observations of Lyman {\ensuremath{\alpha}} Emission in Star-forming Galaxies at 6.5 {\ensuremath{\lesssim}} z {\ensuremath{\lesssim}} 13}}.
\bjtitle{\apj}
\bvolume{975}(\bissue{2}),
\bfpage{208}
(\byear{2024})
\doiurl{10.3847/1538-4357/ad7eb7}
{\href{https://arxiv.org/abs/2408.01507}{{arXiv:2408.01507}}}
{[astro-ph.GA]}
\end{barticle}
\endbibitem

\bibitem[\protect\citeauthoryear{{Witstok} et~al.}{2025}]{Witstok2025}
\begin{barticle}
\bauthor{\bsnm{{Witstok}}, \binits{J.}},
\bauthor{\bsnm{{Jakobsen}}, \binits{P.}},
\bauthor{\bsnm{{Maiolino}}, \binits{R.}},
\bauthor{\bsnm{{Helton}}, \binits{J.M.}},
\bauthor{\bsnm{{Johnson}}, \binits{B.D.}},
\bauthor{\bsnm{{Robertson}}, \binits{B.E.}},
\bauthor{\bsnm{{Tacchella}}, \binits{S.}},
\bauthor{\bsnm{{Cameron}}, \binits{A.J.}},
\bauthor{\bsnm{{Smit}}, \binits{R.}},
\bauthor{\bsnm{{Bunker}}, \binits{A.J.}},
\bauthor{\bsnm{al.}}:
\batitle{{Witnessing the onset of reionization through Lyman-{\ensuremath{\alpha}} emission at redshift 13}}.
\bjtitle{\nat}
\bvolume{639}(\bissue{8056}),
\bfpage{897}--\blpage{901}
(\byear{2025})
\doiurl{10.1038/s41586-025-08779-5}
{\href{https://arxiv.org/abs/2408.16608}{{arXiv:2408.16608}}}
{[astro-ph.GA]}
\end{barticle}
\endbibitem

\bibitem[\protect\citeauthoryear{{Hu} et~al.}{2016}]{Hu2016}
\begin{barticle}
\bauthor{\bsnm{{Hu}}, \binits{E.M.}},
\bauthor{\bsnm{{Cowie}}, \binits{L.L.}},
\bauthor{\bsnm{{Songaila}}, \binits{A.}},
\bauthor{\bsnm{{Barger}}, \binits{A.J.}},
\bauthor{\bsnm{{Rosenwasser}}, \binits{B.}},
\bauthor{\bsnm{{Wold}}, \binits{I.G.B.}}:
\batitle{{An Ultraluminous Ly{\ensuremath{\alpha}} Emitter with a Blue Wing at z = 6.6}}.
\bjtitle{\apjl}
\bvolume{825}(\bissue{1}),
\bfpage{7}
(\byear{2016})
\doiurl{10.3847/2041-8205/825/1/L7}
{\href{https://arxiv.org/abs/1606.03526}{{arXiv:1606.03526}}}
{[astro-ph.GA]}
\end{barticle}
\endbibitem

\bibitem[\protect\citeauthoryear{{Meyer} et~al.}{2021}]{Meyer2021}
\begin{barticle}
\bauthor{\bsnm{{Meyer}}, \binits{R.A.}},
\bauthor{\bsnm{{Laporte}}, \binits{N.}},
\bauthor{\bsnm{{Ellis}}, \binits{R.S.}},
\bauthor{\bsnm{{Verhamme}}, \binits{A.}},
\bauthor{\bsnm{{Garel}}, \binits{T.}}:
\batitle{{Double-peaked Lyman {\ensuremath{\alpha}} emission at z = 6.803: a reionization-era galaxy self-ionizing its local H II bubble}}.
\bjtitle{\mnras}
\bvolume{500}(\bissue{1}),
\bfpage{558}--\blpage{564}
(\byear{2021})
\doiurl{10.1093/mnras/staa3216}
{\href{https://arxiv.org/abs/2010.06241}{{arXiv:2010.06241}}}
{[astro-ph.GA]}
\end{barticle}
\endbibitem

\bibitem[\protect\citeauthoryear{{Gronke} et~al.}{2021}]{Gronke2021}
\begin{barticle}
\bauthor{\bsnm{{Gronke}}, \binits{M.}},
\bauthor{\bsnm{{Ocvirk}}, \binits{P.}},
\bauthor{\bsnm{{Mason}}, \binits{C.}},
\bauthor{\bsnm{{Matthee}}, \binits{J.}},
\bauthor{\bsnm{{Bosman}}, \binits{S.E.I.}},
\bauthor{\bsnm{{Sorce}}, \binits{J.G.}},
\bauthor{\bsnm{{Lewis}}, \binits{J.}},
\bauthor{\bsnm{{Ahn}}, \binits{K.}},
\bauthor{\bsnm{{Aubert}}, \binits{D.}},
\bauthor{\bsnm{{Dawoodbhoy}}, \binits{T.}},
\bauthor{\bsnm{{Iliev}}, \binits{I.T.}},
\bauthor{\bsnm{{Shapiro}}, \binits{P.R.}},
\bauthor{\bsnm{{Yepes}}, \binits{G.}}:
\batitle{{Lyman-{\ensuremath{\alpha}} transmission properties of the intergalactic medium in the CoDaII simulation}}.
\bjtitle{\mnras}
\bvolume{508}(\bissue{3}),
\bfpage{3697}--\blpage{3709}
(\byear{2021})
\doiurl{10.1093/mnras/stab2762}
{\href{https://arxiv.org/abs/2004.14496}{{arXiv:2004.14496}}}
{[astro-ph.GA]}
\end{barticle}
\endbibitem

\bibitem[\protect\citeauthoryear{{Garel} et~al.}{2021}]{Garel2021}
\begin{barticle}
\bauthor{\bsnm{{Garel}}, \binits{T.}},
\bauthor{\bsnm{{Blaizot}}, \binits{J.}},
\bauthor{\bsnm{{Rosdahl}}, \binits{J.}},
\bauthor{\bsnm{{Michel-Dansac}}, \binits{L.}},
\bauthor{\bsnm{{Haehnelt}}, \binits{M.G.}},
\bauthor{\bsnm{{Katz}}, \binits{H.}},
\bauthor{\bsnm{{Kimm}}, \binits{T.}},
\bauthor{\bsnm{{Verhamme}}, \binits{A.}}:
\batitle{{Ly {\ensuremath{\alpha}} as a tracer of cosmic reionization in the SPHINX radiation-hydrodynamics cosmological simulation}}.
\bjtitle{\mnras}
\bvolume{504}(\bissue{2}),
\bfpage{1902}--\blpage{1926}
(\byear{2021})
\doiurl{10.1093/mnras/stab990}
{\href{https://arxiv.org/abs/2104.03339}{{arXiv:2104.03339}}}
{[astro-ph.GA]}
\end{barticle}
\endbibitem

\bibitem[\protect\citeauthoryear{{Padmanabhan} and {Loeb}}{2021}]{Padmanabhan2021}
\begin{barticle}
\bauthor{\bsnm{{Padmanabhan}}, \binits{H.}},
\bauthor{\bsnm{{Loeb}}, \binits{A.}}:
\batitle{{Distinguishing AGN from starbursts as the origin of double-peaked Lyman-alpha emitters in the reionization era}}.
\bjtitle{\aap}
\bvolume{646},
\bfpage{10}
(\byear{2021})
\doiurl{10.1051/0004-6361/202040107}
{\href{https://arxiv.org/abs/2012.00014}{{arXiv:2012.00014}}}
{[astro-ph.GA]}
\end{barticle}
\endbibitem

\bibitem[\protect\citeauthoryear{{Ba{\~n}ados} et~al.}{2023}]{Banados2023}
\begin{barticle}
\bauthor{\bsnm{{Ba{\~n}ados}}, \binits{E.}},
\bauthor{\bsnm{{Schindler}}, \binits{J.-T.}},
\bauthor{\bsnm{{Venemans}}, \binits{B.P.}},
\bauthor{\bsnm{{Connor}}, \binits{T.}},
\bauthor{\bsnm{{Decarli}}, \binits{R.}},
\bauthor{\bsnm{{Farina}}, \binits{E.P.}},
\bauthor{\bsnm{{Mazzucchelli}}, \binits{C.}},
\bauthor{\bsnm{{Meyer}}, \binits{R.A.}},
\bauthor{\bsnm{{Stern}}, \binits{D.}},
\bauthor{\bsnm{{Walter}}, \binits{F.}},
\bauthor{\bsnm{{Fan}}, \binits{X.}},
\bauthor{\bsnm{{Hennawi}}, \binits{J.F.}},
\bauthor{\bsnm{{Khusanova}}, \binits{Y.}},
\bauthor{\bsnm{{Morrell}}, \binits{N.}},
\bauthor{\bsnm{{Nanni}}, \binits{R.}},
\bauthor{\bsnm{{Noirot}}, \binits{G.}},
\bauthor{\bsnm{{Pensabene}}, \binits{A.}},
\bauthor{\bsnm{{Rix}}, \binits{H.-W.}},
\bauthor{\bsnm{{Simon}}, \binits{J.}},
\bauthor{\bsnm{{Verdoes Kleijn}}, \binits{G.A.}},
\bauthor{\bsnm{{Xie}}, \binits{Z.-L.}},
\bauthor{\bsnm{{Yang}}, \binits{D.-M.}},
\bauthor{\bsnm{{Connor}}, \binits{A.}}:
\batitle{{The Pan-STARRS1 z > 5.6 Quasar Survey. II. Discovery of 55 Quasars at 5.6 < z < 6.5}}.
\bjtitle{\apjs}
\bvolume{265}(\bissue{1}),
\bfpage{29}
(\byear{2023})
\doiurl{10.3847/1538-4365/acb3c7}
{\href{https://arxiv.org/abs/2212.04452}{{arXiv:2212.04452}}}
{[astro-ph.GA]}
\end{barticle}
\endbibitem

\bibitem[\protect\citeauthoryear{{Dalla Bont{\`a}} et~al.}{2025}]{DB2025}
\begin{barticle}
\bauthor{\bsnm{{Dalla Bont{\`a}}}, \binits{E.}},
\bauthor{\bsnm{{Peterson}}, \binits{B.M.}},
\bauthor{\bsnm{{Grier}}, \binits{C.J.}},
\bauthor{\bsnm{{Berton}}, \binits{M.}},
\bauthor{\bsnm{{Brandt}}, \binits{W.N.}},
\bauthor{\bsnm{{Ciroi}}, \binits{S.}},
\bauthor{\bsnm{{Corsini}}, \binits{E.M.}},
\bauthor{\bsnm{{Dalla Barba}}, \binits{B.}},
\bauthor{\bsnm{{Davies}}, \binits{R.}},
\bauthor{\bsnm{{Dehghanian}}, \binits{M.}},
\bauthor{\bsnm{{Edelson}}, \binits{R.}},
\bauthor{\bsnm{{Foschini}}, \binits{L.}},
\bauthor{\bsnm{{Gasparri}}, \binits{D.}},
\bauthor{\bsnm{{Ho}}, \binits{L.C.}},
\bauthor{\bsnm{{Horne}}, \binits{K.}},
\bauthor{\bsnm{{Iodice}}, \binits{E.}},
\bauthor{\bsnm{{Morelli}}, \binits{L.}},
\bauthor{\bsnm{{Pizzella}}, \binits{A.}},
\bauthor{\bsnm{{Portaluri}}, \binits{E.}},
\bauthor{\bsnm{{Shen}}, \binits{Y.}},
\bauthor{\bsnm{{Schneider}}, \binits{D.P.}},
\bauthor{\bsnm{{Vestergaard}}, \binits{M.}}:
\batitle{{Estimating masses of supermassive black holes in active galactic nuclei from the H{\ensuremath{\alpha}} emission line}}.
\bjtitle{\aap}
\bvolume{696},
\bfpage{48}
(\byear{2025})
\doiurl{10.1051/0004-6361/202452746}
{\href{https://arxiv.org/abs/2410.21387}{{arXiv:2410.21387}}}
{[astro-ph.GA]}
\end{barticle}
\endbibitem

\bibitem[\protect\citeauthoryear{{{\"U}bler} et~al.}{2023}]{Ubler2023}
\begin{barticle}
\bauthor{\bsnm{{{\"U}bler}}, \binits{H.}},
\bauthor{\bsnm{{Maiolino}}, \binits{R.}},
\bauthor{\bsnm{{Curtis-Lake}}, \binits{E.}},
\bauthor{\bsnm{{P{\'e}rez-Gonz{\'a}lez}}, \binits{P.G.}},
\bauthor{\bsnm{{Curti}}, \binits{M.}},
\bauthor{\bsnm{{Perna}}, \binits{M.}},
\bauthor{\bsnm{{Arribas}}, \binits{S.}},
\bauthor{\bsnm{{Charlot}}, \binits{S.}},
\bauthor{\bsnm{{Marshall}}, \binits{M.A.}},
\bauthor{\bsnm{{D'Eugenio}}, \binits{F.}},
\bauthor{\bsnm{{Scholtz}}, \binits{J.}},
\bauthor{\bsnm{{Bunker}}, \binits{A.}},
\bauthor{\bsnm{{Carniani}}, \binits{S.}},
\bauthor{\bsnm{{Ferruit}}, \binits{P.}},
\bauthor{\bsnm{{Jakobsen}}, \binits{P.}},
\bauthor{\bsnm{{Rix}}, \binits{H.-W.}},
\bauthor{\bsnm{{Rodr{\'\i}guez Del Pino}}, \binits{B.}},
\bauthor{\bsnm{{Willott}}, \binits{C.J.}},
\bauthor{\bsnm{{Boeker}}, \binits{T.}},
\bauthor{\bsnm{{Cresci}}, \binits{G.}},
\bauthor{\bsnm{{Jones}}, \binits{G.C.}},
\bauthor{\bsnm{{Kumari}}, \binits{N.}},
\bauthor{\bsnm{{Rawle}}, \binits{T.}}:
\batitle{{GA-NIFS: A massive black hole in a low-metallicity AGN at z {\ensuremath{\sim}} 5.55 revealed by JWST/NIRSpec IFS}}.
\bjtitle{\aap}
\bvolume{677},
\bfpage{145}
(\byear{2023})
\doiurl{10.1051/0004-6361/202346137}
{\href{https://arxiv.org/abs/2302.06647}{{arXiv:2302.06647}}}
{[astro-ph.GA]}
\end{barticle}
\endbibitem

\bibitem[\protect\citeauthoryear{{Juod{\v{z}}balis} et~al.}{2024}]{Juodzbalis2024}
\begin{barticle}
\bauthor{\bsnm{{Juod{\v{z}}balis}}, \binits{I.}},
\bauthor{\bsnm{{Maiolino}}, \binits{R.}},
\bauthor{\bsnm{{Baker}}, \binits{W.M.}},
\bauthor{\bsnm{{Tacchella}}, \binits{S.}},
\bauthor{\bsnm{{Scholtz}}, \binits{J.}},
\bauthor{\bsnm{{D'Eugenio}}, \binits{F.}},
\bauthor{\bsnm{{Witstok}}, \binits{J.}},
\bauthor{\bsnm{{Schneider}}, \binits{R.}},
\bauthor{\bsnm{{Trinca}}, \binits{A.}},
\bauthor{\bsnm{{Valiante}}, \binits{R.}},
\bauthor{\bsnm{al.}}:
\batitle{{A dormant overmassive black hole in the early Universe}}.
\bjtitle{\nat}
\bvolume{636}(\bissue{8043}),
\bfpage{594}--\blpage{597}
(\byear{2024})
\doiurl{10.1038/s41586-024-08210-5}
{\href{https://arxiv.org/abs/2403.03872}{{arXiv:2403.03872}}}
{[astro-ph.GA]}
\end{barticle}
\endbibitem

\bibitem[\protect\citeauthoryear{{Hutchison} et~al.}{2024}]{Hutchison2024}
\begin{barticle}
\bauthor{\bsnm{{Hutchison}}, \binits{T.A.}},
\bauthor{\bsnm{{Welch}}, \binits{B.D.}},
\bauthor{\bsnm{{Rigby}}, \binits{J.R.}},
\bauthor{\bsnm{{Olivier}}, \binits{G.M.}},
\bauthor{\bsnm{{Birkin}}, \binits{J.E.}},
\bauthor{\bsnm{{Phadke}}, \binits{K.A.}},
\bauthor{\bsnm{{Khullar}}, \binits{G.}},
\bauthor{\bsnm{{Rauscher}}, \binits{B.J.}},
\bauthor{\bsnm{{Sharon}}, \binits{K.}},
\bauthor{\bsnm{{Aravena}}, \binits{M.}},
\bauthor{\bsnm{{Bayliss}}, \binits{M.B.}},
\bauthor{\bsnm{{Elicker}}, \binits{L.A.}},
\bauthor{\bsnm{{Kim}}, \binits{S.}},
\bauthor{\bsnm{{Solimano}}, \binits{M.}},
\bauthor{\bsnm{{Vieira}}, \binits{J.D.}},
\bauthor{\bsnm{{Vizgan}}, \binits{D.}},
\bauthor{\bsnm{{Jwst Templates Early Release Science Team}}}:
\batitle{{TEMPLATES: A Robust Outlier Rejection Method for JWST/NIRSpec Integral Field Spectroscopy}}.
\bjtitle{\pasp}
\bvolume{136}(\bissue{4}),
\bfpage{044503}
(\byear{2024})
\doiurl{10.1088/1538-3873/ad34fd}
{\href{https://arxiv.org/abs/2312.12518}{{arXiv:2312.12518}}}
{[astro-ph.IM]}
\end{barticle}
\endbibitem

\bibitem[\protect\citeauthoryear{{Eilers} et~al.}{2017}]{Eilers2017}
\begin{barticle}
\bauthor{\bsnm{{Eilers}}, \binits{A.-C.}},
\bauthor{\bsnm{{Davies}}, \binits{F.B.}},
\bauthor{\bsnm{{Hennawi}}, \binits{J.F.}},
\bauthor{\bsnm{{Prochaska}}, \binits{J.X.}},
\bauthor{\bsnm{{Luki{\'c}}}, \binits{Z.}},
\bauthor{\bsnm{{Mazzucchelli}}, \binits{C.}}:
\batitle{{Implications of z {\ensuremath{\sim}} 6 Quasar Proximity Zones for the Epoch of Reionization and Quasar Lifetimes}}.
\bjtitle{\apj}
\bvolume{840}(\bissue{1}),
\bfpage{24}
(\byear{2017})
\doiurl{10.3847/1538-4357/aa6c60}
{\href{https://arxiv.org/abs/1703.02539}{{arXiv:1703.02539}}}
{[astro-ph.GA]}
\end{barticle}
\endbibitem

\bibitem[\protect\citeauthoryear{{Eilers} et~al.}{2020}]{Eilers2020}
\begin{barticle}
\bauthor{\bsnm{{Eilers}}, \binits{A.-C.}},
\bauthor{\bsnm{{Hennawi}}, \binits{J.F.}},
\bauthor{\bsnm{{Decarli}}, \binits{R.}},
\bauthor{\bsnm{{Davies}}, \binits{F.B.}},
\bauthor{\bsnm{{Venemans}}, \binits{B.}},
\bauthor{\bsnm{{Walter}}, \binits{F.}},
\bauthor{\bsnm{{Ba{\~n}ados}}, \binits{E.}},
\bauthor{\bsnm{{Fan}}, \binits{X.}},
\bauthor{\bsnm{{Farina}}, \binits{E.P.}},
\bauthor{\bsnm{{Mazzucchelli}}, \binits{C.}},
\bauthor{\bsnm{{Novak}}, \binits{M.}},
\bauthor{\bsnm{{Schindler}}, \binits{J.-T.}},
\bauthor{\bsnm{{Simcoe}}, \binits{R.A.}},
\bauthor{\bsnm{{Wang}}, \binits{F.}},
\bauthor{\bsnm{{Yang}}, \binits{J.}}:
\batitle{{Detecting and Characterizing Young Quasars. I. Systemic Redshifts and Proximity Zone Measurements}}.
\bjtitle{\apj}
\bvolume{900}(\bissue{1}),
\bfpage{37}
(\byear{2020})
\doiurl{10.3847/1538-4357/aba52e}
{\href{https://arxiv.org/abs/2002.01811}{{arXiv:2002.01811}}}
{[astro-ph.GA]}
\end{barticle}
\endbibitem

\bibitem[\protect\citeauthoryear{{Davies} et~al.}{2020}]{Davies2020}
\begin{barticle}
\bauthor{\bsnm{{Davies}}, \binits{F.B.}},
\bauthor{\bsnm{{Hennawi}}, \binits{J.F.}},
\bauthor{\bsnm{{Eilers}}, \binits{A.-C.}}:
\batitle{{Time-dependent behaviour of quasar proximity zones at z {\ensuremath{\sim}} 6}}.
\bjtitle{\mnras}
\bvolume{493}(\bissue{1}),
\bfpage{1330}--\blpage{1343}
(\byear{2020})
\doiurl{10.1093/mnras/stz3303}
{\href{https://arxiv.org/abs/1903.12346}{{arXiv:1903.12346}}}
{[astro-ph.CO]}
\end{barticle}
\endbibitem

\bibitem[\protect\citeauthoryear{{Satyavolu} et~al.}{2023}]{Satyavolu2023}
\begin{barticle}
\bauthor{\bsnm{{Satyavolu}}, \binits{S.}},
\bauthor{\bsnm{{Eilers}}, \binits{A.-C.}},
\bauthor{\bsnm{{Kulkarni}}, \binits{G.}},
\bauthor{\bsnm{{Ryan-Weber}}, \binits{E.}},
\bauthor{\bsnm{{Davies}}, \binits{R.L.}},
\bauthor{\bsnm{{Becker}}, \binits{G.D.}},
\bauthor{\bsnm{{Bosman}}, \binits{S.E.I.}},
\bauthor{\bsnm{{Greig}}, \binits{B.}},
\bauthor{\bsnm{{Mazzucchelli}}, \binits{C.}},
\bauthor{\bsnm{{Ba{\~n}ados}}, \binits{E.}},
\bauthor{\bsnm{al.}}:
\batitle{{New quasar proximity zone size measurements at z 6 using the enlarged XQR-30 sample}}.
\bjtitle{\mnras}
\bvolume{522}(\bissue{4}),
\bfpage{4918}--\blpage{4933}
(\byear{2023})
\doiurl{10.1093/mnras/stad1326}
{\href{https://arxiv.org/abs/2305.00998}{{arXiv:2305.00998}}}
{[astro-ph.GA]}
\end{barticle}
\endbibitem

\bibitem[\protect\citeauthoryear{{Keating} et~al.}{2020}]{Keating2020}
\begin{barticle}
\bauthor{\bsnm{{Keating}}, \binits{L.C.}},
\bauthor{\bsnm{{Weinberger}}, \binits{L.H.}},
\bauthor{\bsnm{{Kulkarni}}, \binits{G.}},
\bauthor{\bsnm{{Haehnelt}}, \binits{M.G.}},
\bauthor{\bsnm{{Chardin}}, \binits{J.}},
\bauthor{\bsnm{{Aubert}}, \binits{D.}}:
\batitle{{Long troughs in the Lyman-{\ensuremath{\alpha}} forest below redshift 6 due to islands of neutral hydrogen}}.
\bjtitle{\mnras}
\bvolume{491}(\bissue{2}),
\bfpage{1736}--\blpage{1745}
(\byear{2020})
\doiurl{10.1093/mnras/stz3083}
{\href{https://arxiv.org/abs/1905.12640}{{arXiv:1905.12640}}}
{[astro-ph.CO]}
\end{barticle}
\endbibitem

\bibitem[\protect\citeauthoryear{{Garaldi} et~al.}{2022}]{Garaldi2022}
\begin{barticle}
\bauthor{\bsnm{{Garaldi}}, \binits{E.}},
\bauthor{\bsnm{{Kannan}}, \binits{R.}},
\bauthor{\bsnm{{Smith}}, \binits{A.}},
\bauthor{\bsnm{{Springel}}, \binits{V.}},
\bauthor{\bsnm{{Pakmor}}, \binits{R.}},
\bauthor{\bsnm{{Vogelsberger}}, \binits{M.}},
\bauthor{\bsnm{{Hernquist}}, \binits{L.}}:
\batitle{{The THESAN project: properties of the intergalactic medium and its connection to reionization-era galaxies}}.
\bjtitle{\mnras}
\bvolume{512}(\bissue{4}),
\bfpage{4909}--\blpage{4933}
(\byear{2022})
\doiurl{10.1093/mnras/stac257}
{\href{https://arxiv.org/abs/2110.01628}{{arXiv:2110.01628}}}
{[astro-ph.CO]}
\end{barticle}
\endbibitem

\bibitem[\protect\citeauthoryear{{Lewis} et~al.}{2022}]{Lewis2022}
\begin{barticle}
\bauthor{\bsnm{{Lewis}}, \binits{J.S.W.}},
\bauthor{\bsnm{{Ocvirk}}, \binits{P.}},
\bauthor{\bsnm{{Sorce}}, \binits{J.G.}},
\bauthor{\bsnm{{Dubois}}, \binits{Y.}},
\bauthor{\bsnm{{Aubert}}, \binits{D.}},
\bauthor{\bsnm{{Conaboy}}, \binits{L.}},
\bauthor{\bsnm{{Shapiro}}, \binits{P.R.}},
\bauthor{\bsnm{{Dawoodbhoy}}, \binits{T.}},
\bauthor{\bsnm{{Teyssier}}, \binits{R.}},
\bauthor{\bsnm{{Yepes}}, \binits{G.}},
\bauthor{\bsnm{{Gottl{\"o}ber}}, \binits{S.}},
\bauthor{\bsnm{{Rasera}}, \binits{Y.}},
\bauthor{\bsnm{{Ahn}}, \binits{K.}},
\bauthor{\bsnm{{Iliev}}, \binits{I.T.}},
\bauthor{\bsnm{{Park}}, \binits{H.}},
\bauthor{\bsnm{{Th{\'e}lie}}, \binits{{\'E}.}}:
\batitle{{The short ionizing photon mean free path at z = 6 in Cosmic Dawn III, a new fully coupled radiation-hydrodynamical simulation of the Epoch of Reionization}}.
\bjtitle{\mnras}
\bvolume{516}(\bissue{3}),
\bfpage{3389}--\blpage{3397}
(\byear{2022})
\doiurl{10.1093/mnras/stac2383}
{\href{https://arxiv.org/abs/2202.05869}{{arXiv:2202.05869}}}
{[astro-ph.CO]}
\end{barticle}
\endbibitem

\bibitem[\protect\citeauthoryear{{Gaikwad} et~al.}{2023}]{Gaikwad2023}
\begin{barticle}
\bauthor{\bsnm{{Gaikwad}}, \binits{P.}},
\bauthor{\bsnm{{Haehnelt}}, \binits{M.G.}},
\bauthor{\bsnm{{Davies}}, \binits{F.B.}},
\bauthor{\bsnm{{Bosman}}, \binits{S.E.I.}},
\bauthor{\bsnm{{Molaro}}, \binits{M.}},
\bauthor{\bsnm{{Kulkarni}}, \binits{G.}},
\bauthor{\bsnm{{D'Odorico}}, \binits{V.}},
\bauthor{\bsnm{{Becker}}, \binits{G.D.}},
\bauthor{\bsnm{{Davies}}, \binits{R.L.}},
\bauthor{\bsnm{{Nasir}}, \binits{F.}},
\bauthor{\bsnm{{Bolton}}, \binits{J.S.}},
\bauthor{\bsnm{{Keating}}, \binits{L.C.}},
\bauthor{\bsnm{{Ir{\v{s}}i{\v{c}}}}, \binits{V.}},
\bauthor{\bsnm{{Puchwein}}, \binits{E.}},
\bauthor{\bsnm{{Zhu}}, \binits{Y.}},
\bauthor{\bsnm{{Asthana}}, \binits{S.}},
\bauthor{\bsnm{{Yang}}, \binits{J.}},
\bauthor{\bsnm{{Lai}}, \binits{S.}},
\bauthor{\bsnm{{Eilers}}, \binits{A.-C.}}:
\batitle{{Measuring the photoionization rate, neutral fraction, and mean free path of H I ionizing photons at 4.9 {\ensuremath{\leq}} z {\ensuremath{\leq}} 6.0 from a large sample of XShooter and ESI spectra}}.
\bjtitle{\mnras}
\bvolume{525}(\bissue{3}),
\bfpage{4093}--\blpage{4120}
(\byear{2023})
\doiurl{10.1093/mnras/stad2566}
{\href{https://arxiv.org/abs/2304.02038}{{arXiv:2304.02038}}}
{[astro-ph.CO]}
\end{barticle}
\endbibitem

\bibitem[\protect\citeauthoryear{{Shen} et~al.}{2007}]{Shen2007}
\begin{barticle}
\bauthor{\bsnm{{Shen}}, \binits{Y.}},
\bauthor{\bsnm{{Strauss}}, \binits{M.A.}},
\bauthor{\bsnm{{Oguri}}, \binits{M.}},
\bauthor{\bsnm{{Hennawi}}, \binits{J.F.}},
\bauthor{\bsnm{{Fan}}, \binits{X.}},
\bauthor{\bsnm{{Richards}}, \binits{G.T.}},
\bauthor{\bsnm{{Hall}}, \binits{P.B.}},
\bauthor{\bsnm{{Gunn}}, \binits{J.E.}},
\bauthor{\bsnm{{Schneider}}, \binits{D.P.}},
\bauthor{\bsnm{{Szalay}}, \binits{A.S.}},
\bauthor{\bsnm{{Thakar}}, \binits{A.R.}},
\bauthor{\bsnm{{Vanden Berk}}, \binits{D.E.}},
\bauthor{\bsnm{{Anderson}}, \binits{S.F.}},
\bauthor{\bsnm{{Bahcall}}, \binits{N.A.}},
\bauthor{\bsnm{{Connolly}}, \binits{A.J.}},
\bauthor{\bsnm{{Knapp}}, \binits{G.R.}}:
\batitle{{Clustering of High-Redshift (z >= 2.9) Quasars from the Sloan Digital Sky Survey}}.
\bjtitle{\aj}
\bvolume{133}(\bissue{5}),
\bfpage{2222}--\blpage{2241}
(\byear{2007})
\doiurl{10.1086/513517}
{\href{https://arxiv.org/abs/astro-ph/0702214}{{arXiv:astro-ph/0702214}}}
{[astro-ph]}
\end{barticle}
\endbibitem

\bibitem[\protect\citeauthoryear{{White} et~al.}{2012}]{White2012}
\begin{barticle}
\bauthor{\bsnm{{White}}, \binits{M.}},
\bauthor{\bsnm{{Myers}}, \binits{A.D.}},
\bauthor{\bsnm{{Ross}}, \binits{N.P.}},
\bauthor{\bsnm{{Schlegel}}, \binits{D.J.}},
\bauthor{\bsnm{{Hennawi}}, \binits{J.F.}},
\bauthor{\bsnm{{Shen}}, \binits{Y.}},
\bauthor{\bsnm{{McGreer}}, \binits{I.}},
\bauthor{\bsnm{{Strauss}}, \binits{M.A.}},
\bauthor{\bsnm{{Bolton}}, \binits{A.S.}},
\bauthor{\bsnm{{Bovy}}, \binits{J.}},
\bauthor{\bsnm{{Fan}}, \binits{X.}},
\bauthor{\bsnm{{Miralda-Escude}}, \binits{J.}},
\bauthor{\bsnm{{Palanque-Delabrouille}}, \binits{N.}},
\bauthor{\bsnm{{Paris}}, \binits{I.}},
\bauthor{\bsnm{{Petitjean}}, \binits{P.}},
\bauthor{\bsnm{{Schneider}}, \binits{D.P.}},
\bauthor{\bsnm{{Viel}}, \binits{M.}},
\bauthor{\bsnm{{Weinberg}}, \binits{D.H.}},
\bauthor{\bsnm{{Yeche}}, \binits{C.}},
\bauthor{\bsnm{{Zehavi}}, \binits{I.}},
\bauthor{\bsnm{{Pan}}, \binits{K.}},
\bauthor{\bsnm{{Snedden}}, \binits{S.}},
\bauthor{\bsnm{{Bizyaev}}, \binits{D.}},
\bauthor{\bsnm{{Brewington}}, \binits{H.}},
\bauthor{\bsnm{{Brinkmann}}, \binits{J.}},
\bauthor{\bsnm{{Malanushenko}}, \binits{V.}},
\bauthor{\bsnm{{Malanushenko}}, \binits{E.}},
\bauthor{\bsnm{{Oravetz}}, \binits{D.}},
\bauthor{\bsnm{{Simmons}}, \binits{A.}},
\bauthor{\bsnm{{Sheldon}}, \binits{A.}},
\bauthor{\bsnm{{Weaver}}, \binits{B.A.}}:
\batitle{{The clustering of intermediate-redshift quasars as measured by the Baryon Oscillation Spectroscopic Survey}}.
\bjtitle{\mnras}
\bvolume{424}(\bissue{2}),
\bfpage{933}--\blpage{950}
(\byear{2012})
\doiurl{10.1111/j.1365-2966.2012.21251.x}
{\href{https://arxiv.org/abs/1203.5306}{{arXiv:1203.5306}}}
{[astro-ph.CO]}
\end{barticle}
\endbibitem

\bibitem[\protect\citeauthoryear{{Eftekharzadeh} et~al.}{2015}]{Eftekharzadeh2015}
\begin{barticle}
\bauthor{\bsnm{{Eftekharzadeh}}, \binits{S.}},
\bauthor{\bsnm{{Myers}}, \binits{A.D.}},
\bauthor{\bsnm{{White}}, \binits{M.}},
\bauthor{\bsnm{{Weinberg}}, \binits{D.H.}},
\bauthor{\bsnm{{Schneider}}, \binits{D.P.}},
\bauthor{\bsnm{{Shen}}, \binits{Y.}},
\bauthor{\bsnm{{Font-Ribera}}, \binits{A.}},
\bauthor{\bsnm{{Ross}}, \binits{N.P.}},
\bauthor{\bsnm{{Paris}}, \binits{I.}},
\bauthor{\bsnm{{Streblyanska}}, \binits{A.}}:
\batitle{{Clustering of intermediate redshift quasars using the final SDSS III-BOSS sample}}.
\bjtitle{\mnras}
\bvolume{453}(\bissue{3}),
\bfpage{2779}--\blpage{2798}
(\byear{2015})
\doiurl{10.1093/mnras/stv1763}
{\href{https://arxiv.org/abs/1507.08380}{{arXiv:1507.08380}}}
{[astro-ph.CO]}
\end{barticle}
\endbibitem

\bibitem[\protect\citeauthoryear{{Laurent} et~al.}{2017}]{Laurent2017}
\begin{barticle}
\bauthor{\bsnm{{Laurent}}, \binits{P.}},
\bauthor{\bsnm{{Eftekharzadeh}}, \binits{S.}},
\bauthor{\bsnm{{Le Goff}}, \binits{J.-M.}},
\bauthor{\bsnm{{Myers}}, \binits{A.}},
\bauthor{\bsnm{{Burtin}}, \binits{E.}},
\bauthor{\bsnm{{White}}, \binits{M.}},
\bauthor{\bsnm{{Ross}}, \binits{A.J.}},
\bauthor{\bsnm{{Tinker}}, \binits{J.}},
\bauthor{\bsnm{{Tojeiro}}, \binits{R.}},
\bauthor{\bsnm{{Bautista}}, \binits{J.}},
\bauthor{\bsnm{{Brinkmann}}, \binits{J.}},
\bauthor{\bsnm{{Comparat}}, \binits{J.}},
\bauthor{\bsnm{{Dawson}}, \binits{K.}},
\bauthor{\bsnm{{du Mas des Bourboux}}, \binits{H.}},
\bauthor{\bsnm{{Kneib}}, \binits{J.-P.}},
\bauthor{\bsnm{{McGreer}}, \binits{I.D.}},
\bauthor{\bsnm{{Palanque-Delabrouille}}, \binits{N.}},
\bauthor{\bsnm{{Percival}}, \binits{W.J.}},
\bauthor{\bsnm{{Prada}}, \binits{F.}},
\bauthor{\bsnm{{Rossi}}, \binits{G.}},
\bauthor{\bsnm{{Schneider}}, \binits{D.P.}},
\bauthor{\bsnm{{Weinberg}}, \binits{D.}},
\bauthor{\bsnm{{Y{\`e}che}}, \binits{C.}},
\bauthor{\bsnm{{Zarrouk}}, \binits{P.}},
\bauthor{\bsnm{{Zhao}}, \binits{G.-B.}}:
\batitle{{Clustering of quasars in SDSS-IV eBOSS: study of potential systematics and bias determination}}.
\bjtitle{\jcap}
\bvolume{2017}(\bissue{7}),
\bfpage{017}
(\byear{2017})
\doiurl{10.1088/1475-7516/2017/07/017}
{\href{https://arxiv.org/abs/1705.04718}{{arXiv:1705.04718}}}
{[astro-ph.CO]}
\end{barticle}
\endbibitem

\bibitem[\protect\citeauthoryear{{Pizzati} et~al.}{2024a}]{Pizzati2024}
\begin{barticle}
\bauthor{\bsnm{{Pizzati}}, \binits{E.}},
\bauthor{\bsnm{{Hennawi}}, \binits{J.F.}},
\bauthor{\bsnm{{Schaye}}, \binits{J.}},
\bauthor{\bsnm{{Schaller}}, \binits{M.}}:
\batitle{{Revisiting the extreme clustering of z {\ensuremath{\approx}} 4 quasars with large volume cosmological simulations}}.
\bjtitle{\mnras}
\bvolume{528}(\bissue{3}),
\bfpage{4466}--\blpage{4489}
(\byear{2024})
\doiurl{10.1093/mnras/stae329}
{\href{https://arxiv.org/abs/2311.17181}{{arXiv:2311.17181}}}
{[astro-ph.GA]}
\end{barticle}
\endbibitem

\bibitem[\protect\citeauthoryear{{Pizzati} et~al.}{2024b}]{Pizzati2024a}
\begin{barticle}
\bauthor{\bsnm{{Pizzati}}, \binits{E.}},
\bauthor{\bsnm{{Hennawi}}, \binits{J.F.}},
\bauthor{\bsnm{{Schaye}}, \binits{J.}},
\bauthor{\bsnm{{Schaller}}, \binits{M.}},
\bauthor{\bsnm{{Eilers}}, \binits{A.-C.}},
\bauthor{\bsnm{{Wang}}, \binits{F.}},
\bauthor{\bsnm{{Frenk}}, \binits{C.S.}},
\bauthor{\bsnm{{Elbers}}, \binits{W.}},
\bauthor{\bsnm{{Helly}}, \binits{J.C.}},
\bauthor{\bsnm{{Mackenzie}}, \binits{R.}},
\bauthor{\bsnm{{Matthee}}, \binits{J.}},
\bauthor{\bsnm{{Bordoloi}}, \binits{R.}},
\bauthor{\bsnm{{Kashino}}, \binits{D.}},
\bauthor{\bsnm{{Naidu}}, \binits{R.P.}},
\bauthor{\bsnm{{Yue}}, \binits{M.}}:
\batitle{{A unified model for the clustering of quasars and galaxies at z = 6}}.
\bjtitle{\mnras}
\bvolume{534}(\bissue{4}),
\bfpage{3155}--\blpage{3175}
(\byear{2024})
\doiurl{10.1093/mnras/stae2307}
{\href{https://arxiv.org/abs/2403.12140}{{arXiv:2403.12140}}}
{[astro-ph.GA]}
\end{barticle}
\endbibitem

\bibitem[\protect\citeauthoryear{{Eilers} et~al.}{2024}]{Eilers2024}
\begin{barticle}
\bauthor{\bsnm{{Eilers}}, \binits{A.-C.}},
\bauthor{\bsnm{{Mackenzie}}, \binits{R.}},
\bauthor{\bsnm{{Pizzati}}, \binits{E.}},
\bauthor{\bsnm{{Matthee}}, \binits{J.}},
\bauthor{\bsnm{{Hennawi}}, \binits{J.F.}},
\bauthor{\bsnm{{Zhang}}, \binits{H.}},
\bauthor{\bsnm{{Bordoloi}}, \binits{R.}},
\bauthor{\bsnm{{Kashino}}, \binits{D.}},
\bauthor{\bsnm{{Lilly}}, \binits{S.J.}},
\bauthor{\bsnm{{Naidu}}, \binits{R.P.}},
\bauthor{\bsnm{{Simcoe}}, \binits{R.A.}},
\bauthor{\bsnm{{Yue}}, \binits{M.}},
\bauthor{\bsnm{{Frenk}}, \binits{C.S.}},
\bauthor{\bsnm{{Helly}}, \binits{J.C.}},
\bauthor{\bsnm{{Schaller}}, \binits{M.}},
\bauthor{\bsnm{{Schaye}}, \binits{J.}}:
\batitle{{EIGER. VI. The Correlation Function, Host Halo Mass, and Duty Cycle of Luminous Quasars at z {\ensuremath{\gtrsim}} 6}}.
\bjtitle{\apj}
\bvolume{974}(\bissue{2}),
\bfpage{275}
(\byear{2024})
\doiurl{10.3847/1538-4357/ad778b}
{\href{https://arxiv.org/abs/2403.07986}{{arXiv:2403.07986}}}
{[astro-ph.GA]}
\end{barticle}
\endbibitem

\bibitem[\protect\citeauthoryear{{{\v{D}}urov{\v{c}}{\'\i}kov{\'a}} et~al.}{2024}]{Durovcikova2024}
\begin{barticle}
\bauthor{\bsnm{{{\v{D}}urov{\v{c}}{\'\i}kov{\'a}}}, \binits{D.}},
\bauthor{\bsnm{{Eilers}}, \binits{A.-C.}},
\bauthor{\bsnm{{Chen}}, \binits{H.}},
\bauthor{\bsnm{{Satyavolu}}, \binits{S.}},
\bauthor{\bsnm{{Kulkarni}}, \binits{G.}},
\bauthor{\bsnm{{Simcoe}}, \binits{R.A.}},
\bauthor{\bsnm{{Keating}}, \binits{L.C.}},
\bauthor{\bsnm{{Haehnelt}}, \binits{M.G.}},
\bauthor{\bsnm{{Ba{\~n}ados}}, \binits{E.}}:
\batitle{{Chronicling the Reionization History at 6 {\ensuremath{\lesssim}} z {\ensuremath{\lesssim}} 7 with Emergent Quasar Damping Wings}}.
\bjtitle{\apj}
\bvolume{969}(\bissue{2}),
\bfpage{162}
(\byear{2024})
\doiurl{10.3847/1538-4357/ad4888}
{\href{https://arxiv.org/abs/2401.10328}{{arXiv:2401.10328}}}
{[astro-ph.CO]}
\end{barticle}
\endbibitem

\bibitem[\protect\citeauthoryear{{Davies} et~al.}{2019}]{Davies2019}
\begin{barticle}
\bauthor{\bsnm{{Davies}}, \binits{F.B.}},
\bauthor{\bsnm{{Hennawi}}, \binits{J.F.}},
\bauthor{\bsnm{{Eilers}}, \binits{A.-C.}}:
\batitle{{Evidence for Low Radiative Efficiency or Highly Obscured Growth of z > 7 Quasars}}.
\bjtitle{\apjl}
\bvolume{884}(\bissue{1}),
\bfpage{19}
(\byear{2019})
\doiurl{10.3847/2041-8213/ab42e3}
{\href{https://arxiv.org/abs/1906.10130}{{arXiv:1906.10130}}}
{[astro-ph.GA]}
\end{barticle}
\endbibitem

\bibitem[\protect\citeauthoryear{{Wang} et~al.}{2026}]{Wang2026}
\begin{botherref}
\oauthor{\bsnm{{Wang}}, \binits{F.}},
\oauthor{\bsnm{{Champagne}}, \binits{J.B.}},
\oauthor{\bsnm{{Huang}}, \binits{J.}},
\oauthor{\bsnm{{Yang}}, \binits{J.}},
\oauthor{\bsnm{{Hennawi}}, \binits{J.F.}},
\oauthor{\bsnm{{Fan}}, \binits{X.}},
\oauthor{\bsnm{{Zhang}}, \binits{H.}},
\oauthor{\bsnm{{Costa}}, \binits{T.}},
\oauthor{\bsnm{{Decarli}}, \binits{R.}},
\oauthor{\bsnm{{Habouzit}}, \binits{M.}},
\oauthor{\bsnm{{Sun}}, \binits{F.}},
\oauthor{\bsnm{{Banados}}, \binits{E.}},
\oauthor{\bsnm{{Jin}}, \binits{X.}},
\oauthor{\bsnm{{Kakiichi}}, \binits{K.}},
\oauthor{\bsnm{{Meyer}}, \binits{R.A.}},
\oauthor{\bsnm{{Wu}}, \binits{Y.}},
\oauthor{\bsnm{{Belladitta}}, \binits{S.}},
\oauthor{\bsnm{{Blecha}}, \binits{L.}},
\oauthor{\bsnm{{Bosman}}, \binits{S.E.I.}},
\oauthor{\bsnm{{Cai}}, \binits{Z.}},
\oauthor{\bsnm{{Connor}}, \binits{T.}},
\oauthor{\bsnm{{Davies}}, \binits{F.B.}},
\oauthor{\bsnm{{Eilers}}, \binits{A.-C.}},
\oauthor{\bsnm{{Haiman}}, \binits{Z.}},
\oauthor{\bsnm{{Jun}}, \binits{H.D.}},
\oauthor{\bsnm{{Li}}, \binits{M.}},
\oauthor{\bsnm{{Li}}, \binits{Z.}},
\oauthor{\bsnm{{Liu}}, \binits{W.}},
\oauthor{\bsnm{{Lupi}}, \binits{A.}},
\oauthor{\bsnm{{Lyu}}, \binits{J.}},
\oauthor{\bsnm{{Mazzucchelli}}, \binits{C.}},
\oauthor{\bsnm{{Onoue}}, \binits{M.}},
\oauthor{\bsnm{{Pudoka}}, \binits{M.}},
\oauthor{\bsnm{{Rojas-Ruiz}}, \binits{S.}},
\oauthor{\bsnm{{Schindler}}, \binits{J.-T.}},
\oauthor{\bsnm{{Shen}}, \binits{Y.}},
\oauthor{\bsnm{{Tee}}, \binits{W.L.}},
\oauthor{\bsnm{{Trakhtenbrot}}, \binits{B.}},
\oauthor{\bsnm{{Trebitsch}}, \binits{M.}},
\oauthor{\bsnm{{Vestergaard}}, \binits{M.}},
\oauthor{\bsnm{{Volonteri}}, \binits{M.}},
\oauthor{\bsnm{{Walter}}, \binits{F.}},
\oauthor{\bsnm{{Zhang}}, \binits{H.}},
\oauthor{\bsnm{{Zou}}, \binits{S.}}:
{ASPIRE: The Environments and Dark Matter Halos of Luminous Quasars in the Epoch of Reionization}.
arXiv e-prints,
2602--04979
(2026)
\doiurl{10.48550/arXiv.2602.04979}
{\href{https://arxiv.org/abs/2602.04979}{{arXiv:2602.04979}}}
{[astro-ph.GA]}
\end{botherref}
\endbibitem

\bibitem[\protect\citeauthoryear{{Huang} et~al.}{2026}]{Huang2026}
\begin{botherref}
\oauthor{\bsnm{{Huang}}, \binits{J.}},
\oauthor{\bsnm{{Hennawi}}, \binits{J.}},
\oauthor{\bsnm{{Pizzati}}, \binits{E.}},
\oauthor{\bsnm{{Wang}}, \binits{F.}},
\oauthor{\bsnm{{Yang}}, \binits{J.}},
\oauthor{\bsnm{{Champagne}}, \binits{J.B.}},
\oauthor{\bsnm{{Fan}}, \binits{X.}},
\oauthor{\bsnm{{Ba{\~n}ados}}, \binits{E.}},
\oauthor{\bsnm{{Jin}}, \binits{X.}},
\oauthor{\bsnm{{Kakiichi}}, \binits{K.}},
\oauthor{\bsnm{{Meyer}}, \binits{R.A.}},
\oauthor{\bsnm{{Sun}}, \binits{F.}},
\oauthor{\bsnm{{Wu}}, \binits{Y.}},
\oauthor{\bsnm{{Zhang}}, \binits{H.}},
\oauthor{\bsnm{{Mazzucchelli}}, \binits{C.}},
\oauthor{\bsnm{{Eilers}}, \binits{A.-C.}},
\oauthor{\bsnm{{Pudoka}}, \binits{M.}},
\oauthor{\bsnm{{Zhang}}, \binits{H.}},
\oauthor{\bsnm{{Schindler}}, \binits{J.-T.}},
\oauthor{\bsnm{{Schaller}}, \binits{M.}},
\oauthor{\bsnm{{Schaye}}, \binits{J.}},
\oauthor{\bsnm{{Snyder}}, \binits{B.}},
\oauthor{\bsnm{{Kang}}, \binits{Y.}},
\oauthor{\bsnm{{Onorato}}, \binits{S.}}:
{Clustering of z\raisebox{-0.5ex}\textasciitilde6.6 Quasars and [O III] Emitters Constrains Host Halo Masses and Duty Cycles in 25 ASPIRE Fields}.
arXiv e-prints,
2602--04974
(2026)
\doiurl{10.48550/arXiv.2602.04974}
{\href{https://arxiv.org/abs/2602.04974}{{arXiv:2602.04974}}}
{[astro-ph.GA]}
\end{botherref}
\endbibitem

\bibitem[\protect\citeauthoryear{{Satyavolu} et~al.}{2023}]{Satyavolu2023b}
\begin{barticle}
\bauthor{\bsnm{{Satyavolu}}, \binits{S.}},
\bauthor{\bsnm{{Kulkarni}}, \binits{G.}},
\bauthor{\bsnm{{Keating}}, \binits{L.C.}},
\bauthor{\bsnm{{Haehnelt}}, \binits{M.G.}}:
\batitle{{The need for obscured supermassive black hole growth to explain quasar proximity zones in the epoch of reionization}}.
\bjtitle{\mnras}
\bvolume{521}(\bissue{2}),
\bfpage{3108}--\blpage{3126}
(\byear{2023})
\doiurl{10.1093/mnras/stad729}
{\href{https://arxiv.org/abs/2209.08103}{{arXiv:2209.08103}}}
{[astro-ph.GA]}
\end{barticle}
\endbibitem

\bibitem[\protect\citeauthoryear{{Taylor} et~al.}{2023}]{Taylor2023}
\begin{barticle}
\bauthor{\bsnm{{Taylor}}, \binits{A.J.}},
\bauthor{\bsnm{{Barger}}, \binits{A.J.}},
\bauthor{\bsnm{{Cowie}}, \binits{L.L.}},
\bauthor{\bsnm{{Hasinger}}, \binits{G.}},
\bauthor{\bsnm{{Hu}}, \binits{E.M.}},
\bauthor{\bsnm{{Songaila}}, \binits{A.}}:
\batitle{{HEROES: The Hawaii eROSITA Ecliptic Pole Survey Catalog}}.
\bjtitle{\apjs}
\bvolume{266}(\bissue{2}),
\bfpage{24}
(\byear{2023})
\doiurl{10.3847/1538-4365/accd70}
{\href{https://arxiv.org/abs/2302.11581}{{arXiv:2302.11581}}}
{[astro-ph.GA]}
\end{barticle}
\endbibitem

\bibitem[\protect\citeauthoryear{{Kulas} et~al.}{2012}]{Kulas2012}
\begin{barticle}
\bauthor{\bsnm{{Kulas}}, \binits{K.R.}},
\bauthor{\bsnm{{Shapley}}, \binits{A.E.}},
\bauthor{\bsnm{{Kollmeier}}, \binits{J.A.}},
\bauthor{\bsnm{{Zheng}}, \binits{Z.}},
\bauthor{\bsnm{{Steidel}}, \binits{C.C.}},
\bauthor{\bsnm{{Hainline}}, \binits{K.N.}}:
\batitle{{The Kinematics of Multiple-peaked Ly{\ensuremath{\alpha}} Emission in Star-forming Galaxies at z \raisebox{-0.5ex}\textasciitilde 2-3}}.
\bjtitle{\apj}
\bvolume{745}(\bissue{1}),
\bfpage{33}
(\byear{2012})
\doiurl{10.1088/0004-637X/745/1/33}
{\href{https://arxiv.org/abs/1107.4367}{{arXiv:1107.4367}}}
{[astro-ph.CO]}
\end{barticle}
\endbibitem

\bibitem[\protect\citeauthoryear{{Vitte} et~al.}{2025}]{Vitte2025}
\begin{barticle}
\bauthor{\bsnm{{Vitte}}, \binits{E.}},
\bauthor{\bsnm{{Verhamme}}, \binits{A.}},
\bauthor{\bsnm{{Hibon}}, \binits{P.}},
\bauthor{\bsnm{{Leclercq}}, \binits{F.}},
\bauthor{\bsnm{{Alcalde Pampliega}}, \binits{B.}},
\bauthor{\bsnm{{Kerutt}}, \binits{J.}},
\bauthor{\bsnm{{Kusakabe}}, \binits{H.}},
\bauthor{\bsnm{{Matthee}}, \binits{J.}},
\bauthor{\bsnm{{Guo}}, \binits{Y.}},
\bauthor{\bsnm{{Bacon}}, \binits{R.}},
\bauthor{\bsnm{al.}}:
\batitle{{The MUSE eXtremely Deep Field: Classifying the spectral shapes of Ly{\ensuremath{\alpha}}-emitting galaxies}}.
\bjtitle{\aap}
\bvolume{694},
\bfpage{100}
(\byear{2025})
\doiurl{10.1051/0004-6361/202450426}
{\href{https://arxiv.org/abs/2411.14327}{{arXiv:2411.14327}}}
{[astro-ph.GA]}
\end{barticle}
\endbibitem

\bibitem[\protect\citeauthoryear{{Matsuoka} et~al.}{2025}]{Matsuoka2025}
\begin{barticle}
\bauthor{\bsnm{{Matsuoka}}, \binits{Y.}},
\bauthor{\bsnm{{Onoue}}, \binits{M.}},
\bauthor{\bsnm{{Iwasawa}}, \binits{K.}},
\bauthor{\bsnm{{Aoki}}, \binits{K.}},
\bauthor{\bsnm{{Strauss}}, \binits{M.A.}},
\bauthor{\bsnm{{Silverman}}, \binits{J.D.}},
\bauthor{\bsnm{{Ding}}, \binits{X.}},
\bauthor{\bsnm{{Phillips}}, \binits{C.L.}},
\bauthor{\bsnm{{Akiyama}}, \binits{M.}},
\bauthor{\bsnm{{Arita}}, \binits{J.}},
\bauthor{\bsnm{{Imanishi}}, \binits{M.}},
\bauthor{\bsnm{{Izumi}}, \binits{T.}},
\bauthor{\bsnm{{Kashikawa}}, \binits{N.}},
\bauthor{\bsnm{{Kawaguchi}}, \binits{T.}},
\bauthor{\bsnm{{Kikuta}}, \binits{S.}},
\bauthor{\bsnm{{Kohno}}, \binits{K.}},
\bauthor{\bsnm{{Lee}}, \binits{C.-H.}},
\bauthor{\bsnm{{Nagao}}, \binits{T.}},
\bauthor{\bsnm{{Takahashi}}, \binits{A.}},
\bauthor{\bsnm{{Toba}}, \binits{Y.}}:
\batitle{{SHELLQs. Bridging the Gap: JWST Unveils Obscured Quasars in the Most Luminous Galaxies at z > 6}}.
\bjtitle{\apj}
\bvolume{988}(\bissue{1}),
\bfpage{57}
(\byear{2025})
\doiurl{10.3847/1538-4357/addf4e}
{\href{https://arxiv.org/abs/2505.04825}{{arXiv:2505.04825}}}
{[astro-ph.GA]}
\end{barticle}
\endbibitem

\bibitem[\protect\citeauthoryear{{Nakajima} et~al.}{2022}]{Nakajima2022}
\begin{barticle}
\bauthor{\bsnm{{Nakajima}}, \binits{K.}},
\bauthor{\bsnm{{Ouchi}}, \binits{M.}},
\bauthor{\bsnm{{Xu}}, \binits{Y.}},
\bauthor{\bsnm{{Rauch}}, \binits{M.}},
\bauthor{\bsnm{{Harikane}}, \binits{Y.}},
\bauthor{\bsnm{{Nishigaki}}, \binits{M.}},
\bauthor{\bsnm{{Isobe}}, \binits{Y.}},
\bauthor{\bsnm{{Kusakabe}}, \binits{H.}},
\bauthor{\bsnm{{Nagao}}, \binits{T.}},
\bauthor{\bsnm{{Ono}}, \binits{Y.}},
\bauthor{\bsnm{{Onodera}}, \binits{M.}},
\bauthor{\bsnm{{Sugahara}}, \binits{Y.}},
\bauthor{\bsnm{{Kim}}, \binits{J.H.}},
\bauthor{\bsnm{{Komiyama}}, \binits{Y.}},
\bauthor{\bsnm{{Lee}}, \binits{C.-H.}},
\bauthor{\bsnm{{Zahedy}}, \binits{F.S.}}
\bjtitle{\apjs}
\bvolume{262}(\bissue{1}),
\bfpage{3}
(\byear{2022})
\doiurl{10.3847/1538-4365/ac7710}
{\href{https://arxiv.org/abs/2206.02824}{{arXiv:2206.02824}}}
{[astro-ph.GA]}
\end{barticle}
\endbibitem

\bibitem[\protect\citeauthoryear{{Bian} et~al.}{2018}]{Bian2018}
\begin{barticle}
\bauthor{\bsnm{{Bian}}, \binits{F.}},
\bauthor{\bsnm{{Kewley}}, \binits{L.J.}},
\bauthor{\bsnm{{Dopita}}, \binits{M.A.}}:
\batitle{{{\textquotedblleft}Direct{\textquotedblright} Gas-phase Metallicity in Local Analogs of High-redshift Galaxies: Empirical Metallicity Calibrations for High-redshift Star-forming Galaxies}}.
\bjtitle{\apj}
\bvolume{859}(\bissue{2}),
\bfpage{175}
(\byear{2018})
\doiurl{10.3847/1538-4357/aabd74}
{\href{https://arxiv.org/abs/1805.08224}{{arXiv:1805.08224}}}
{[astro-ph.GA]}
\end{barticle}
\endbibitem

\bibitem[\protect\citeauthoryear{{Stark} et~al.}{2011}]{Stark2011}
\begin{barticle}
\bauthor{\bsnm{{Stark}}, \binits{D.P.}},
\bauthor{\bsnm{{Ellis}}, \binits{R.S.}},
\bauthor{\bsnm{{Ouchi}}, \binits{M.}}:
\batitle{{Keck Spectroscopy of Faint 3>z>7 Lyman Break Galaxies: A High Fraction of Line Emitters at Redshift Six}}.
\bjtitle{\apjl}
\bvolume{728}(\bissue{1}),
\bfpage{2}
(\byear{2011})
\doiurl{10.1088/2041-8205/728/1/L2}
{\href{https://arxiv.org/abs/1009.5471}{{arXiv:1009.5471}}}
{[astro-ph.CO]}
\end{barticle}
\endbibitem

\bibitem[\protect\citeauthoryear{{Pentericci} et~al.}{2018}]{Pentericci2018}
\begin{barticle}
\bauthor{\bsnm{{Pentericci}}, \binits{L.}},
\bauthor{\bsnm{{Vanzella}}, \binits{E.}},
\bauthor{\bsnm{{Castellano}}, \binits{M.}},
\bauthor{\bsnm{{Fontana}}, \binits{A.}},
\bauthor{\bsnm{{De Barros}}, \binits{S.}},
\bauthor{\bsnm{{Grazian}}, \binits{A.}},
\bauthor{\bsnm{{Marchi}}, \binits{F.}},
\bauthor{\bsnm{{Bradac}}, \binits{M.}},
\bauthor{\bsnm{{Conselice}}, \binits{C.J.}},
\bauthor{\bsnm{{Cristiani}}, \binits{S.}},
\bauthor{\bsnm{{Dickinson}}, \binits{M.}},
\bauthor{\bsnm{{Finkelstein}}, \binits{S.L.}},
\bauthor{\bsnm{{Giallongo}}, \binits{E.}},
\bauthor{\bsnm{{Guaita}}, \binits{L.}},
\bauthor{\bsnm{{Koekemoer}}, \binits{A.M.}},
\bauthor{\bsnm{{Maiolino}}, \binits{R.}},
\bauthor{\bsnm{{Santini}}, \binits{P.}},
\bauthor{\bsnm{{Tilvi}}, \binits{V.}}:
\batitle{{CANDELSz7: a large spectroscopic survey of CANDELS galaxies in the reionization epoch}}.
\bjtitle{\aap}
\bvolume{619},
\bfpage{147}
(\byear{2018})
\doiurl{10.1051/0004-6361/201732465}
{\href{https://arxiv.org/abs/1808.01847}{{arXiv:1808.01847}}}
{[astro-ph.GA]}
\end{barticle}
\endbibitem

\bibitem[\protect\citeauthoryear{{Roberts-Borsani} et~al.}{2016}]{Roberts-Borsani2016}
\begin{barticle}
\bauthor{\bsnm{{Roberts-Borsani}}, \binits{G.W.}},
\bauthor{\bsnm{{Bouwens}}, \binits{R.J.}},
\bauthor{\bsnm{{Oesch}}, \binits{P.A.}},
\bauthor{\bsnm{{Labbe}}, \binits{I.}},
\bauthor{\bsnm{{Smit}}, \binits{R.}},
\bauthor{\bsnm{{Illingworth}}, \binits{G.D.}},
\bauthor{\bsnm{{van Dokkum}}, \binits{P.}},
\bauthor{\bsnm{{Holden}}, \binits{B.}},
\bauthor{\bsnm{{Gonzalez}}, \binits{V.}},
\bauthor{\bsnm{{Stefanon}}, \binits{M.}},
\bauthor{\bsnm{{Holwerda}}, \binits{B.}},
\bauthor{\bsnm{{Wilkins}}, \binits{S.}}:
\batitle{{z {\ensuremath{\gtrsim}} 7 Galaxies with Red Spitzer/IRAC [3.6]-[4.5] Colors in the Full CANDELS Data Set: The Brightest-Known Galaxies at z \raisebox{-0.5ex}\textasciitilde 7-9 and a Probable Spectroscopic Confirmation at z = 7.48}}.
\bjtitle{\apj}
\bvolume{823}(\bissue{2}),
\bfpage{143}
(\byear{2016})
\doiurl{10.3847/0004-637X/823/2/143}
{\href{https://arxiv.org/abs/1506.00854}{{arXiv:1506.00854}}}
{[astro-ph.GA]}
\end{barticle}
\endbibitem

\bibitem[\protect\citeauthoryear{{Laporte} et~al.}{2017}]{Laporte2017}
\begin{barticle}
\bauthor{\bsnm{{Laporte}}, \binits{N.}},
\bauthor{\bsnm{{Nakajima}}, \binits{K.}},
\bauthor{\bsnm{{Ellis}}, \binits{R.S.}},
\bauthor{\bsnm{{Zitrin}}, \binits{A.}},
\bauthor{\bsnm{{Stark}}, \binits{D.P.}},
\bauthor{\bsnm{{Mainali}}, \binits{R.}},
\bauthor{\bsnm{{Roberts-Borsani}}, \binits{G.W.}}:
\batitle{{A Spectroscopic Search for AGN Activity in the Reionization Era}}.
\bjtitle{\apj}
\bvolume{851}(\bissue{1}),
\bfpage{40}
(\byear{2017})
\doiurl{10.3847/1538-4357/aa96a8}
{\href{https://arxiv.org/abs/1708.05173}{{arXiv:1708.05173}}}
{[astro-ph.GA]}
\end{barticle}
\endbibitem

\bibitem[\protect\citeauthoryear{{{\"U}bler} et~al.}{2024}]{Ubler2024}
\begin{barticle}
\bauthor{\bsnm{{{\"U}bler}}, \binits{H.}},
\bauthor{\bsnm{{Maiolino}}, \binits{R.}},
\bauthor{\bsnm{{P{\'e}rez-Gonz{\'a}lez}}, \binits{P.G.}},
\bauthor{\bsnm{{D'Eugenio}}, \binits{F.}},
\bauthor{\bsnm{{Perna}}, \binits{M.}},
\bauthor{\bsnm{{Curti}}, \binits{M.}},
\bauthor{\bsnm{{Arribas}}, \binits{S.}},
\bauthor{\bsnm{{Bunker}}, \binits{A.}},
\bauthor{\bsnm{{Carniani}}, \binits{S.}},
\bauthor{\bsnm{{Charlot}}, \binits{S.}},
\bauthor{\bsnm{{Rodr{\'\i}guez Del Pino}}, \binits{B.}},
\bauthor{\bsnm{{Baker}}, \binits{W.}},
\bauthor{\bsnm{{B{\"o}ker}}, \binits{T.}},
\bauthor{\bsnm{{Cresci}}, \binits{G.}},
\bauthor{\bsnm{{Dunlop}}, \binits{J.}},
\bauthor{\bsnm{{Grogin}}, \binits{N.A.}},
\bauthor{\bsnm{{Jones}}, \binits{G.C.}},
\bauthor{\bsnm{{Kumari}}, \binits{N.}},
\bauthor{\bsnm{{Lamperti}}, \binits{I.}},
\bauthor{\bsnm{{Laporte}}, \binits{N.}},
\bauthor{\bsnm{{Marshall}}, \binits{M.A.}},
\bauthor{\bsnm{{Mazzolari}}, \binits{G.}},
\bauthor{\bsnm{{Parlanti}}, \binits{E.}},
\bauthor{\bsnm{{Rawle}}, \binits{T.}},
\bauthor{\bsnm{{Scholtz}}, \binits{J.}},
\bauthor{\bsnm{{Venturi}}, \binits{G.}},
\bauthor{\bsnm{{Witstok}}, \binits{J.}}:
\batitle{{GA-NIFS: JWST discovers an offset AGN 740 million years after the big bang}}.
\bjtitle{\mnras}
\bvolume{531}(\bissue{1}),
\bfpage{355}--\blpage{365}
(\byear{2024})
\doiurl{10.1093/mnras/stae943}
{\href{https://arxiv.org/abs/2312.03589}{{arXiv:2312.03589}}}
{[astro-ph.GA]}
\end{barticle}
\endbibitem

\bibitem[\protect\citeauthoryear{{Bunker} et~al.}{2023}]{Bunker2023}
\begin{barticle}
\bauthor{\bsnm{{Bunker}}, \binits{A.J.}},
\bauthor{\bsnm{{Saxena}}, \binits{A.}},
\bauthor{\bsnm{{Cameron}}, \binits{A.J.}},
\bauthor{\bsnm{{Willott}}, \binits{C.J.}},
\bauthor{\bsnm{{Curtis-Lake}}, \binits{E.}},
\bauthor{\bsnm{{Jakobsen}}, \binits{P.}},
\bauthor{\bsnm{{Carniani}}, \binits{S.}},
\bauthor{\bsnm{{Smit}}, \binits{R.}},
\bauthor{\bsnm{{Maiolino}}, \binits{R.}},
\bauthor{\bsnm{{Witstok}}, \binits{J.}},
\bauthor{\bsnm{{Curti}}, \binits{M.}},
\bauthor{\bsnm{{D'Eugenio}}, \binits{F.}},
\bauthor{\bsnm{{Jones}}, \binits{G.C.}},
\bauthor{\bsnm{{Ferruit}}, \binits{P.}},
\bauthor{\bsnm{{Arribas}}, \binits{S.}},
\bauthor{\bsnm{{Charlot}}, \binits{S.}},
\bauthor{\bsnm{{Chevallard}}, \binits{J.}},
\bauthor{\bsnm{{Giardino}}, \binits{G.}},
\bauthor{\bsnm{{de Graaff}}, \binits{A.}},
\bauthor{\bsnm{{Looser}}, \binits{T.J.}},
\bauthor{\bsnm{{L{\"u}tzgendorf}}, \binits{N.}},
\bauthor{\bsnm{{Maseda}}, \binits{M.V.}},
\bauthor{\bsnm{{Rawle}}, \binits{T.}},
\bauthor{\bsnm{{Rix}}, \binits{H.-W.}},
\bauthor{\bsnm{{Del Pino}}, \binits{B.R.}},
\bauthor{\bsnm{{Alberts}}, \binits{S.}},
\bauthor{\bsnm{{Egami}}, \binits{E.}},
\bauthor{\bsnm{{Eisenstein}}, \binits{D.J.}},
\bauthor{\bsnm{{Endsley}}, \binits{R.}},
\bauthor{\bsnm{{Hainline}}, \binits{K.}},
\bauthor{\bsnm{{Hausen}}, \binits{R.}},
\bauthor{\bsnm{{Johnson}}, \binits{B.D.}},
\bauthor{\bsnm{{Rieke}}, \binits{G.}},
\bauthor{\bsnm{{Rieke}}, \binits{M.}},
\bauthor{\bsnm{{Robertson}}, \binits{B.E.}},
\bauthor{\bsnm{{Shivaei}}, \binits{I.}},
\bauthor{\bsnm{{Stark}}, \binits{D.P.}},
\bauthor{\bsnm{{Sun}}, \binits{F.}},
\bauthor{\bsnm{{Tacchella}}, \binits{S.}},
\bauthor{\bsnm{{Tang}}, \binits{M.}},
\bauthor{\bsnm{{Williams}}, \binits{C.C.}},
\bauthor{\bsnm{{Willmer}}, \binits{C.N.A.}},
\bauthor{\bsnm{{Baker}}, \binits{W.M.}},
\bauthor{\bsnm{{Baum}}, \binits{S.}},
\bauthor{\bsnm{{Bhatawdekar}}, \binits{R.}},
\bauthor{\bsnm{{Bowler}}, \binits{R.}},
\bauthor{\bsnm{{Boyett}}, \binits{K.}},
\bauthor{\bsnm{{Chen}}, \binits{Z.}},
\bauthor{\bsnm{{Circosta}}, \binits{C.}},
\bauthor{\bsnm{{Helton}}, \binits{J.M.}},
\bauthor{\bsnm{{Ji}}, \binits{Z.}},
\bauthor{\bsnm{{Kumari}}, \binits{N.}},
\bauthor{\bsnm{{Lyu}}, \binits{J.}},
\bauthor{\bsnm{{Nelson}}, \binits{E.}},
\bauthor{\bsnm{{Parlanti}}, \binits{E.}},
\bauthor{\bsnm{{Perna}}, \binits{M.}},
\bauthor{\bsnm{{Sandles}}, \binits{L.}},
\bauthor{\bsnm{{Scholtz}}, \binits{J.}},
\bauthor{\bsnm{{Suess}}, \binits{K.A.}},
\bauthor{\bsnm{{Topping}}, \binits{M.W.}},
\bauthor{\bsnm{{{\"U}bler}}, \binits{H.}},
\bauthor{\bsnm{{Wallace}}, \binits{I.E.B.}},
\bauthor{\bsnm{{Whitler}}, \binits{L.}}:
\batitle{{JADES NIRSpec Spectroscopy of GN-z11: Lyman-{\ensuremath{\alpha}} emission and possible enhanced nitrogen abundance in a z = 10.60 luminous galaxy}}.
\bjtitle{\aap}
\bvolume{677},
\bfpage{88}
(\byear{2023})
\doiurl{10.1051/0004-6361/202346159}
{\href{https://arxiv.org/abs/2302.07256}{{arXiv:2302.07256}}}
{[astro-ph.GA]}
\end{barticle}
\endbibitem

\bibitem[\protect\citeauthoryear{{Scholtz} et~al.}{2023}]{Scholtz2023}
\begin{botherref}
\oauthor{\bsnm{{Scholtz}}, \binits{J.}},
\oauthor{\bsnm{{Maiolino}}, \binits{R.}},
\oauthor{\bsnm{{D'Eugenio}}, \binits{F.}},
\oauthor{\bsnm{{Curtis-Lake}}, \binits{E.}},
\oauthor{\bsnm{{Carniani}}, \binits{S.}},
\oauthor{\bsnm{{Charlot}}, \binits{S.}},
\oauthor{\bsnm{{Curti}}, \binits{M.}},
\oauthor{\bsnm{{Silcock}}, \binits{M.S.}},
\oauthor{\bsnm{{Arribas}}, \binits{S.}},
\oauthor{\bsnm{{Baker}}, \binits{W.}},
\oauthor{\bsnm{{Bhatawdekar}}, \binits{R.}},
\oauthor{\bsnm{{Boyett}}, \binits{K.}},
\oauthor{\bsnm{{Bunker}}, \binits{A.J.}},
\oauthor{\bsnm{{Chevallard}}, \binits{J.}},
\oauthor{\bsnm{{Circosta}}, \binits{C.}},
\oauthor{\bsnm{{Eisenstein}}, \binits{D.J.}},
\oauthor{\bsnm{{Hainline}}, \binits{K.}},
\oauthor{\bsnm{{Hausen}}, \binits{R.}},
\oauthor{\bsnm{{Ji}}, \binits{X.}},
\oauthor{\bsnm{{Ji}}, \binits{Z.}},
\oauthor{\bsnm{{Johnson}}, \binits{B.D.}},
\oauthor{\bsnm{{Kumari}}, \binits{N.}},
\oauthor{\bsnm{{Looser}}, \binits{T.J.}},
\oauthor{\bsnm{{Lyu}}, \binits{J.}},
\oauthor{\bsnm{{Maseda}}, \binits{M.V.}},
\oauthor{\bsnm{{Parlanti}}, \binits{E.}},
\oauthor{\bsnm{{Perna}}, \binits{M.}},
\oauthor{\bsnm{{Rieke}}, \binits{M.}},
\oauthor{\bsnm{{Robertson}}, \binits{B.}},
\oauthor{\bsnm{{Rodr{\'\i}guez Del Pino}}, \binits{B.}},
\oauthor{\bsnm{{Sun}}, \binits{F.}},
\oauthor{\bsnm{{Tacchella}}, \binits{S.}},
\oauthor{\bsnm{{{\"U}bler}}, \binits{H.}},
\oauthor{\bsnm{{Venturi}}, \binits{G.}},
\oauthor{\bsnm{{Williams}}, \binits{C.C.}},
\oauthor{\bsnm{{Willmer}}, \binits{C.N.A.}},
\oauthor{\bsnm{{Willott}}, \binits{C.}},
\oauthor{\bsnm{{Witstok}}, \binits{J.}}:
{JADES: A large population of obscured, narrow line AGN at high redshift}.
arXiv e-prints,
2311--18731
(2023)
\doiurl{10.48550/arXiv.2311.18731}
{\href{https://arxiv.org/abs/2311.18731}{{arXiv:2311.18731}}}
{[astro-ph.GA]}
\end{botherref}
\endbibitem

\bibitem[\protect\citeauthoryear{{Maiolino} et~al.}{2024}]{Maiolino2024}
\begin{barticle}
\bauthor{\bsnm{{Maiolino}}, \binits{R.}},
\bauthor{\bsnm{{Scholtz}}, \binits{J.}},
\bauthor{\bsnm{{Witstok}}, \binits{J.}},
\bauthor{\bsnm{{Carniani}}, \binits{S.}},
\bauthor{\bsnm{{D'Eugenio}}, \binits{F.}},
\bauthor{\bsnm{{de Graaff}}, \binits{A.}},
\bauthor{\bsnm{{{\"U}bler}}, \binits{H.}},
\bauthor{\bsnm{{Tacchella}}, \binits{S.}},
\bauthor{\bsnm{{Curtis-Lake}}, \binits{E.}},
\bauthor{\bsnm{{Arribas}}, \binits{S.}},
\bauthor{\bsnm{{Bunker}}, \binits{A.}},
\bauthor{\bsnm{{Charlot}}, \binits{S.}},
\bauthor{\bsnm{{Chevallard}}, \binits{J.}},
\bauthor{\bsnm{{Curti}}, \binits{M.}},
\bauthor{\bsnm{{Looser}}, \binits{T.J.}},
\bauthor{\bsnm{{Maseda}}, \binits{M.V.}},
\bauthor{\bsnm{{Rawle}}, \binits{T.D.}},
\bauthor{\bsnm{{Rodr{\'\i}guez del Pino}}, \binits{B.}},
\bauthor{\bsnm{{Willott}}, \binits{C.J.}},
\bauthor{\bsnm{{Egami}}, \binits{E.}},
\bauthor{\bsnm{{Eisenstein}}, \binits{D.J.}},
\bauthor{\bsnm{{Hainline}}, \binits{K.N.}},
\bauthor{\bsnm{{Robertson}}, \binits{B.}},
\bauthor{\bsnm{{Williams}}, \binits{C.C.}},
\bauthor{\bsnm{{Willmer}}, \binits{C.N.A.}},
\bauthor{\bsnm{{Baker}}, \binits{W.M.}},
\bauthor{\bsnm{{Boyett}}, \binits{K.}},
\bauthor{\bsnm{{DeCoursey}}, \binits{C.}},
\bauthor{\bsnm{{Fabian}}, \binits{A.C.}},
\bauthor{\bsnm{{Helton}}, \binits{J.M.}},
\bauthor{\bsnm{{Ji}}, \binits{Z.}},
\bauthor{\bsnm{{Jones}}, \binits{G.C.}},
\bauthor{\bsnm{{Kumari}}, \binits{N.}},
\bauthor{\bsnm{{Laporte}}, \binits{N.}},
\bauthor{\bsnm{{Nelson}}, \binits{E.J.}},
\bauthor{\bsnm{{Perna}}, \binits{M.}},
\bauthor{\bsnm{{Sandles}}, \binits{L.}},
\bauthor{\bsnm{{Shivaei}}, \binits{I.}},
\bauthor{\bsnm{{Sun}}, \binits{F.}}:
\batitle{{A small and vigorous black hole in the early Universe}}.
\bjtitle{\nat}
\bvolume{627}(\bissue{8002}),
\bfpage{59}--\blpage{63}
(\byear{2024})
\doiurl{10.1038/s41586-024-07052-5}
{\href{https://arxiv.org/abs/2305.12492}{{arXiv:2305.12492}}}
{[astro-ph.GA]}
\end{barticle}
\endbibitem

\bibitem[\protect\citeauthoryear{{Cohon} et~al.}{2025}]{Cohon2025}
\begin{botherref}
\oauthor{\bsnm{{Cohon}}, \binits{J.}},
\oauthor{\bsnm{{Cain}}, \binits{C.}},
\oauthor{\bsnm{{Windhorst}}, \binits{R.}},
\oauthor{\bsnm{{D'Aloisio}}, \binits{A.}},
\oauthor{\bsnm{{Carleton}}, \binits{T.}},
\oauthor{\bsnm{{Zhu}}, \binits{Y.}}:
{A long time ago in an LAE far, far away: a signpost of early reionization or a nascent AGN at $z=13$?}
arXiv e-prints,
2508--05739
(2025)
\doiurl{10.48550/arXiv.2508.05739}
{\href{https://arxiv.org/abs/2508.05739}{{arXiv:2508.05739}}}
{[astro-ph.GA]}
\end{botherref}
\endbibitem

\bibitem[\protect\citeauthoryear{{Oke} and {Gunn}}{1983}]{OkeGunn1983}
\begin{barticle}
\bauthor{\bsnm{{Oke}}, \binits{J.B.}},
\bauthor{\bsnm{{Gunn}}, \binits{J.E.}}:
\batitle{{Secondary standard stars for absolute spectrophotometry.}}
\bjtitle{\apj}
\bvolume{266},
\bfpage{713}--\blpage{717}
(\byear{1983})
\doiurl{10.1086/160817}
\end{barticle}
\endbibitem

\bibitem[\protect\citeauthoryear{{Rigby} et~al.}{2023}]{Rigby2023}
\begin{botherref}
\oauthor{\bsnm{{Rigby}}, \binits{J.R.}},
\oauthor{\bsnm{{Vieira}}, \binits{J.D.}},
\oauthor{\bsnm{{Phadke}}, \binits{K.A.}},
\oauthor{\bsnm{{Hutchison}}, \binits{T.A.}},
\oauthor{\bsnm{{Welch}}, \binits{B.}},
\oauthor{\bsnm{{Cathey}}, \binits{J.}},
\oauthor{\bsnm{{Spilker}}, \binits{J.S.}},
\oauthor{\bsnm{{Gonzalez}}, \binits{A.H.}},
\oauthor{\bsnm{{Adhikari}}, \binits{P.}},
\oauthor{\bsnm{{Aravena}}, \binits{M.}},
\oauthor{\bsnm{{Bayliss}}, \binits{M.B.}},
\oauthor{\bsnm{{Birkin}}, \binits{J.E.}},
\oauthor{\bsnm{{Bursk}}, \binits{E.}},
\oauthor{\bsnm{{Chapman}}, \binits{S.C.}},
\oauthor{\bsnm{{Dahle}}, \binits{H.}},
\oauthor{\bsnm{{Elicker}}, \binits{L.A.}},
\oauthor{\bsnm{{Fischer}}, \binits{T.C.}},
\oauthor{\bsnm{{Florian}}, \binits{M.K.}},
\oauthor{\bsnm{{Gladders}}, \binits{M.D.}},
\oauthor{\bsnm{{Hayward}}, \binits{C.C.}},
\oauthor{\bsnm{{Hewald}}, \binits{R.}},
\oauthor{\bsnm{{Kettler}}, \binits{L.A.}},
\oauthor{\bsnm{{Khullar}}, \binits{G.}},
\oauthor{\bsnm{{Kim}}, \binits{S.}},
\oauthor{\bsnm{{Law}}, \binits{D.R.}},
\oauthor{\bsnm{{Mahler}}, \binits{G.}},
\oauthor{\bsnm{{Malhotra}}, \binits{S.}},
\oauthor{\bsnm{{Murphy}}, \binits{E.J.}},
\oauthor{\bsnm{{Narayanan}}, \binits{D.}},
\oauthor{\bsnm{{Olivier}}, \binits{G.M.}},
\oauthor{\bsnm{{Rhoads}}, \binits{J.E.}},
\oauthor{\bsnm{{Sharon}}, \binits{K.}},
\oauthor{\bsnm{{Solimano}}, \binits{M.}},
\oauthor{\bsnm{{Thiruvengadam}}, \binits{A.}},
\oauthor{\bsnm{{Vizgan}}, \binits{D.}},
\oauthor{\bsnm{{Younker}}, \binits{N.}}:
{JWST Early Release Science Program TEMPLATES: Targeting Extremely Magnified Panchromatic Lensed Arcs and their Extended Star formation}.
arXiv e-prints,
2312--10465
(2023)
\doiurl{10.48550/arXiv.2312.10465}
{\href{https://arxiv.org/abs/2312.10465}{{arXiv:2312.10465}}}
{[astro-ph.GA]}
\end{botherref}
\endbibitem

\bibitem[\protect\citeauthoryear{{Loiacono} et~al.}{2024}]{Loiacono2024}
\begin{barticle}
\bauthor{\bsnm{{Loiacono}}, \binits{F.}},
\bauthor{\bsnm{{Decarli}}, \binits{R.}},
\bauthor{\bsnm{{Mignoli}}, \binits{M.}},
\bauthor{\bsnm{{Farina}}, \binits{E.P.}},
\bauthor{\bsnm{{Ba{\~n}ados}}, \binits{E.}},
\bauthor{\bsnm{{Bosman}}, \binits{S.}},
\bauthor{\bsnm{{Eilers}}, \binits{A.-C.}},
\bauthor{\bsnm{{Schindler}}, \binits{J.-T.}},
\bauthor{\bsnm{{Strauss}}, \binits{M.A.}},
\bauthor{\bsnm{{Vestergaard}}, \binits{M.}},
\bauthor{\bsnm{{Wang}}, \binits{F.}},
\bauthor{\bsnm{{Blecha}}, \binits{L.}},
\bauthor{\bsnm{{Carilli}}, \binits{C.L.}},
\bauthor{\bsnm{{Comastri}}, \binits{A.}},
\bauthor{\bsnm{{Connor}}, \binits{T.}},
\bauthor{\bsnm{{Costa}}, \binits{T.}},
\bauthor{\bsnm{{Dotti}}, \binits{M.}},
\bauthor{\bsnm{{Fan}}, \binits{X.}},
\bauthor{\bsnm{{Gilli}}, \binits{R.}},
\bauthor{\bsnm{{Jun}}, \binits{H.D.}},
\bauthor{\bsnm{{Liu}}, \binits{W.}},
\bauthor{\bsnm{{Lupi}}, \binits{A.}},
\bauthor{\bsnm{{Marshall}}, \binits{M.A.}},
\bauthor{\bsnm{{Mazzucchelli}}, \binits{C.}},
\bauthor{\bsnm{{Meyer}}, \binits{R.A.}},
\bauthor{\bsnm{{Neeleman}}, \binits{M.}},
\bauthor{\bsnm{{Overzier}}, \binits{R.}},
\bauthor{\bsnm{{Pensabene}}, \binits{A.}},
\bauthor{\bsnm{{Riechers}}, \binits{D.A.}},
\bauthor{\bsnm{{Trakhtenbrot}}, \binits{B.}},
\bauthor{\bsnm{{Trebitsch}}, \binits{M.}},
\bauthor{\bsnm{{Venemans}}, \binits{B.}},
\bauthor{\bsnm{{Walter}}, \binits{F.}},
\bauthor{\bsnm{{Yang}}, \binits{J.}}:
\batitle{{A quasar-galaxy merger at z {\ensuremath{\sim}} 6.2: Black hole mass and quasar properties from the NIRSpec spectrum}}.
\bjtitle{\aap}
\bvolume{685},
\bfpage{121}
(\byear{2024})
\doiurl{10.1051/0004-6361/202348535}
{\href{https://arxiv.org/abs/2402.13319}{{arXiv:2402.13319}}}
{[astro-ph.GA]}
\end{barticle}
\endbibitem

\bibitem[\protect\citeauthoryear{{Perna} et~al.}{2023}]{Perna2023}
\begin{barticle}
\bauthor{\bsnm{{Perna}}, \binits{M.}},
\bauthor{\bsnm{{Arribas}}, \binits{S.}},
\bauthor{\bsnm{{Marshall}}, \binits{M.}},
\bauthor{\bsnm{{D'Eugenio}}, \binits{F.}},
\bauthor{\bsnm{{{\"U}bler}}, \binits{H.}},
\bauthor{\bsnm{{Bunker}}, \binits{A.}},
\bauthor{\bsnm{{Charlot}}, \binits{S.}},
\bauthor{\bsnm{{Carniani}}, \binits{S.}},
\bauthor{\bsnm{{Jakobsen}}, \binits{P.}},
\bauthor{\bsnm{{Maiolino}}, \binits{R.}},
\bauthor{\bsnm{al.}}:
\batitle{{GA-NIFS: The ultra-dense, interacting environment of a dual AGN at z {\ensuremath{\sim}} 3.3 revealed by JWST/NIRSpec IFS}}.
\bjtitle{\aap}
\bvolume{679},
\bfpage{89}
(\byear{2023})
\doiurl{10.1051/0004-6361/202346649}
{\href{https://arxiv.org/abs/2304.06756}{{arXiv:2304.06756}}}
{[astro-ph.GA]}
\end{barticle}
\endbibitem

\bibitem[\protect\citeauthoryear{{Bianchin} et~al.}{2024}]{Bianchin2024}
\begin{barticle}
\bauthor{\bsnm{{Bianchin}}, \binits{M.}},
\bauthor{\bsnm{{U}}, \binits{V.}},
\bauthor{\bsnm{{Song}}, \binits{Y.}},
\bauthor{\bsnm{{Lai}}, \binits{T.S.-Y.}},
\bauthor{\bsnm{{Remigio}}, \binits{R.P.}},
\bauthor{\bsnm{{Barcos-Mu{\~n}oz}}, \binits{L.}},
\bauthor{\bsnm{{D{\'\i}az-Santos}}, \binits{T.}},
\bauthor{\bsnm{{Armus}}, \binits{L.}},
\bauthor{\bsnm{{Inami}}, \binits{H.}},
\bauthor{\bsnm{{Larson}}, \binits{K.L.}},
\bauthor{\bsnm{al.}}:
\batitle{{GOALS-JWST: Gas Dynamics and Excitation in NGC 7469 Revealed by NIRSpec}}.
\bjtitle{\apj}
\bvolume{965}(\bissue{2}),
\bfpage{103}
(\byear{2024})
\doiurl{10.3847/1538-4357/ad2a50}
{\href{https://arxiv.org/abs/2308.00209}{{arXiv:2308.00209}}}
{[astro-ph.GA]}
\end{barticle}
\endbibitem

\bibitem[\protect\citeauthoryear{{Davies} et~al.}{2019}]{Davies2019_outflows}
\begin{barticle}
\bauthor{\bsnm{{Davies}}, \binits{R.L.}},
\bauthor{\bsnm{{F{\"o}rster Schreiber}}, \binits{N.M.}},
\bauthor{\bsnm{{{\"U}bler}}, \binits{H.}},
\bauthor{\bsnm{{Genzel}}, \binits{R.}},
\bauthor{\bsnm{{Lutz}}, \binits{D.}},
\bauthor{\bsnm{{Renzini}}, \binits{A.}},
\bauthor{\bsnm{{Tacchella}}, \binits{S.}},
\bauthor{\bsnm{{Tacconi}}, \binits{L.J.}},
\bauthor{\bsnm{{Belli}}, \binits{S.}},
\bauthor{\bsnm{{Burkert}}, \binits{A.}},
\bauthor{\bsnm{{Carollo}}, \binits{C.M.}},
\bauthor{\bsnm{{Davies}}, \binits{R.I.}},
\bauthor{\bsnm{{Herrera-Camus}}, \binits{R.}},
\bauthor{\bsnm{{Lilly}}, \binits{S.J.}},
\bauthor{\bsnm{{Mancini}}, \binits{C.}},
\bauthor{\bsnm{{Naab}}, \binits{T.}},
\bauthor{\bsnm{{Nelson}}, \binits{E.J.}},
\bauthor{\bsnm{{Price}}, \binits{S.H.}},
\bauthor{\bsnm{{Shimizu}}, \binits{T.T.}},
\bauthor{\bsnm{{Sternberg}}, \binits{A.}},
\bauthor{\bsnm{{Wisnioski}}, \binits{E.}},
\bauthor{\bsnm{{Wuyts}}, \binits{S.}}:
\batitle{{Kiloparsec Scale Properties of Star Formation Driven Outflows at z {\ensuremath{\sim}} 2.3 in the SINS/zC-SINF AO Survey}}.
\bjtitle{\apj}
\bvolume{873}(\bissue{2}),
\bfpage{122}
(\byear{2019})
\doiurl{10.3847/1538-4357/ab06f1}
{\href{https://arxiv.org/abs/1808.10700}{{arXiv:1808.10700}}}
{[astro-ph.GA]}
\end{barticle}
\endbibitem

\bibitem[\protect\citeauthoryear{{Llerena} et~al.}{2023}]{Llerena2023}
\begin{barticle}
\bauthor{\bsnm{{Llerena}}, \binits{M.}},
\bauthor{\bsnm{{Amor{\'\i}n}}, \binits{R.}},
\bauthor{\bsnm{{Pentericci}}, \binits{L.}},
\bauthor{\bsnm{{Calabr{\`o}}}, \binits{A.}},
\bauthor{\bsnm{{Shapley}}, \binits{A.E.}},
\bauthor{\bsnm{{Boutsia}}, \binits{K.}},
\bauthor{\bsnm{{P{\'e}rez-Montero}}, \binits{E.}},
\bauthor{\bsnm{{V{\'\i}lchez}}, \binits{J.M.}},
\bauthor{\bsnm{{Nakajima}}, \binits{K.}}:
\batitle{{Ionized gas kinematics and chemical abundances of low-mass star-forming galaxies at z {\ensuremath{\sim}} 3}}.
\bjtitle{\aap}
\bvolume{676},
\bfpage{53}
(\byear{2023})
\doiurl{10.1051/0004-6361/202346232}
{\href{https://arxiv.org/abs/2303.01536}{{arXiv:2303.01536}}}
{[astro-ph.GA]}
\end{barticle}
\endbibitem

\bibitem[\protect\citeauthoryear{{Matthee} et~al.}{2024}]{Matthee2024}
\begin{barticle}
\bauthor{\bsnm{{Matthee}}, \binits{J.}},
\bauthor{\bsnm{{Naidu}}, \binits{R.P.}},
\bauthor{\bsnm{{Brammer}}, \binits{G.}},
\bauthor{\bsnm{{Chisholm}}, \binits{J.}},
\bauthor{\bsnm{{Eilers}}, \binits{A.-C.}},
\bauthor{\bsnm{{Goulding}}, \binits{A.}},
\bauthor{\bsnm{{Greene}}, \binits{J.}},
\bauthor{\bsnm{{Kashino}}, \binits{D.}},
\bauthor{\bsnm{{Labbe}}, \binits{I.}},
\bauthor{\bsnm{{Lilly}}, \binits{S.J.}},
\bauthor{\bsnm{{Mackenzie}}, \binits{R.}},
\bauthor{\bsnm{{Oesch}}, \binits{P.A.}},
\bauthor{\bsnm{{Weibel}}, \binits{A.}},
\bauthor{\bsnm{{Wuyts}}, \binits{S.}},
\bauthor{\bsnm{{Xiao}}, \binits{M.}},
\bauthor{\bsnm{{Bordoloi}}, \binits{R.}},
\bauthor{\bsnm{{Bouwens}}, \binits{R.}},
\bauthor{\bsnm{{van Dokkum}}, \binits{P.}},
\bauthor{\bsnm{{Illingworth}}, \binits{G.}},
\bauthor{\bsnm{{Kramarenko}}, \binits{I.}},
\bauthor{\bsnm{{Maseda}}, \binits{M.V.}},
\bauthor{\bsnm{{Mason}}, \binits{C.}},
\bauthor{\bsnm{{Meyer}}, \binits{R.A.}},
\bauthor{\bsnm{{Nelson}}, \binits{E.J.}},
\bauthor{\bsnm{{Reddy}}, \binits{N.A.}},
\bauthor{\bsnm{{Shivaei}}, \binits{I.}},
\bauthor{\bsnm{{Simcoe}}, \binits{R.A.}},
\bauthor{\bsnm{{Yue}}, \binits{M.}}:
\batitle{{Little Red Dots: An Abundant Population of Faint Active Galactic Nuclei at z {\ensuremath{\sim}} 5 Revealed by the EIGER and FRESCO JWST Surveys}}.
\bjtitle{\apj}
\bvolume{963}(\bissue{2}),
\bfpage{129}
(\byear{2024})
\doiurl{10.3847/1538-4357/ad2345}
{\href{https://arxiv.org/abs/2306.05448}{{arXiv:2306.05448}}}
{[astro-ph.GA]}
\end{barticle}
\endbibitem

\bibitem[\protect\citeauthoryear{{Kokorev} et~al.}{2024}]{Kokorev2024}
\begin{barticle}
\bauthor{\bsnm{{Kokorev}}, \binits{V.}},
\bauthor{\bsnm{{Caputi}}, \binits{K.I.}},
\bauthor{\bsnm{{Greene}}, \binits{J.E.}},
\bauthor{\bsnm{{Dayal}}, \binits{P.}},
\bauthor{\bsnm{{Trebitsch}}, \binits{M.}},
\bauthor{\bsnm{{Cutler}}, \binits{S.E.}},
\bauthor{\bsnm{{Fujimoto}}, \binits{S.}},
\bauthor{\bsnm{{Labb{\'e}}}, \binits{I.}},
\bauthor{\bsnm{{Miller}}, \binits{T.B.}},
\bauthor{\bsnm{{Iani}}, \binits{E.}},
\bauthor{\bsnm{{Navarro-Carrera}}, \binits{R.}},
\bauthor{\bsnm{{Rinaldi}}, \binits{P.}}:
\batitle{{A Census of Photometrically Selected Little Red Dots at 4 < z < 9 in JWST Blank Fields}}.
\bjtitle{\apj}
\bvolume{968}(\bissue{1}),
\bfpage{38}
(\byear{2024})
\doiurl{10.3847/1538-4357/ad4265}
{\href{https://arxiv.org/abs/2401.09981}{{arXiv:2401.09981}}}
{[astro-ph.GA]}
\end{barticle}
\endbibitem

\bibitem[\protect\citeauthoryear{{Kocevski} et~al.}{2025}]{Kocevski2025}
\begin{barticle}
\bauthor{\bsnm{{Kocevski}}, \binits{D.D.}},
\bauthor{\bsnm{{Finkelstein}}, \binits{S.L.}},
\bauthor{\bsnm{{Barro}}, \binits{G.}},
\bauthor{\bsnm{{Taylor}}, \binits{A.J.}},
\bauthor{\bsnm{{Calabr{\`o}}}, \binits{A.}},
\bauthor{\bsnm{{Laloux}}, \binits{B.}},
\bauthor{\bsnm{{Buchner}}, \binits{J.}},
\bauthor{\bsnm{{Trump}}, \binits{J.R.}},
\bauthor{\bsnm{{Leung}}, \binits{G.C.K.}},
\bauthor{\bsnm{{Yang}}, \binits{G.}},
\bauthor{\bsnm{{Dickinson}}, \binits{M.}},
\bauthor{\bsnm{{P{\'e}rez-Gonz{\'a}lez}}, \binits{P.G.}},
\bauthor{\bsnm{{Pacucci}}, \binits{F.}},
\bauthor{\bsnm{{Inayoshi}}, \binits{K.}},
\bauthor{\bsnm{{Somerville}}, \binits{R.S.}},
\bauthor{\bsnm{{McGrath}}, \binits{E.J.}},
\bauthor{\bsnm{{Akins}}, \binits{H.B.}},
\bauthor{\bsnm{{Bagley}}, \binits{M.B.}},
\bauthor{\bsnm{{Bowler}}, \binits{R.A.A.}},
\bauthor{\bsnm{{Bisigello}}, \binits{L.}},
\bauthor{\bsnm{{Carnall}}, \binits{A.}},
\bauthor{\bsnm{{Casey}}, \binits{C.M.}},
\bauthor{\bsnm{{Cheng}}, \binits{Y.}},
\bauthor{\bsnm{{Cleri}}, \binits{N.J.}},
\bauthor{\bsnm{{Costantin}}, \binits{L.}},
\bauthor{\bsnm{{Cullen}}, \binits{F.}},
\bauthor{\bsnm{{Davis}}, \binits{K.}},
\bauthor{\bsnm{{Donnan}}, \binits{C.T.}},
\bauthor{\bsnm{{Dunlop}}, \binits{J.S.}},
\bauthor{\bsnm{{Ellis}}, \binits{R.S.}},
\bauthor{\bsnm{{Ferguson}}, \binits{H.C.}},
\bauthor{\bsnm{{Fujimoto}}, \binits{S.}},
\bauthor{\bsnm{{Fontana}}, \binits{A.}},
\bauthor{\bsnm{{Giavalisco}}, \binits{M.}},
\bauthor{\bsnm{{Grazian}}, \binits{A.}},
\bauthor{\bsnm{{Grogin}}, \binits{N.A.}},
\bauthor{\bsnm{{Hathi}}, \binits{N.P.}},
\bauthor{\bsnm{{Hirschmann}}, \binits{M.}},
\bauthor{\bsnm{{Huertas-Company}}, \binits{M.}},
\bauthor{\bsnm{{Holwerda}}, \binits{B.W.}},
\bauthor{\bsnm{{Illingworth}}, \binits{G.}},
\bauthor{\bsnm{{Juneau}}, \binits{S.}},
\bauthor{\bsnm{{Kartaltepe}}, \binits{J.S.}},
\bauthor{\bsnm{{Koekemoer}}, \binits{A.M.}},
\bauthor{\bsnm{{Li}}, \binits{W.}},
\bauthor{\bsnm{{Lucas}}, \binits{R.A.}},
\bauthor{\bsnm{{Magee}}, \binits{D.}},
\bauthor{\bsnm{{Mason}}, \binits{C.}},
\bauthor{\bsnm{{McLeod}}, \binits{D.J.}},
\bauthor{\bsnm{{McLure}}, \binits{R.J.}},
\bauthor{\bsnm{{Napolitano}}, \binits{L.}},
\bauthor{\bsnm{{Papovich}}, \binits{C.}},
\bauthor{\bsnm{{Pirzkal}}, \binits{N.}},
\bauthor{\bsnm{{Rodighiero}}, \binits{G.}},
\bauthor{\bsnm{{Santini}}, \binits{P.}},
\bauthor{\bsnm{{Wilkins}}, \binits{S.M.}},
\bauthor{\bsnm{{Yung}}, \binits{L.Y.A.}}:
\batitle{{The Rise of Faint, Red Active Galactic Nuclei at z > 4: A Sample of Little Red Dots in the JWST Extragalactic Legacy Fields}}.
\bjtitle{\apj}
\bvolume{986}(\bissue{2}),
\bfpage{126}
(\byear{2025})
\doiurl{10.3847/1538-4357/adbc7d}
{\href{https://arxiv.org/abs/2404.03576}{{arXiv:2404.03576}}}
{[astro-ph.GA]}
\end{barticle}
\endbibitem

\bibitem[\protect\citeauthoryear{{de Graaff} et~al.}{2025}]{deGraaff2025}
\begin{botherref}
\oauthor{\bsnm{{de Graaff}}, \binits{A.}},
\oauthor{\bsnm{{Hviding}}, \binits{R.E.}},
\oauthor{\bsnm{{Naidu}}, \binits{R.P.}},
\oauthor{\bsnm{{Greene}}, \binits{J.E.}},
\oauthor{\bsnm{{Miller}}, \binits{T.B.}},
\oauthor{\bsnm{{Leja}}, \binits{J.}},
\oauthor{\bsnm{{Matthee}}, \binits{J.}},
\oauthor{\bsnm{{Brammer}}, \binits{G.}},
\oauthor{\bsnm{{Katz}}, \binits{H.}},
\oauthor{\bsnm{{Bezanson}}, \binits{R.}},
\oauthor{\bsnm{{Boogaard}}, \binits{L.A.}},
\oauthor{\bsnm{{Bose}}, \binits{S.}},
\oauthor{\bsnm{{Chisholm}}, \binits{J.}},
\oauthor{\bsnm{{Cleri}}, \binits{N.J.}},
\oauthor{\bsnm{{Dayal}}, \binits{P.}},
\oauthor{\bsnm{{Feldmann}}, \binits{R.}},
\oauthor{\bsnm{{Fudamoto}}, \binits{Y.}},
\oauthor{\bsnm{{Fujimoto}}, \binits{S.}},
\oauthor{\bsnm{{Furtak}}, \binits{L.J.}},
\oauthor{\bsnm{{Glazebrook}}, \binits{K.}},
\oauthor{\bsnm{{Gottumukkala}}, \binits{R.}},
\oauthor{\bsnm{{Heintz}}, \binits{K.E.}},
\oauthor{\bsnm{{Kokorev}}, \binits{V.}},
\oauthor{\bsnm{{Labbe}}, \binits{I.}},
\oauthor{\bsnm{{Maseda}}, \binits{M.V.}},
\oauthor{\bsnm{{McConachie}}, \binits{I.}},
\oauthor{\bsnm{{Nanayakkara}}, \binits{T.}},
\oauthor{\bsnm{{Nelson}}, \binits{E.}},
\oauthor{\bsnm{{Nowaczyk}}, \binits{P.}},
\oauthor{\bsnm{{Oesch}}, \binits{P.A.}},
\oauthor{\bsnm{{Rix}}, \binits{H.-W.}},
\oauthor{\bsnm{{Setton}}, \binits{D.J.}},
\oauthor{\bsnm{{Torralba}}, \binits{A.}},
\oauthor{\bsnm{{Walter}}, \binits{F.}},
\oauthor{\bsnm{{Wang}}, \binits{B.}},
\oauthor{\bsnm{{Weibel}}, \binits{A.}},
\oauthor{\bsnm{{van der Wel}}, \binits{A.}}:
{Little Red Dots host Black Hole Stars: A unified family of gas-reddened AGN revealed by JWST/NIRSpec spectroscopy}.
arXiv e-prints,
2511--21820
(2025)
\doiurl{10.48550/arXiv.2511.21820}
{\href{https://arxiv.org/abs/2511.21820}{{arXiv:2511.21820}}}
{[astro-ph.GA]}
\end{botherref}
\endbibitem

\bibitem[\protect\citeauthoryear{{Rusakov} et~al.}{2026}]{Rusakov2026}
\begin{barticle}
\bauthor{\bsnm{{Rusakov}}, \binits{V.}},
\bauthor{\bsnm{{Watson}}, \binits{D.}},
\bauthor{\bsnm{{Nikopoulos}}, \binits{G.P.}},
\bauthor{\bsnm{{Brammer}}, \binits{G.}},
\bauthor{\bsnm{{Gottumukkala}}, \binits{R.}},
\bauthor{\bsnm{{Harvey}}, \binits{T.}},
\bauthor{\bsnm{{Heintz}}, \binits{K.E.}},
\bauthor{\bsnm{{Damgaard}}, \binits{R.}},
\bauthor{\bsnm{{Sim}}, \binits{S.A.}},
\bauthor{\bsnm{{Sneppen}}, \binits{A.}},
\bauthor{\bsnm{al.}}:
\batitle{{Little red dots as young supermassive black holes in dense ionized cocoons}}.
\bjtitle{\nat}
\bvolume{649}(\bissue{8097}),
\bfpage{574}--\blpage{579}
(\byear{2026})
\doiurl{10.1038/s41586-025-09900-4}
{\href{https://arxiv.org/abs/2503.16595}{{arXiv:2503.16595}}}
{[astro-ph.GA]}
\end{barticle}
\endbibitem

\bibitem[\protect\citeauthoryear{{Chang} et~al.}{2026}]{Chang2026}
\begin{barticle}
\bauthor{\bsnm{{Chang}}, \binits{S.-J.}},
\bauthor{\bsnm{{Gronke}}, \binits{M.}},
\bauthor{\bsnm{{Matthee}}, \binits{J.}},
\bauthor{\bsnm{{Mason}}, \binits{C.}}:
\batitle{{Impact of resonance, Raman, and Thomson scattering on hydrogen line formation in Little Red Dots}}.
\bjtitle{\mnras}
\bvolume{545}(\bissue{4}),
\bfpage{2131}
(\byear{2026})
\doiurl{10.1093/mnras/staf2131}
{\href{https://arxiv.org/abs/2508.08768}{{arXiv:2508.08768}}}
{[astro-ph.GA]}
\end{barticle}
\endbibitem

\bibitem[\protect\citeauthoryear{{Setton} et~al.}{2025}]{Setton2025}
\begin{barticle}
\bauthor{\bsnm{{Setton}}, \binits{D.J.}},
\bauthor{\bsnm{{Greene}}, \binits{J.E.}},
\bauthor{\bsnm{{de Graaff}}, \binits{A.}},
\bauthor{\bsnm{{Ma}}, \binits{Y.}},
\bauthor{\bsnm{{Leja}}, \binits{J.}},
\bauthor{\bsnm{{Matthee}}, \binits{J.}},
\bauthor{\bsnm{{Bezanson}}, \binits{R.}},
\bauthor{\bsnm{{Boogaard}}, \binits{L.A.}},
\bauthor{\bsnm{{Cleri}}, \binits{N.J.}},
\bauthor{\bsnm{{Katz}}, \binits{H.}},
\bauthor{\bsnm{al.}}:
\batitle{{Little Red Dots at an Inflection Point: Ubiquitous V-shaped Turnover Consistently Occurs at the Balmer Limit}}.
\bjtitle{\apj}
\bvolume{995}(\bissue{1}),
\bfpage{118}
(\byear{2025})
\doiurl{10.3847/1538-4357/ae1500}
{\href{https://arxiv.org/abs/2411.03424}{{arXiv:2411.03424}}}
{[astro-ph.GA]}
\end{barticle}
\endbibitem

\bibitem[\protect\citeauthoryear{{Hviding} et~al.}{2025}]{Hviding2025}
\begin{barticle}
\bauthor{\bsnm{{Hviding}}, \binits{R.E.}},
\bauthor{\bsnm{{de Graaff}}, \binits{A.}},
\bauthor{\bsnm{{Miller}}, \binits{T.B.}},
\bauthor{\bsnm{{Setton}}, \binits{D.J.}},
\bauthor{\bsnm{{Greene}}, \binits{J.E.}},
\bauthor{\bsnm{{Labb{\'e}}}, \binits{I.}},
\bauthor{\bsnm{{Brammer}}, \binits{G.}},
\bauthor{\bsnm{{Bezanson}}, \binits{R.}},
\bauthor{\bsnm{{Boogaard}}, \binits{L.A.}},
\bauthor{\bsnm{{Cleri}}, \binits{N.J.}},
\bauthor{\bsnm{{Leja}}, \binits{J.}},
\bauthor{\bsnm{{Maseda}}, \binits{M.V.}},
\bauthor{\bsnm{{McConachie}}, \binits{I.}},
\bauthor{\bsnm{{Matthee}}, \binits{J.}},
\bauthor{\bsnm{{Naidu}}, \binits{R.P.}},
\bauthor{\bsnm{{Oesch}}, \binits{P.A.}},
\bauthor{\bsnm{{Wang}}, \binits{B.}},
\bauthor{\bsnm{{Whitaker}}, \binits{K.E.}},
\bauthor{\bsnm{{Williams}}, \binits{C.C.}}:
\batitle{{RUBIES: A spectroscopic census of little red dots: All point sources with v-shaped continua have broad lines}}.
\bjtitle{\aap}
\bvolume{702},
\bfpage{57}
(\byear{2025})
\doiurl{10.1051/0004-6361/202555816}
{\href{https://arxiv.org/abs/2506.05459}{{arXiv:2506.05459}}}
{[astro-ph.GA]}
\end{barticle}
\endbibitem

\bibitem[\protect\citeauthoryear{{Torralba} et~al.}{2026}]{Torralba-Torregrosa2026}
\begin{barticle}
\bauthor{\bsnm{{Torralba}}, \binits{A.}},
\bauthor{\bsnm{{Matthee}}, \binits{J.}},
\bauthor{\bsnm{{Pezzulli}}, \binits{G.}},
\bauthor{\bsnm{{Urrutia}}, \binits{T.}},
\bauthor{\bsnm{{Gronke}}, \binits{M.}},
\bauthor{\bsnm{{Mascia}}, \binits{S.}},
\bauthor{\bsnm{{D'Eugenio}}, \binits{F.}},
\bauthor{\bsnm{{Di Cesare}}, \binits{C.}},
\bauthor{\bsnm{{Eilers}}, \binits{A.-C.}},
\bauthor{\bsnm{{Greene}}, \binits{J.E.}},
\bauthor{\bsnm{al.}}:
\batitle{{A weak Ly{\ensuremath{\alpha}} halo for an extremely bright little red dot: Indications of enshrouded supermassive black hole growth}}.
\bjtitle{\aap}
\bvolume{705},
\bfpage{147}
(\byear{2026})
\doiurl{10.1051/0004-6361/202555596}
{\href{https://arxiv.org/abs/2505.09542}{{arXiv:2505.09542}}}
{[astro-ph.GA]}
\end{barticle}
\endbibitem

\bibitem[\protect\citeauthoryear{{Greene} et~al.}{2026}]{Greene2026}
\begin{barticle}
\bauthor{\bsnm{{Greene}}, \binits{J.E.}},
\bauthor{\bsnm{{Setton}}, \binits{D.J.}},
\bauthor{\bsnm{{Furtak}}, \binits{L.J.}},
\bauthor{\bsnm{{Naidu}}, \binits{R.P.}},
\bauthor{\bsnm{{Volonteri}}, \binits{M.}},
\bauthor{\bsnm{{Dayal}}, \binits{P.}},
\bauthor{\bsnm{{Labbe}}, \binits{I.}},
\bauthor{\bsnm{{van Dokkum}}, \binits{P.}},
\bauthor{\bsnm{{Bezanson}}, \binits{R.}},
\bauthor{\bsnm{{Brammer}}, \binits{G.}},
\bauthor{\bsnm{al.}}:
\batitle{{What You See Is What You Get: Empirically Measured Bolometric Luminosities of Little Red Dots}}.
\bjtitle{\apj}
\bvolume{996}(\bissue{2}),
\bfpage{129}
(\byear{2026})
\doiurl{10.3847/1538-4357/ae1836}
{\href{https://arxiv.org/abs/2509.05434}{{arXiv:2509.05434}}}
{[astro-ph.GA]}
\end{barticle}
\endbibitem

\bibitem[\protect\citeauthoryear{{Dong} et~al.}{2008}]{Dong2008}
\begin{barticle}
\bauthor{\bsnm{{Dong}}, \binits{X.}},
\bauthor{\bsnm{{Wang}}, \binits{T.}},
\bauthor{\bsnm{{Wang}}, \binits{J.}},
\bauthor{\bsnm{{Yuan}}, \binits{W.}},
\bauthor{\bsnm{{Zhou}}, \binits{H.}},
\bauthor{\bsnm{{Dai}}, \binits{H.}},
\bauthor{\bsnm{{Zhang}}, \binits{K.}}:
\batitle{{Broad-line Balmer decrements in blue active galactic nuclei}}.
\bjtitle{\mnras}
\bvolume{383}(\bissue{2}),
\bfpage{581}--\blpage{592}
(\byear{2008})
\doiurl{10.1111/j.1365-2966.2007.12560.x}
{\href{https://arxiv.org/abs/0710.1458}{{arXiv:0710.1458}}}
{[astro-ph]}
\end{barticle}
\endbibitem

\bibitem[\protect\citeauthoryear{{Gaskell} and {Ferland}}{1984}]{GaskellFerland1984}
\begin{barticle}
\bauthor{\bsnm{{Gaskell}}, \binits{C.M.}},
\bauthor{\bsnm{{Ferland}}, \binits{G.J.}}:
\batitle{{Theoretical hydrogen-line ratios for the narrow-line regions of active galactic nuclei}}.
\bjtitle{\pasp}
\bvolume{96},
\bfpage{393}--\blpage{397}
(\byear{1984})
\doiurl{10.1086/131352}
\end{barticle}
\endbibitem

\bibitem[\protect\citeauthoryear{{Stern} and {Laor}}{2012}]{Stern2012}
\begin{barticle}
\bauthor{\bsnm{{Stern}}, \binits{J.}},
\bauthor{\bsnm{{Laor}}, \binits{A.}}:
\batitle{{Type 1 AGN at low z- I. Emission properties}}.
\bjtitle{\mnras}
\bvolume{423}(\bissue{1}),
\bfpage{600}--\blpage{631}
(\byear{2012})
\doiurl{10.1111/j.1365-2966.2012.20901.x}
{\href{https://arxiv.org/abs/1203.3158}{{arXiv:1203.3158}}}
{[astro-ph.CO]}
\end{barticle}
\endbibitem

\bibitem[\protect\citeauthoryear{{Dalla Bont{\`a}} et~al.}{2020}]{DB2020}
\begin{barticle}
\bauthor{\bsnm{{Dalla Bont{\`a}}}, \binits{E.}},
\bauthor{\bsnm{{Peterson}}, \binits{B.M.}},
\bauthor{\bsnm{{Bentz}}, \binits{M.C.}},
\bauthor{\bsnm{{Brandt}}, \binits{W.N.}},
\bauthor{\bsnm{{Ciroi}}, \binits{S.}},
\bauthor{\bsnm{{De Rosa}}, \binits{G.}},
\bauthor{\bsnm{{Fonseca Alvarez}}, \binits{G.}},
\bauthor{\bsnm{{Grier}}, \binits{C.J.}},
\bauthor{\bsnm{{Hall}}, \binits{P.B.}},
\bauthor{\bsnm{{Hern{\'a}ndez Santisteban}}, \binits{J.V.}},
\bauthor{\bsnm{{Ho}}, \binits{L.C.}},
\bauthor{\bsnm{{Homayouni}}, \binits{Y.}},
\bauthor{\bsnm{{Horne}}, \binits{K.}},
\bauthor{\bsnm{{Kochanek}}, \binits{C.S.}},
\bauthor{\bsnm{{Li}}, \binits{J.I.-H.}},
\bauthor{\bsnm{{Morelli}}, \binits{L.}},
\bauthor{\bsnm{{Pizzella}}, \binits{A.}},
\bauthor{\bsnm{{Pogge}}, \binits{R.W.}},
\bauthor{\bsnm{{Schneider}}, \binits{D.P.}},
\bauthor{\bsnm{{Shen}}, \binits{Y.}},
\bauthor{\bsnm{{Trump}}, \binits{J.R.}},
\bauthor{\bsnm{{Vestergaard}}, \binits{M.}}:
\batitle{{The Sloan Digital Sky Survey Reverberation Mapping Project: Estimating Masses of Black Holes in Quasars with Single-epoch Spectroscopy}}.
\bjtitle{\apj}
\bvolume{903}(\bissue{2}),
\bfpage{112}
(\byear{2020})
\doiurl{10.3847/1538-4357/abbc1c}
{\href{https://arxiv.org/abs/2007.02963}{{arXiv:2007.02963}}}
{[astro-ph.GA]}
\end{barticle}
\endbibitem

\bibitem[\protect\citeauthoryear{{Pennell} et~al.}{2017}]{Pennell2017}
\begin{barticle}
\bauthor{\bsnm{{Pennell}}, \binits{A.}},
\bauthor{\bsnm{{Runnoe}}, \binits{J.C.}},
\bauthor{\bsnm{{Brotherton}}, \binits{M.S.}}:
\batitle{{Updating quasar bolometric luminosity corrections - III. [O III] bolometric corrections}}.
\bjtitle{\mnras}
\bvolume{468}(\bissue{2}),
\bfpage{1433}--\blpage{1441}
(\byear{2017})
\doiurl{10.1093/mnras/stx556}
{\href{https://arxiv.org/abs/1703.03431}{{arXiv:1703.03431}}}
{[astro-ph.GA]}
\end{barticle}
\endbibitem

\bibitem[\protect\citeauthoryear{{Richards} et~al.}{2006}]{Richards2006}
\begin{barticle}
\bauthor{\bsnm{{Richards}}, \binits{G.T.}},
\bauthor{\bsnm{{Lacy}}, \binits{M.}},
\bauthor{\bsnm{{Storrie-Lombardi}}, \binits{L.J.}},
\bauthor{\bsnm{{Hall}}, \binits{P.B.}},
\bauthor{\bsnm{{Gallagher}}, \binits{S.C.}},
\bauthor{\bsnm{{Hines}}, \binits{D.C.}},
\bauthor{\bsnm{{Fan}}, \binits{X.}},
\bauthor{\bsnm{{Papovich}}, \binits{C.}},
\bauthor{\bsnm{{Vanden Berk}}, \binits{D.E.}},
\bauthor{\bsnm{{Trammell}}, \binits{G.B.}},
\bauthor{\bsnm{al.}}:
\batitle{{Spectral Energy Distributions and Multiwavelength Selection of Type 1 Quasars}}.
\bjtitle{\apjs}
\bvolume{166}(\bissue{2}),
\bfpage{470}--\blpage{497}
(\byear{2006})
\doiurl{10.1086/506525}
{\href{https://arxiv.org/abs/astro-ph/0601558}{{arXiv:astro-ph/0601558}}}
{[astro-ph]}
\end{barticle}
\endbibitem

\bibitem[\protect\citeauthoryear{{Runnoe} et~al.}{2012}]{Runnoe2012}
\begin{barticle}
\bauthor{\bsnm{{Runnoe}}, \binits{J.C.}},
\bauthor{\bsnm{{Brotherton}}, \binits{M.S.}},
\bauthor{\bsnm{{Shang}}, \binits{Z.}}:
\batitle{{Updating quasar bolometric luminosity corrections}}.
\bjtitle{\mnras}
\bvolume{422}(\bissue{1}),
\bfpage{478}--\blpage{493}
(\byear{2012})
\doiurl{10.1111/j.1365-2966.2012.20620.x}
{\href{https://arxiv.org/abs/1201.5155}{{arXiv:1201.5155}}}
{[astro-ph.CO]}
\end{barticle}
\endbibitem

\bibitem[\protect\citeauthoryear{{Mazzolari} et~al.}{2024}]{Mazzolari2024}
\begin{botherref}
\oauthor{\bsnm{{Mazzolari}}, \binits{G.}},
\oauthor{\bsnm{{{\"U}bler}}, \binits{H.}},
\oauthor{\bsnm{{Maiolino}}, \binits{R.}},
\oauthor{\bsnm{{Ji}}, \binits{X.}},
\oauthor{\bsnm{{Nakajima}}, \binits{K.}},
\oauthor{\bsnm{{Feltre}}, \binits{A.}},
\oauthor{\bsnm{{Scholtz}}, \binits{J.}},
\oauthor{\bsnm{{D'Eugenio}}, \binits{F.}},
\oauthor{\bsnm{{Curti}}, \binits{M.}},
\oauthor{\bsnm{{Mignoli}}, \binits{M.}},
\oauthor{\bsnm{{Marconi}}, \binits{A.}}:
{New AGN diagnostic diagrams based on the [OIII]$\lambda 4363$ auroral line}.
arXiv e-prints,
2404--10811
(2024)
\doiurl{10.48550/arXiv.2404.10811}
{\href{https://arxiv.org/abs/2404.10811}{{arXiv:2404.10811}}}
{[astro-ph.GA]}
\end{botherref}
\endbibitem

\bibitem[\protect\citeauthoryear{{Izotov} et~al.}{2018}]{Izotov2018}
\begin{barticle}
\bauthor{\bsnm{{Izotov}}, \binits{Y.I.}},
\bauthor{\bsnm{{Worseck}}, \binits{G.}},
\bauthor{\bsnm{{Schaerer}}, \binits{D.}},
\bauthor{\bsnm{{Guseva}}, \binits{N.G.}},
\bauthor{\bsnm{{Thuan}}, \binits{T.X.}},
\bauthor{\bsnm{{Fricke}}, \binits{V.} \bsuffix{A.}},
\bauthor{\bsnm{{Orlitov{\'a}}}, \binits{I.}}:
\batitle{{Low-redshift Lyman continuum leaking galaxies with high [O III]/[O II] ratios}}.
\bjtitle{\mnras}
\bvolume{478}(\bissue{4}),
\bfpage{4851}--\blpage{4865}
(\byear{2018})
\doiurl{10.1093/mnras/sty1378}
{\href{https://arxiv.org/abs/1805.09865}{{arXiv:1805.09865}}}
{[astro-ph.GA]}
\end{barticle}
\endbibitem

\bibitem[\protect\citeauthoryear{{Flury} et~al.}{2022}]{Flury2022b}
\begin{barticle}
\bauthor{\bsnm{{Flury}}, \binits{S.R.}},
\bauthor{\bsnm{{Jaskot}}, \binits{A.E.}},
\bauthor{\bsnm{{Ferguson}}, \binits{H.C.}},
\bauthor{\bsnm{{Worseck}}, \binits{G.}},
\bauthor{\bsnm{{Makan}}, \binits{K.}},
\bauthor{\bsnm{{Chisholm}}, \binits{J.}},
\bauthor{\bsnm{{Saldana-Lopez}}, \binits{A.}},
\bauthor{\bsnm{{Schaerer}}, \binits{D.}},
\bauthor{\bsnm{{McCandliss}}, \binits{S.R.}},
\bauthor{\bsnm{{Xu}}, \binits{X.}},
\bauthor{\bsnm{al.}}:
\batitle{{The Low-redshift Lyman Continuum Survey. II. New Insights into LyC Diagnostics}}.
\bjtitle{\apj}
\bvolume{930}(\bissue{2}),
\bfpage{126}
(\byear{2022})
\doiurl{10.3847/1538-4357/ac61e4}
{\href{https://arxiv.org/abs/2203.15649}{{arXiv:2203.15649}}}
{[astro-ph.GA]}
\end{barticle}
\endbibitem

\bibitem[\protect\citeauthoryear{{Osterbrock} and {Ferland}}{2006}]{Osterbrock2006}
\begin{bbook}
\bauthor{\bsnm{{Osterbrock}}, \binits{D.E.}},
\bauthor{\bsnm{{Ferland}}, \binits{G.J.}}:
\bbtitle{{Astrophysics of Gaseous Nebulae and Active Galactic Nuclei}},
(\byear{2006})
\end{bbook}
\endbibitem

\bibitem[\protect\citeauthoryear{{Matthee} et~al.}{2023}]{Matthee2023}
\begin{barticle}
\bauthor{\bsnm{{Matthee}}, \binits{J.}},
\bauthor{\bsnm{{Mackenzie}}, \binits{R.}},
\bauthor{\bsnm{{Simcoe}}, \binits{R.A.}},
\bauthor{\bsnm{{Kashino}}, \binits{D.}},
\bauthor{\bsnm{{Lilly}}, \binits{S.J.}},
\bauthor{\bsnm{{Bordoloi}}, \binits{R.}},
\bauthor{\bsnm{{Eilers}}, \binits{A.-C.}}:
\batitle{{EIGER. II. First Spectroscopic Characterization of the Young Stars and Ionized Gas Associated with Strong H{\ensuremath{\beta}} and [O III] Line Emission in Galaxies at z = 5-7 with JWST}}.
\bjtitle{\apj}
\bvolume{950}(\bissue{1}),
\bfpage{67}
(\byear{2023})
\doiurl{10.3847/1538-4357/acc846}
{\href{https://arxiv.org/abs/2211.08255}{{arXiv:2211.08255}}}
{[astro-ph.GA]}
\end{barticle}
\endbibitem

\bibitem[\protect\citeauthoryear{{Meyer} et~al.}{2024}]{Meyer2024}
\begin{botherref}
\oauthor{\bsnm{{Meyer}}, \binits{R.A.}},
\oauthor{\bsnm{{Oesch}}, \binits{P.A.}},
\oauthor{\bsnm{{Giovinazzo}}, \binits{E.}},
\oauthor{\bsnm{{Weibel}}, \binits{A.}},
\oauthor{\bsnm{{Brammer}}, \binits{G.}},
\oauthor{\bsnm{{Matthee}}, \binits{J.}},
\oauthor{\bsnm{{Naidu}}, \binits{R.P.}},
\oauthor{\bsnm{{Bouwens}}, \binits{R.J.}},
\oauthor{\bsnm{{Chisholm}}, \binits{J.}},
\oauthor{\bsnm{{Covelo-Paz}}, \binits{A.}},
\oauthor{\bsnm{{Fudamoto}}, \binits{Y.}},
\oauthor{\bsnm{{Maseda}}, \binits{M.}},
\oauthor{\bsnm{{Nelson}}, \binits{E.}},
\oauthor{\bsnm{{Shivaei}}, \binits{I.}},
\oauthor{\bsnm{{Xiao}}, \binits{M.}},
\oauthor{\bsnm{{Herard-Demanche}}, \binits{T.}},
\oauthor{\bsnm{{Illingworth}}, \binits{G.D.}},
\oauthor{\bsnm{{Kerutt}}, \binits{J.}},
\oauthor{\bsnm{{Kramarenko}}, \binits{I.}},
\oauthor{\bsnm{{Labbe}}, \binits{I.}},
\oauthor{\bsnm{{Leonova}}, \binits{E.}},
\oauthor{\bsnm{{Magee}}, \binits{D.}},
\oauthor{\bsnm{{Matharu}}, \binits{J.}},
\oauthor{\bsnm{{Prieto Lyon}}, \binits{G.}},
\oauthor{\bsnm{{Reddy}}, \binits{N.}},
\oauthor{\bsnm{{Schaerer}}, \binits{D.}},
\oauthor{\bsnm{{Shapley}}, \binits{A.}},
\oauthor{\bsnm{{Stefanon}}, \binits{M.}},
\oauthor{\bsnm{{Wozniak}}, \binits{M.A.}},
\oauthor{\bsnm{{Wuyts}}, \binits{S.}}:
{JWST FRESCO: a comprehensive census of H$\beta$+[OIII] emitters at 6.8<z<9.0 in the GOODS fields}.
arXiv e-prints,
2405--05111
(2024)
\doiurl{10.48550/arXiv.2405.05111}
{\href{https://arxiv.org/abs/2405.05111}{{arXiv:2405.05111}}}
{[astro-ph.GA]}
\end{botherref}
\endbibitem

\bibitem[\protect\citeauthoryear{{Chabrier}}{2003}]{Chabrier2003}
\begin{barticle}
\bauthor{\bsnm{{Chabrier}}, \binits{G.}}:
\batitle{{Galactic Stellar and Substellar Initial Mass Function}}.
\bjtitle{\pasp}
\bvolume{115}(\bissue{809}),
\bfpage{763}--\blpage{795}
(\byear{2003})
\doiurl{10.1086/376392}
{\href{https://arxiv.org/abs/astro-ph/0304382}{{arXiv:astro-ph/0304382}}}
{[astro-ph]}
\end{barticle}
\endbibitem

\bibitem[\protect\citeauthoryear{{Kennicutt} and {Evans}}{2012}]{Kennicutt2012}
\begin{barticle}
\bauthor{\bsnm{{Kennicutt}}, \binits{R.C.}},
\bauthor{\bsnm{{Evans}}, \binits{N.J.}}:
\batitle{{Star Formation in the Milky Way and Nearby Galaxies}}.
\bjtitle{\araa}
\bvolume{50},
\bfpage{531}--\blpage{608}
(\byear{2012})
\doiurl{10.1146/annurev-astro-081811-125610}
{\href{https://arxiv.org/abs/1204.3552}{{arXiv:1204.3552}}}
{[astro-ph.GA]}
\end{barticle}
\endbibitem

\bibitem[\protect\citeauthoryear{{Carnall} et~al.}{2018}]{Carnall2018}
\begin{barticle}
\bauthor{\bsnm{{Carnall}}, \binits{A.C.}},
\bauthor{\bsnm{{McLure}}, \binits{R.J.}},
\bauthor{\bsnm{{Dunlop}}, \binits{J.S.}},
\bauthor{\bsnm{{Dav{\'e}}}, \binits{R.}}:
\batitle{{Inferring the star formation histories of massive quiescent galaxies with BAGPIPES: evidence for multiple quenching mechanisms}}.
\bjtitle{\mnras}
\bvolume{480}(\bissue{4}),
\bfpage{4379}--\blpage{4401}
(\byear{2018})
\doiurl{10.1093/mnras/sty2169}
{\href{https://arxiv.org/abs/1712.04452}{{arXiv:1712.04452}}}
{[astro-ph.GA]}
\end{barticle}
\endbibitem

\bibitem[\protect\citeauthoryear{{Torralba-Torregrosa} et~al.}{2024}]{Torralba-Torregrosa2024}
\begin{botherref}
\oauthor{\bsnm{{Torralba-Torregrosa}}, \binits{A.}},
\oauthor{\bsnm{{Matthee}}, \binits{J.}},
\oauthor{\bsnm{{Naidu}}, \binits{R.P.}},
\oauthor{\bsnm{{Mackenzie}}, \binits{R.}},
\oauthor{\bsnm{{Pezzulli}}, \binits{G.}},
\oauthor{\bsnm{{Hutter}}, \binits{A.}},
\oauthor{\bsnm{{Arnalte-Mur}}, \binits{P.}},
\oauthor{\bsnm{{Gurung-L{\'o}pez}}, \binits{S.}},
\oauthor{\bsnm{{Tacchella}}, \binits{S.}},
\oauthor{\bsnm{{Oesch}}, \binits{P.}},
\oauthor{\bsnm{{Kashino}}, \binits{D.}},
\oauthor{\bsnm{{Conroy}}, \binits{C.}},
\oauthor{\bsnm{{Sobral}}, \binits{D.}}:
{Anatomy of an ionized bubble: NIRCam grism spectroscopy of the $z=6.6$ double-peaked Lyman-$\alpha$ emitter COLA1 and its environment}.
arXiv e-prints,
2404--10040
(2024)
\doiurl{10.48550/arXiv.2404.10040}
{\href{https://arxiv.org/abs/2404.10040}{{arXiv:2404.10040}}}
{[astro-ph.GA]}
\end{botherref}
\endbibitem

\bibitem[\protect\citeauthoryear{{Tacchella} et~al.}{2022}]{Tacchella2022}
\begin{barticle}
\bauthor{\bsnm{{Tacchella}}, \binits{S.}},
\bauthor{\bsnm{{Finkelstein}}, \binits{S.L.}},
\bauthor{\bsnm{{Bagley}}, \binits{M.}},
\bauthor{\bsnm{{Dickinson}}, \binits{M.}},
\bauthor{\bsnm{{Ferguson}}, \binits{H.C.}},
\bauthor{\bsnm{{Giavalisco}}, \binits{M.}},
\bauthor{\bsnm{{Graziani}}, \binits{L.}},
\bauthor{\bsnm{{Grogin}}, \binits{N.A.}},
\bauthor{\bsnm{{Hathi}}, \binits{N.}},
\bauthor{\bsnm{{Hutchison}}, \binits{T.A.}},
\bauthor{\bsnm{{Jung}}, \binits{I.}},
\bauthor{\bsnm{{Koekemoer}}, \binits{A.M.}},
\bauthor{\bsnm{{Larson}}, \binits{R.L.}},
\bauthor{\bsnm{{Papovich}}, \binits{C.}},
\bauthor{\bsnm{{Pirzkal}}, \binits{N.}},
\bauthor{\bsnm{{Rojas-Ruiz}}, \binits{S.}},
\bauthor{\bsnm{{Song}}, \binits{M.}},
\bauthor{\bsnm{{Schneider}}, \binits{R.}},
\bauthor{\bsnm{{Somerville}}, \binits{R.S.}},
\bauthor{\bsnm{{Wilkins}}, \binits{S.M.}},
\bauthor{\bsnm{{Yung}}, \binits{L.Y.A.}}:
\batitle{{On the Stellar Populations of Galaxies at z = 9-11: The Growth of Metals and Stellar Mass at Early Times}}.
\bjtitle{\apj}
\bvolume{927}(\bissue{2}),
\bfpage{170}
(\byear{2022})
\doiurl{10.3847/1538-4357/ac4cad}
{\href{https://arxiv.org/abs/2111.05351}{{arXiv:2111.05351}}}
{[astro-ph.GA]}
\end{barticle}
\endbibitem

\bibitem[\protect\citeauthoryear{{Calzetti} et~al.}{2000}]{Calzetti2000}
\begin{barticle}
\bauthor{\bsnm{{Calzetti}}, \binits{D.}},
\bauthor{\bsnm{{Armus}}, \binits{L.}},
\bauthor{\bsnm{{Bohlin}}, \binits{R.C.}},
\bauthor{\bsnm{{Kinney}}, \binits{A.L.}},
\bauthor{\bsnm{{Koornneef}}, \binits{J.}},
\bauthor{\bsnm{{Storchi-Bergmann}}, \binits{T.}}:
\batitle{{The Dust Content and Opacity of Actively Star-forming Galaxies}}.
\bjtitle{\apj}
\bvolume{533}(\bissue{2}),
\bfpage{682}--\blpage{695}
(\byear{2000})
\doiurl{10.1086/308692}
{\href{https://arxiv.org/abs/astro-ph/9911459}{{arXiv:astro-ph/9911459}}}
{[astro-ph]}
\end{barticle}
\endbibitem

\bibitem[\protect\citeauthoryear{{Cappellari} et~al.}{2006}]{Cappellari2006}
\begin{barticle}
\bauthor{\bsnm{{Cappellari}}, \binits{M.}},
\bauthor{\bsnm{{Bacon}}, \binits{R.}},
\bauthor{\bsnm{{Bureau}}, \binits{M.}},
\bauthor{\bsnm{{Damen}}, \binits{M.C.}},
\bauthor{\bsnm{{Davies}}, \binits{R.L.}},
\bauthor{\bsnm{{de Zeeuw}}, \binits{P.T.}},
\bauthor{\bsnm{{Emsellem}}, \binits{E.}},
\bauthor{\bsnm{{Falc{\'o}n-Barroso}}, \binits{J.}},
\bauthor{\bsnm{{Krajnovi{\'c}}}, \binits{D.}},
\bauthor{\bsnm{{Kuntschner}}, \binits{H.}},
\bauthor{\bsnm{{McDermid}}, \binits{R.M.}},
\bauthor{\bsnm{{Peletier}}, \binits{R.F.}},
\bauthor{\bsnm{{Sarzi}}, \binits{M.}},
\bauthor{\bsnm{{van den Bosch}}, \binits{R.C.E.}},
\bauthor{\bsnm{{van de Ven}}, \binits{G.}}:
\batitle{{The SAURON project - IV. The mass-to-light ratio, the virial mass estimator and the Fundamental Plane of elliptical and lenticular galaxies}}.
\bjtitle{\mnras}
\bvolume{366}(\bissue{4}),
\bfpage{1126}--\blpage{1150}
(\byear{2006})
\doiurl{10.1111/j.1365-2966.2005.09981.x}
{\href{https://arxiv.org/abs/astro-ph/0505042}{{arXiv:astro-ph/0505042}}}
{[astro-ph]}
\end{barticle}
\endbibitem

\bibitem[\protect\citeauthoryear{{van der Wel} et~al.}{2022}]{vanderWel2022}
\begin{barticle}
\bauthor{\bsnm{{van der Wel}}, \binits{A.}},
\bauthor{\bsnm{{van Houdt}}, \binits{J.}},
\bauthor{\bsnm{{Bezanson}}, \binits{R.}},
\bauthor{\bsnm{{Franx}}, \binits{M.}},
\bauthor{\bsnm{{D'Eugenio}}, \binits{F.}},
\bauthor{\bsnm{{Straatman}}, \binits{C.}},
\bauthor{\bsnm{{Bell}}, \binits{E.F.}},
\bauthor{\bsnm{{Muzzin}}, \binits{A.}},
\bauthor{\bsnm{{Sobral}}, \binits{D.}},
\bauthor{\bsnm{{Maseda}}, \binits{M.V.}},
\bauthor{\bsnm{al.}}:
\batitle{{The Mass Scale of High-redshift Galaxies: Virial Mass Estimates Calibrated with Stellar Dynamical Models from LEGA-C}}.
\bjtitle{\apj}
\bvolume{936}(\bissue{1}),
\bfpage{9}
(\byear{2022})
\doiurl{10.3847/1538-4357/ac83c5}
{\href{https://arxiv.org/abs/2208.12605}{{arXiv:2208.12605}}}
{[astro-ph.GA]}
\end{barticle}
\endbibitem

\bibitem[\protect\citeauthoryear{{Bezanson} et~al.}{2018}]{Bezanson2018}
\begin{barticle}
\bauthor{\bsnm{{Bezanson}}, \binits{R.}},
\bauthor{\bsnm{{van der Wel}}, \binits{A.}},
\bauthor{\bsnm{{Straatman}}, \binits{C.}},
\bauthor{\bsnm{{Pacifici}}, \binits{C.}},
\bauthor{\bsnm{{Wu}}, \binits{P.-F.}},
\bauthor{\bsnm{{Bari{\v{s}}i{\'c}}}, \binits{I.}},
\bauthor{\bsnm{{Bell}}, \binits{E.F.}},
\bauthor{\bsnm{{Conroy}}, \binits{C.}},
\bauthor{\bsnm{{D'Eugenio}}, \binits{F.}},
\bauthor{\bsnm{{Franx}}, \binits{M.}},
\bauthor{\bsnm{al.}}:
\batitle{{1D Kinematics from Stars and Ionized Gas at z {\ensuremath{\sim}} 0.8 from the LEGA-C Spectroscopic Survey of Massive Galaxies}}.
\bjtitle{\apjl}
\bvolume{868}(\bissue{2}),
\bfpage{36}
(\byear{2018})
\doiurl{10.3847/2041-8213/aaf16b}
{\href{https://arxiv.org/abs/1811.07900}{{arXiv:1811.07900}}}
{[astro-ph.GA]}
\end{barticle}
\endbibitem

\bibitem[\protect\citeauthoryear{{Fiore} et~al.}{2017}]{Fiore2017}
\begin{barticle}
\bauthor{\bsnm{{Fiore}}, \binits{F.}},
\bauthor{\bsnm{{Feruglio}}, \binits{C.}},
\bauthor{\bsnm{{Shankar}}, \binits{F.}},
\bauthor{\bsnm{{Bischetti}}, \binits{M.}},
\bauthor{\bsnm{{Bongiorno}}, \binits{A.}},
\bauthor{\bsnm{{Brusa}}, \binits{M.}},
\bauthor{\bsnm{{Carniani}}, \binits{S.}},
\bauthor{\bsnm{{Cicone}}, \binits{C.}},
\bauthor{\bsnm{{Duras}}, \binits{F.}},
\bauthor{\bsnm{{Lamastra}}, \binits{A.}},
\bauthor{\bsnm{{Mainieri}}, \binits{V.}},
\bauthor{\bsnm{{Marconi}}, \binits{A.}},
\bauthor{\bsnm{{Menci}}, \binits{N.}},
\bauthor{\bsnm{{Maiolino}}, \binits{R.}},
\bauthor{\bsnm{{Piconcelli}}, \binits{E.}},
\bauthor{\bsnm{{Vietri}}, \binits{G.}},
\bauthor{\bsnm{{Zappacosta}}, \binits{L.}}:
\batitle{{AGN wind scaling relations and the co-evolution of black holes and galaxies}}.
\bjtitle{\aap}
\bvolume{601},
\bfpage{143}
(\byear{2017})
\doiurl{10.1051/0004-6361/201629478}
{\href{https://arxiv.org/abs/1702.04507}{{arXiv:1702.04507}}}
{[astro-ph.GA]}
\end{barticle}
\endbibitem

\bibitem[\protect\citeauthoryear{Prochaska et~al.}{2020}]{pypeit:joss_pub}
\begin{barticle}
\bauthor{\bsnm{Prochaska}, \binits{J.X.}},
\bauthor{\bsnm{Hennawi}, \binits{J.F.}},
\bauthor{\bsnm{Westfall}, \binits{K.B.}},
\bauthor{\bsnm{Cooke}, \binits{R.J.}},
\bauthor{\bsnm{Wang}, \binits{F.}},
\bauthor{\bsnm{Hsyu}, \binits{T.}},
\bauthor{\bsnm{Davies}, \binits{F.B.}},
\bauthor{\bsnm{Farina}, \binits{E.P.}},
\bauthor{\bsnm{Pelliccia}, \binits{D.}}:
\batitle{Pypeit: The python spectroscopic data reduction pipeline}.
\bjtitle{Journal of Open Source Software}
\bvolume{5}(\bissue{56}),
\bfpage{2308}
(\byear{2020})
\doiurl{10.21105/joss.02308}
\end{barticle}
\endbibitem

\bibitem[\protect\citeauthoryear{{Prochaska} et~al.}{2020}]{pypeit:zenodo}
\begin{botherref}
\oauthor{\bsnm{{Prochaska}}, \binits{J.X.}},
\oauthor{\bsnm{{Hennawi}}, \binits{J.}},
\oauthor{\bsnm{{Cooke}}, \binits{R.}},
\oauthor{\bsnm{{Westfall}}, \binits{K.}},
\oauthor{\bsnm{{Wang}}, \binits{F.}},
\oauthor{\bsnm{{EmAstro}}},
\oauthor{\bsnm{{Tiffanyhsyu}}},
\oauthor{\bsnm{{Wasserman}}, \binits{A.}},
\oauthor{\bsnm{{Villaume}}, \binits{A.}},
\oauthor{\bsnm{{Marijana777}}},
\oauthor{\bsnm{{Schindler}}, \binits{J.}},
\oauthor{\bsnm{{Young}}, \binits{D.}},
\oauthor{\bsnm{{Simha}}, \binits{S.}},
\oauthor{\bsnm{{Wilde}}, \binits{M.}},
\oauthor{\bsnm{{Tejos}}, \binits{N.}},
\oauthor{\bsnm{{Isbell}}, \binits{J.}},
\oauthor{\bsnm{{Fl{\"o}rs}}, \binits{A.}},
\oauthor{\bsnm{{Sandford}}, \binits{N.}},
\oauthor{\bsnm{{Vasovi{\'c}}}, \binits{Z.}},
\oauthor{\bsnm{{Betts}}, \binits{E.}},
\oauthor{\bsnm{{Holden}}, \binits{B.}}:
{pypeit/PypeIt: Release 1.0.0}.
Zenodo
(2020).
\doiurl{10.5281/zenodo.3743493}
\end{botherref}
\endbibitem

\bibitem[\protect\citeauthoryear{{Verhamme} et~al.}{2017}]{Verhamme2017}
\begin{barticle}
\bauthor{\bsnm{{Verhamme}}, \binits{A.}},
\bauthor{\bsnm{{Orlitov{\'a}}}, \binits{I.}},
\bauthor{\bsnm{{Schaerer}}, \binits{D.}},
\bauthor{\bsnm{{Izotov}}, \binits{Y.}},
\bauthor{\bsnm{{Worseck}}, \binits{G.}},
\bauthor{\bsnm{{Thuan}}, \binits{T.X.}},
\bauthor{\bsnm{{Guseva}}, \binits{N.}}:
\batitle{{Lyman-{\ensuremath{\alpha}} spectral properties of five newly discovered Lyman continuum emitters}}.
\bjtitle{\aap}
\bvolume{597},
\bfpage{13}
(\byear{2017})
\doiurl{10.1051/0004-6361/201629264}
{\href{https://arxiv.org/abs/1609.03477}{{arXiv:1609.03477}}}
{[astro-ph.GA]}
\end{barticle}
\endbibitem

\bibitem[\protect\citeauthoryear{{Simmonds} et~al.}{2024}]{Simmonds2024}
\begin{barticle}
\bauthor{\bsnm{{Simmonds}}, \binits{C.}},
\bauthor{\bsnm{{Tacchella}}, \binits{S.}},
\bauthor{\bsnm{{Hainline}}, \binits{K.}},
\bauthor{\bsnm{{Johnson}}, \binits{B.D.}},
\bauthor{\bsnm{{Pusk{\'a}s}}, \binits{D.}},
\bauthor{\bsnm{{Robertson}}, \binits{B.}},
\bauthor{\bsnm{{Baker}}, \binits{W.M.}},
\bauthor{\bsnm{{Bhatawdekar}}, \binits{R.}},
\bauthor{\bsnm{{Boyett}}, \binits{K.}},
\bauthor{\bsnm{{Bunker}}, \binits{A.J.}},
\bauthor{\bsnm{{Cargile}}, \binits{P.A.}},
\bauthor{\bsnm{{Carniani}}, \binits{S.}},
\bauthor{\bsnm{{Chevallard}}, \binits{J.}},
\bauthor{\bsnm{{Curti}}, \binits{M.}},
\bauthor{\bsnm{{Curtis-Lake}}, \binits{E.}},
\bauthor{\bsnm{{Ji}}, \binits{Z.}},
\bauthor{\bsnm{{Jones}}, \binits{G.C.}},
\bauthor{\bsnm{{Kumari}}, \binits{N.}},
\bauthor{\bsnm{{Laseter}}, \binits{I.}},
\bauthor{\bsnm{{Maiolino}}, \binits{R.}},
\bauthor{\bsnm{{Maseda}}, \binits{M.V.}},
\bauthor{\bsnm{{Rinaldi}}, \binits{P.}},
\bauthor{\bsnm{{Stoffers}}, \binits{A.}},
\bauthor{\bsnm{{{\"U}bler}}, \binits{H.}},
\bauthor{\bsnm{{Villanueva}}, \binits{N.C.}},
\bauthor{\bsnm{{Williams}}, \binits{C.C.}},
\bauthor{\bsnm{{Willott}}, \binits{C.}},
\bauthor{\bsnm{{Witstok}}, \binits{J.}},
\bauthor{\bsnm{{Zhu}}, \binits{Y.}}:
\batitle{{Ionizing properties of galaxies in JADES for a stellar mass complete sample: resolving the cosmic ionizing photon budget crisis at the Epoch of Reionization}}.
\bjtitle{\mnras}
\bvolume{535}(\bissue{4}),
\bfpage{2998}--\blpage{3019}
(\byear{2024})
\doiurl{10.1093/mnras/stae2537}
{\href{https://arxiv.org/abs/2409.01286}{{arXiv:2409.01286}}}
{[astro-ph.GA]}
\end{barticle}
\endbibitem

\bibitem[\protect\citeauthoryear{{Pahl} et~al.}{2025}]{Pahl2025}
\begin{barticle}
\bauthor{\bsnm{{Pahl}}, \binits{A.}},
\bauthor{\bsnm{{Topping}}, \binits{M.W.}},
\bauthor{\bsnm{{Shapley}}, \binits{A.}},
\bauthor{\bsnm{{Sanders}}, \binits{R.}},
\bauthor{\bsnm{{Reddy}}, \binits{N.A.}},
\bauthor{\bsnm{{Clarke}}, \binits{L.}},
\bauthor{\bsnm{{Kehoe}}, \binits{E.}},
\bauthor{\bsnm{{Bento}}, \binits{T.}},
\bauthor{\bsnm{{Brammer}}, \binits{G.}}:
\batitle{{A Spectroscopic Analysis of the Ionizing Photon Production Efficiency in JADES and CEERS: Implications for the Ionizing Photon Budget}}.
\bjtitle{\apj}
\bvolume{981}(\bissue{2}),
\bfpage{134}
(\byear{2025})
\doiurl{10.3847/1538-4357/adb1ab}
{\href{https://arxiv.org/abs/2407.03399}{{arXiv:2407.03399}}}
{[astro-ph.GA]}
\end{barticle}
\endbibitem

\bibitem[\protect\citeauthoryear{{Worseck} et~al.}{2014}]{Worseck2014}
\begin{barticle}
\bauthor{\bsnm{{Worseck}}, \binits{G.}},
\bauthor{\bsnm{{Prochaska}}, \binits{J.X.}},
\bauthor{\bsnm{{O'Meara}}, \binits{J.M.}},
\bauthor{\bsnm{{Becker}}, \binits{G.D.}},
\bauthor{\bsnm{{Ellison}}, \binits{S.L.}},
\bauthor{\bsnm{{Lopez}}, \binits{S.}},
\bauthor{\bsnm{{Meiksin}}, \binits{A.}},
\bauthor{\bsnm{{M{\'e}nard}}, \binits{B.}},
\bauthor{\bsnm{{Murphy}}, \binits{M.T.}},
\bauthor{\bsnm{{Fumagalli}}, \binits{M.}}:
\batitle{{The Giant Gemini GMOS survey of z$_{em}$ > 4.4 quasars - I. Measuring the mean free path across cosmic time}}.
\bjtitle{\mnras}
\bvolume{445}(\bissue{2}),
\bfpage{1745}--\blpage{1760}
(\byear{2014})
\doiurl{10.1093/mnras/stu1827}
{\href{https://arxiv.org/abs/1402.4154}{{arXiv:1402.4154}}}
{[astro-ph.CO]}
\end{barticle}
\endbibitem

\bibitem[\protect\citeauthoryear{{Becker} et~al.}{2015}]{Becker2015_rev}
\begin{barticle}
\bauthor{\bsnm{{Becker}}, \binits{G.D.}},
\bauthor{\bsnm{{Bolton}}, \binits{J.S.}},
\bauthor{\bsnm{{Lidz}}, \binits{A.}}:
\batitle{{Reionisation and High-Redshift Galaxies: The View from Quasar Absorption Lines}}.
\bjtitle{\pasa}
\bvolume{32},
\bfpage{045}
(\byear{2015})
\doiurl{10.1017/pasa.2015.45}
{\href{https://arxiv.org/abs/1510.03368}{{arXiv:1510.03368}}}
{[astro-ph.CO]}
\end{barticle}
\endbibitem

\bibitem[\protect\citeauthoryear{{Cristiani} et~al.}{2016}]{Cristiani2016}
\begin{barticle}
\bauthor{\bsnm{{Cristiani}}, \binits{S.}},
\bauthor{\bsnm{{Serrano}}, \binits{L.M.}},
\bauthor{\bsnm{{Fontanot}}, \binits{F.}},
\bauthor{\bsnm{{Vanzella}}, \binits{E.}},
\bauthor{\bsnm{{Monaco}}, \binits{P.}}:
\batitle{{The spectral slope and escape fraction of bright quasars at z {\ensuremath{\sim}} 3.8: the contribution to the cosmic UV background}}.
\bjtitle{\mnras}
\bvolume{462}(\bissue{3}),
\bfpage{2478}--\blpage{2485}
(\byear{2016})
\doiurl{10.1093/mnras/stw1810}
{\href{https://arxiv.org/abs/1603.09351}{{arXiv:1603.09351}}}
{[astro-ph.CO]}
\end{barticle}
\endbibitem

\bibitem[\protect\citeauthoryear{{Grazian} et~al.}{2018}]{Grazian2018}
\begin{barticle}
\bauthor{\bsnm{{Grazian}}, \binits{A.}},
\bauthor{\bsnm{{Giallongo}}, \binits{E.}},
\bauthor{\bsnm{{Boutsia}}, \binits{K.}},
\bauthor{\bsnm{{Cristiani}}, \binits{S.}},
\bauthor{\bsnm{{Vanzella}}, \binits{E.}},
\bauthor{\bsnm{{Scarlata}}, \binits{C.}},
\bauthor{\bsnm{{Santini}}, \binits{P.}},
\bauthor{\bsnm{{Pentericci}}, \binits{L.}},
\bauthor{\bsnm{{Merlin}}, \binits{E.}},
\bauthor{\bsnm{{Menci}}, \binits{N.}},
\bauthor{\bsnm{{Fontanot}}, \binits{F.}},
\bauthor{\bsnm{{Fontana}}, \binits{A.}},
\bauthor{\bsnm{{Fiore}}, \binits{F.}},
\bauthor{\bsnm{{Civano}}, \binits{F.}},
\bauthor{\bsnm{{Castellano}}, \binits{M.}},
\bauthor{\bsnm{{Brusa}}, \binits{M.}},
\bauthor{\bsnm{{Bonchi}}, \binits{A.}},
\bauthor{\bsnm{{Carini}}, \binits{R.}},
\bauthor{\bsnm{{Cusano}}, \binits{F.}},
\bauthor{\bsnm{{Faccini}}, \binits{M.}},
\bauthor{\bsnm{{Garilli}}, \binits{B.}},
\bauthor{\bsnm{{Marchetti}}, \binits{A.}},
\bauthor{\bsnm{{Rossi}}, \binits{A.}},
\bauthor{\bsnm{{Speziali}}, \binits{R.}}:
\batitle{{The contribution of faint AGNs to the ionizing background at z 4}}.
\bjtitle{\aap}
\bvolume{613},
\bfpage{44}
(\byear{2018})
\doiurl{10.1051/0004-6361/201732385}
{\href{https://arxiv.org/abs/1802.01953}{{arXiv:1802.01953}}}
{[astro-ph.GA]}
\end{barticle}
\endbibitem

\bibitem[\protect\citeauthoryear{{Romano} et~al.}{2019}]{Romano2019}
\begin{barticle}
\bauthor{\bsnm{{Romano}}, \binits{M.}},
\bauthor{\bsnm{{Grazian}}, \binits{A.}},
\bauthor{\bsnm{{Giallongo}}, \binits{E.}},
\bauthor{\bsnm{{Cristiani}}, \binits{S.}},
\bauthor{\bsnm{{Fontanot}}, \binits{F.}},
\bauthor{\bsnm{{Boutsia}}, \binits{K.}},
\bauthor{\bsnm{{Fiore}}, \binits{F.}},
\bauthor{\bsnm{{Menci}}, \binits{N.}}:
\batitle{{Lyman continuum escape fraction and mean free path of hydrogen ionizing photons for bright z {\ensuremath{\sim}} 4 QSOs from SDSS DR14}}.
\bjtitle{\aap}
\bvolume{632},
\bfpage{45}
(\byear{2019})
\doiurl{10.1051/0004-6361/201935550}
{\href{https://arxiv.org/abs/1910.02775}{{arXiv:1910.02775}}}
{[astro-ph.GA]}
\end{barticle}
\endbibitem

\bibitem[\protect\citeauthoryear{{Wang} et~al.}{2019}]{Wang2019}
\begin{barticle}
\bauthor{\bsnm{{Wang}}, \binits{F.}},
\bauthor{\bsnm{{Yang}}, \binits{J.}},
\bauthor{\bsnm{{Fan}}, \binits{X.}},
\bauthor{\bsnm{{Wu}}, \binits{X.-B.}},
\bauthor{\bsnm{{Yue}}, \binits{M.}},
\bauthor{\bsnm{{Li}}, \binits{J.-T.}},
\bauthor{\bsnm{{Bian}}, \binits{F.}},
\bauthor{\bsnm{{Jiang}}, \binits{L.}},
\bauthor{\bsnm{{Ba{\~n}ados}}, \binits{E.}},
\bauthor{\bsnm{{Schindler}}, \binits{J.-T.}},
\bauthor{\bsnm{{Findlay}}, \binits{J.R.}},
\bauthor{\bsnm{{Davies}}, \binits{F.B.}},
\bauthor{\bsnm{{Decarli}}, \binits{R.}},
\bauthor{\bsnm{{Farina}}, \binits{E.P.}},
\bauthor{\bsnm{{Green}}, \binits{R.}},
\bauthor{\bsnm{{Hennawi}}, \binits{J.F.}},
\bauthor{\bsnm{{Huang}}, \binits{Y.-H.}},
\bauthor{\bsnm{{Mazzuccheli}}, \binits{C.}},
\bauthor{\bsnm{{McGreer}}, \binits{I.D.}},
\bauthor{\bsnm{{Venemans}}, \binits{B.}},
\bauthor{\bsnm{{Walter}}, \binits{F.}},
\bauthor{\bsnm{{Dye}}, \binits{S.}},
\bauthor{\bsnm{{Lyke}}, \binits{B.W.}},
\bauthor{\bsnm{{Myers}}, \binits{A.D.}},
\bauthor{\bsnm{{Nunez}}, \binits{E.H.}}:
\batitle{{Exploring Reionization-era Quasars. III. Discovery of 16 Quasars at 6.4 {\ensuremath{\lesssim}} z {\ensuremath{\lesssim}} 6.9 with DESI Legacy Imaging Surveys and the UKIRT Hemisphere Survey and Quasar Luminosity Function at z {\ensuremath{\sim}} 6.7}}.
\bjtitle{\apj}
\bvolume{884}(\bissue{1}),
\bfpage{30}
(\byear{2019})
\doiurl{10.3847/1538-4357/ab2be5}
{\href{https://arxiv.org/abs/1810.11926}{{arXiv:1810.11926}}}
{[astro-ph.GA]}
\end{barticle}
\endbibitem

\bibitem[\protect\citeauthoryear{{Taylor} et~al.}{2025}]{Taylor2025}
\begin{barticle}
\bauthor{\bsnm{{Taylor}}, \binits{A.J.}},
\bauthor{\bsnm{{Barger}}, \binits{A.J.}},
\bauthor{\bsnm{{Cowie}}, \binits{L.L.}},
\bauthor{\bsnm{{Hu}}, \binits{E.M.}},
\bauthor{\bsnm{{Songaila}}, \binits{A.}}:
\batitle{{Spectroscopic Ultraluminous Ly{\ensuremath{\alpha}} Luminosity Functions at z = 5.7 and z = 6.6 from HEROES: Evidence for Ionized Bubbles}}.
\bjtitle{\apj}
\bvolume{989}(\bissue{1}),
\bfpage{31}
(\byear{2025})
\doiurl{10.3847/1538-4357/ade4c7}
{\href{https://arxiv.org/abs/2506.13854}{{arXiv:2506.13854}}}
{[astro-ph.GA]}
\end{barticle}
\endbibitem

\bibitem[\protect\citeauthoryear{{Kerutt} et~al.}{2022}]{Kerutt2022}
\begin{barticle}
\bauthor{\bsnm{{Kerutt}}, \binits{J.}},
\bauthor{\bsnm{{Wisotzki}}, \binits{L.}},
\bauthor{\bsnm{{Verhamme}}, \binits{A.}},
\bauthor{\bsnm{{Schmidt}}, \binits{K.B.}},
\bauthor{\bsnm{{Leclercq}}, \binits{F.}},
\bauthor{\bsnm{{Herenz}}, \binits{E.C.}},
\bauthor{\bsnm{{Urrutia}}, \binits{T.}},
\bauthor{\bsnm{{Garel}}, \binits{T.}},
\bauthor{\bsnm{{Hashimoto}}, \binits{T.}},
\bauthor{\bsnm{{Maseda}}, \binits{M.}},
\bauthor{\bsnm{{Matthee}}, \binits{J.}},
\bauthor{\bsnm{{Kusakabe}}, \binits{H.}},
\bauthor{\bsnm{{Schaye}}, \binits{J.}},
\bauthor{\bsnm{{Richard}}, \binits{J.}},
\bauthor{\bsnm{{Guiderdoni}}, \binits{B.}},
\bauthor{\bsnm{{Mauerhofer}}, \binits{V.}},
\bauthor{\bsnm{{Nanayakkara}}, \binits{T.}},
\bauthor{\bsnm{{Vitte}}, \binits{E.}}:
\batitle{{Equivalent widths of Lyman {\ensuremath{\alpha}} emitters in MUSE-Wide and MUSE-Deep}}.
\bjtitle{\aap}
\bvolume{659},
\bfpage{183}
(\byear{2022})
\doiurl{10.1051/0004-6361/202141900}
{\href{https://arxiv.org/abs/2202.06642}{{arXiv:2202.06642}}}
{[astro-ph.GA]}
\end{barticle}
\endbibitem

\end{thebibliography}

\end{document}